\documentclass[longauth]{aa}  
\usepackage{xcolor}
\usepackage{graphicx}
\usepackage{txfonts}
\usepackage{tikz,lipsum,lmodern}
\usepackage[most]{tcolorbox}
\usepackage{subcaption}
\usepackage{tabularx} 
\usepackage{booktabs}
\usepackage[rightcaption]{sidecap}
\usepackage[colorlinks=true,linkcolor=purple,citecolor=blue,filecolor=black,urlcolor=cyan,final=true]{hyperref}
\usepackage{lineno}

\makeatletter
\renewcommand*\aa@pageof{, page \thepage{} of \pageref*{LastPage}}
\makeatother

\newcommand{\revise}{}

\begin{document}

   \title{On the reason for the widespread energetic storm particle event of 13 March 2023} 
   %
   
   \subtitle{}

    \author{N.~Dresing\inst{1}
            \and 
        I.~C.~Jebaraj\inst{1}
            \and
        N.~Wijsen\inst{2}
        \and
        E.~Palmerio\inst{3}
        \and 
        L.~Rodr\'iguez-Garc\'ia\inst{4,5}
            \and
        C.~Palmroos\inst{1}
            \and 
        J.~Gieseler\inst{1}
            \and
        M.~Jarry\inst{6,7}
            \and
        E.~Asvestari\inst{8}
            \and
        J.~G.~Mitchell\inst{9}
            \and
        C.~M.~S.~Cohen\inst{10}
            \and  
        C.~O.~Lee\inst{11}
            \and
        W.~Wei\inst{11}
        \and 
        R.~Ramstad\inst{19}
        \and
        E.~Riihonen\inst{1} 
        \and
        P.~Oleynik\inst{1}
            \and
        A.~Kouloumvakos\inst{12}
            \and
        A.~Warmuth\inst{13} 
            \and
        B.~S\'anchez-Cano\inst{14}
        \and
        B.~Ehresmann\inst{15}
        \and
        P.~Dunn\inst{11}
        \and 
        O.~Dudnik\inst{16,17}
        \and
        C.~Mac~Cormack\inst{18,7}
        }
        
   \institute{Department of Physics and Astronomy,
            University of Turku, Turku, Finland\\  
            \email{nina.dresing@utu.fi}
        \and
            Center for mathematical Plasma Astrophysics, KU Leuven, Kortrijk/Leuven, Belgium 
        \and
            Predictive Science Inc., San Diego, CA, USA 
        \and 
            European Space Astronomy Centre, European Space Agency, Villanueva de la Ca{\~n}ada, Madrid, Spain 
        \and 
            Universidad de Alcalá, Space Research Group (SRG-UAH), Alcalá de Henares, Madrid, Spain
        \and
            Institut de Recherche en Astrophysique et Plan{\'e}tologie (IRAP), CNRS, Universit{\'e} de Toulouse III-Paul Sabatier, France 
        \and 
            Institute for Astronomy, Astrophysics, Space Applications and Remote Sensing (IAASARS), National Observatory of Athens, Penteli, Greece
        \and
            Faculty of Science, University of Helsinki, Helsinki, Finland 
        \and
            Heliophysics Science Division, NASA Goddard Space Flight Center, Greenbelt, MD, USA 
        \and 
            California Institute of Technology, Pasadena, CA, USA 
        \and
            Space Sciences Laboratory, University of California--Berkeley, Berkeley, CA, USA 
        \and
            The Johns Hopkins University Applied Physics Laboratory, Laurel, MD, USA 
        \and
            Leibniz-Institut f\"ur Astrophysik Potsdam (AIP), Potsdam, Germany 
        \and
            School of Physics and Astronomy, University of Leicester, Leicester, UK 
        \and 
            Southwest Research Institute, Boulder, CO, USA 
        \and 
            Space Research Center of Polish Academy of Sciences, Warsaw, Poland 
        \and 
            Institute of Radio Astronomy of National Academy of Sciences of Ukraine, Kharkiv, Ukraine 
        \and
            Department of Physics, The Catholic University of America, Washington, DC, USA 
        \and
            Laboratory for Atmospheric and Space Physics, University of Colorado Boulder, CO, USA 
        }

   \date{}

 
  \abstract
   {On 13 March 2023, when the Parker Solar Probe spacecraft was situated on the far side of the Sun as seen from Earth, a large solar eruption took place, which created a strong solar energetic particle (SEP) event observed by multiple spacecraft (S/C) all around the Sun. The energetic event was observed at six well-separated locations in the heliosphere, provided by Parker Solar Probe, Solar Orbiter, BepiColombo, STEREO~A, near-Earth S/C, and MAVEN at Mars. Clear signatures of an in-situ shock crossing and a related energetic storm particle (ESP) event were observed at all inner-heliospheric S/C, suggesting that the interplanetary coronal mass ejection (CME)-driven shock extended all around the Sun. However, the solar event was accompanied by a series of pre-event CMEs.}
   {We aim to characterize this extreme widespread SEP event and to provide an explanation for the unusual observation of a circumsolar interplanetary shock and corresponding circumsolar ESP event.}
   {We analyse data from seven space missions, namely Parker Solar Probe, Solar Orbiter, BepiColombo, STEREO~A, SOHO, Wind, and MAVEN to characterize the solar eruption at the Sun, the energetic particle event, and the interplanetary context at each observer location as well as the magnetic connectivity of each observer to the Sun. We then employ magnetohydrodynamic simulations of the solar wind in which we inject various CMEs that were launched before as well as contemporaneously with the solar eruption under study. In particular, we test two different scenarios that could have produced the observed global ESP event: 1) a single circumsolar blast-wave-like shock launched by the associated solar eruption, and 2) the combination of multiple CMEs driving shocks into different directions.}
   {By comparing the simulations of the two scenarios with observations we find that both settings are able to explain the observations. However, the blast-wave scenario performs slightly better in terms of the predicted shock arrival times at the various observers.}
   {Our work demonstrates that a circumsolar ESP event, driven by a single solar eruption into the inner heliosphere, is a realistic scenario.}


   \maketitle
  
%
\section{Introduction} \label{sec:intro}

Solar energetic particle (SEP) events that show particle spreading almost all around the Sun are called widespread events since the era of the Solar TErrestrial RElations Observatory \citep[STEREO;][]{Kaiser2008} spacecraft (S/C) \citep[e.g.][]{Dresing2014, Prise2014, Lario2016, Xie2019, Rodriguez-Garcia2021}. \citet{Dresing2014} defined a widespread event by a SEP distribution extending up to an S/C, whose magnetic footpoint at the Sun is longitudinally separated from the associated flare location by at least 80$^{\circ}$. Depending on the particle species, only 15--20 such events are observed during a solar cycle \citep[e.g.,][]{Richardson2014}.

There are likely multiple mechanisms capable of producing this extreme longitudinal spread of energetic particles \citep{Dresing2014}. Currently, the most favored scenarios are coronal mass ejection (CME)-driven shocks with a very wide angular extent that provide a strongly extended SEP source region \citep[e.g.,][]{Gomez-Herrero2015}, or very efficient transport in the interplanetary (IP) medium perpendicular to the mean magnetic field \citep[e.g.,][]{Dresing2012, Droege2014, Droege2016, Laitinen2016, Strauss2017}. However, the potential mixture of those effects in the same event was also suggested \citep[e.g., ][]{Dresing2014, Lario2014, Rodriguez-Garcia2021, Kollhoff2021, Kouloumvakos2022}.

Also, the role of extreme ultraviolet (EUV) waves (fast-magnetosonic waves or shock waves) was investigated in terms of the spreading of energetic particles close to the Sun when they reach the magnetic footpoint locations of certain observers \citep[][]{Prise2014}. This scenario was often found incapable of explaining the particle arrivals at certain observer positions \citep[][]{Lario2014, Miteva2014}. However, a reason for some of those timing discrepancies could be that the shock propagation is faster higher in the corona than close to the solar surface, where it is visible in EUV images \citep{Zhu2018, Kouloumvakos2023, Zhuang2024}. Furthermore, a strong field line spread close to the flaring region in the lower corona has also been proposed to contribute to widespread SEP events \citep[cf.,][]{Klein2008}, and reconnection of field lines during the erupting CME flux rope provides another mechanism potentially capable of injecting flare-accelerated SEPs far from the flare site \citep[][]{Masson2013, Klein2024}. 

The locations of the available S/C with respect to the solar source location certainly impact our ability to detect, study, and prove the widespread nature of an event \citep[][]{Dresing2024}. Furthermore, due to the limited spatial coverage of observing S/C, especially in the past, detailed analyses of the longitudinal particle distributions were not possible, which led researchers to characterize the particle spread with simple Gaussian functions \citep[e.g., ][]{Lario2013, Dresing2014, Dresing2018, Richardson2014}. However, the possibility of non-Gaussian or asymmetric SEP distributions, potentially exhibiting local intensity variations, has been proposed \citep[][]{Klassen2016}. The reason could be longitudinally varying transport conditions \citep{Dalla2024}, for example, caused by the presence of different solar wind streams that may even include preceding CMEs \citep[e.g.,][]{Palmerio2021, Rodriguez-Garcia2024}. Moreover, the presence of stream interaction regions (SIRs) can lead to adiabatic de- or acceleration as well as further particle acceleration at SIR-associated shocks \citep[e.g.,][]{Wijsen2019} or even affect the efficiency of CME-driven shocks \citep[][]{Lario2022}. 

The presence of multiple distinct and almost simultaneous SEP injections close to the Sun is another potential scenario for creating wide particle spreads with potential non-Gaussian distributions. One such asymmetry was recently observed thanks to the expanded S/C fleet that is now in place built by new missions such as Parker Solar Probe \citep[Parker;][]{Fox2016}, Solar Orbiter \citep[][]{Muller2020}, and BepiColombo \citep[][]{Benkhoff2021} on its cruise to Mercury, in combination with established missions such as STEREO, the SOlar and Heliospheric Observatory \citep[SOHO;][]{Domingo1995}, and Wind \citep[][]{Ogilvie1997}. \citet{Dresing2023} found that several distinct SEP injections must have been present during the 17 April 2021 widespread SEP event, directed into significantly different directions spanning a longitudinal sector of 110$^{\circ}$. These formed an asymmetric SEP distribution that became evident thanks to observations at Parker, which did not fit an expected Gaussian-like longitudinal trend. 

The rising phase of solar cycle 25 has produced several energetic and widespread SEP events \citep[e.g.,][]{Kollhoff2021, Dresing2023, Khoo2024}, which were observed with this new S/C fleet. 
The new instrumentation carried by these missions, constantly changing S/C constellations, as well as the growing number of observers allow us to tackle long-standing questions in connection with widespread SEP events, and SEP events, in general, such as the particle acceleration mechanisms \citep[e.g.,][]{Klein2022, Jebaraj2023a,  Kouloumvakos2024, Ding2024}, their injection into the open heliospheric magnetic field \citep[e.g.,][]{Gomez-Herrero2021, Kouloumvakos2022, Lario2022}, and their propagation through the IP medium \citep[e.g.,][]{Wimmer2023, Wijsen23}.
Another highlight of the new S/C fleet is the passage of CME-driven shocks at close distances to the Sun \citep[e.g.,][]{Jebaraj2023b, Jebaraj2024}, which has provided unprecedented insights into shock-acceleration mechanisms. 

\begin{figure*}[ht!]  
    \centering
    \includegraphics[width=0.6\textwidth]{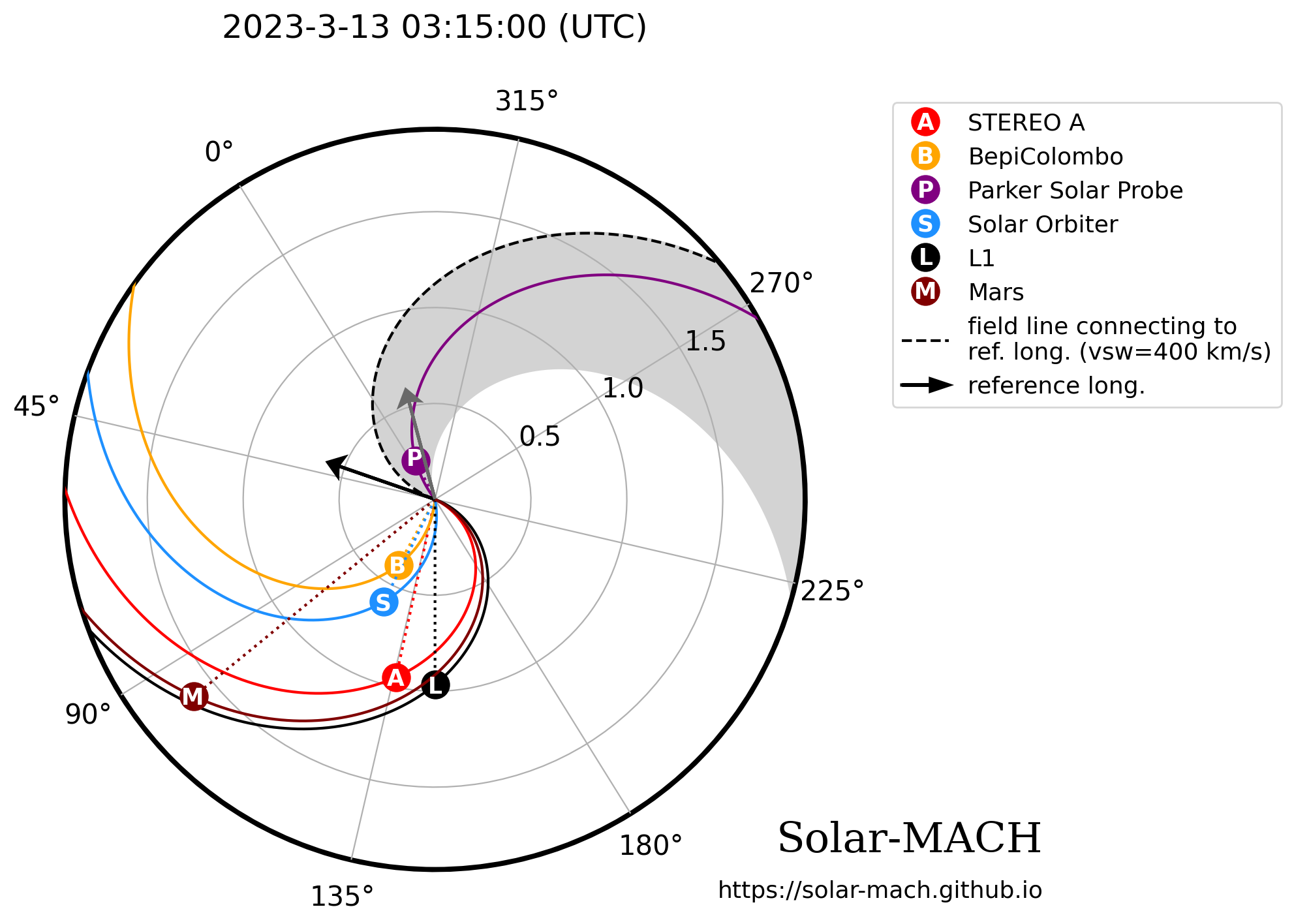} \\
    \includegraphics[width=0.8\textwidth]{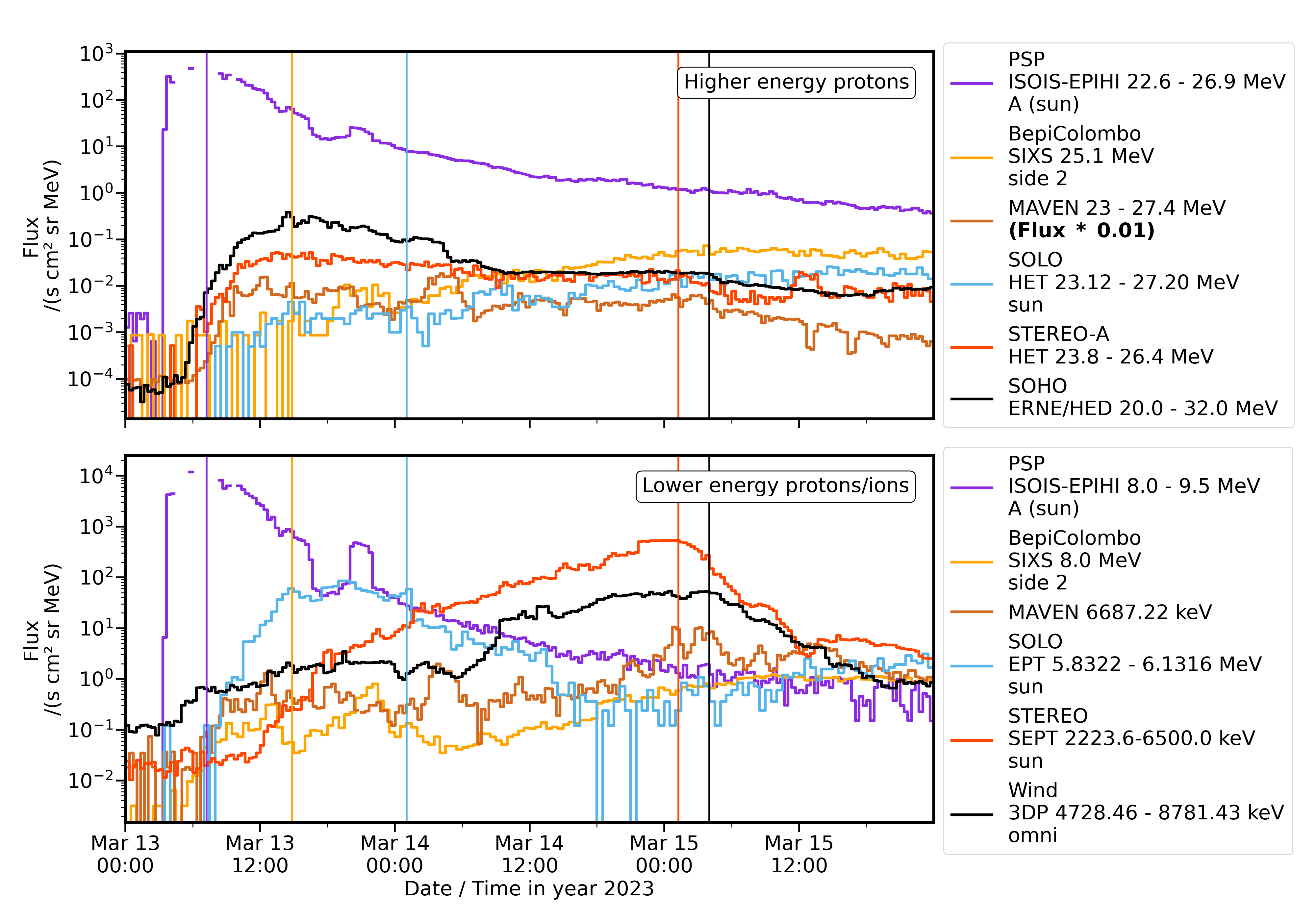}
        \caption{Top: S/C constellation in the ecliptic plane, including nominal Parker spiral field lines connecting the S/C with the Sun using measured solar wind speed (see Table~\ref{tab:coordinates}). The shaded area and two arrows mark the longitudinal sector of the potential particle injection of the 13 March 2023 event. Bottom: Intensities \revise{(20~min averages)} of ${\sim}25$ MeV (top) and ${\sim}8$ MeV (bottom) ions/protons observed by Parker (PSP, purple), BepiColombo (yellow), Solar Orbiter (SOLO, blue), STEREO~A (red), SOHO/Wind (black), and MAVEN (brown, scaled by a factor of $0.01$). The vertical lines denote the times of IP shock crossings at the various S/C (using the same color coding). } 
        \label{fig:solar-mach}
\end{figure*}

In this work, we study a multi-S/C SEP event that occurred on 13 March 2023. The SEP event was detected by six well-separated observers, namely Parker, BepiColombo, Solar Orbiter, STEREO~A, near-Earth missions such as SOHO and Wind, and by the Mars Atmosphere and Volatile Evolution \citep[MAVEN;][]{Jakosky2015} probe at Mars. The lower panel of Fig.~\ref{fig:solar-mach} shows ${\sim}25$~MeV (top) and ${\sim}8$~MeV (bottom) proton measurements at the six locations, which were spanning a longitudinal sector of 153° (.~\ref{fig:solar-mach}, top). Although the exact location of the associated flare is unknown due to missing imaging observations of the Sun's far side, the S/C constellation results in all observers, except Parker, being longitudinally far separated from the flaring active region (AR), making this event an unambiguous widespread event with a suggested spread of particles all around the Sun. Small onset delays and the presence of high-energy SEPs with comparably high intensities\footnote{This event is listed as event \#40 in the SERPENTINE Solar cycle 25 SEP Events Catalog \citep[][and references therein]{Dresing2024}.} at far-separated observer positions make this widespread event an extreme one. 
Even more remarkable is the observation of an associated energetic storm particle (ESP) event detected by the farthest separated observers. Such ESP events, characterized by a continuous increase of SEP intensities, reaching their maximum when a CME-driven shock passes the S/C, are usually caused by the same CME associated with the parent solar eruption.

In this paper, we study two scenarios that could explain this remarkable widespread SEP and ESP event, including i) a circumsolar shock wave and ii) a pre-event CME-driven shock being responsible for the ESP components on the front side, that is the side of Earth. Because the pre-event CME was not associated with its own SEP event, this scenario \revise{assumes} that its shock would re-accelerate or further accelerate the SEPs of the 13 March 2023 solar eruption.
The manuscript is organized as follows. In Sect.~\ref{sec:overview} we provide an overview of the event and explore the source location of the solar eruption and the magnetic connectivity of the various S/C with this source region. Section~\ref{sec:in_situ_obs} discusses the in-situ energetic particle and IP context observations in comparison with magnetohydrodynamic (MHD) simulation results. 
Section~\ref{sec:discussion} provides a discussion of the two potential widespread ESP event scenarios and summarizes the work.
%


\section{Event overview} \label{sec:overview}
\subsection{S/C constellation and the associated solar eruption}\label{subsec:constellation}
The top panel of Fig.~\ref{fig:solar-mach} shows the positions of all six observer locations as well as magnetic field lines connecting these S/C with the Sun assuming a nominal Parker spiral field line according to the measured solar wind speeds at 03:15~UT on 13 March 2023 (where measurements were available; cf.\ Tab.~\ref{tab:coordinates}). 
Parker was located on the far side of the Sun as seen from Earth, at a radial distance of 49 solar radii. As can be seen in the bottom part of Fig.~\ref{fig:solar-mach}, which shows multi-S/C observations of $\sim$25~MeV proton intensities (top panel), Parker observes by far the most prominent SEP event marked by its prompt rise time profile, its earlier onset, and very high peak intensity. It is therefore likely that the source region of the SEP event was situated on the far side as seen from Earth, magnetically well-connected with Parker. However, the SEP event was clearly observed all around the Sun, with higher intensities detected by SOHO and STEREO~A compared to Solar Orbiter and BepiColombo, even with Solar Orbiter and BepiColombo being situated closer to the Sun. 

The figure also shows MAVEN observations marking a clear event reaching also Mars. Information on the instrumentation used in this study can be found in Appendix~\ref{app:instrumentation}.
The vertical lines in Fig.~\ref{fig:solar-mach} (bottom) mark IP shock crossings observed by all inner-heliospheric S/C situated within 1~au (using the S/C color coding), which will be discussed in more detail in Sect.~\ref{subsec:IP_shock}. 
The lower panel of Fig.~\ref{fig:solar-mach} shows energetic ion/proton measurements around 8~MeV as observed by the five inner-heliospheric observers and by MAVEN at Mars. 
The intensity--time profiles at these lower energies are different (except for Parker) when compared with the higher energies, reaching their peaks in accordance with the arrival times of the corresponding IP shocks (most clearly seen at STEREO~A and Wind, and less clear at BepiColombo and Solar Orbiter). This suggests the presence of an ESP event even at the front-sided inner-heliospheric observers of the 13 March 2023 event.
The gray-shaded longitudinal sector in Fig.~\ref{fig:solar-mach} (top) marks the region encompassing the active regions (ARs) that we identified as the most likely candidates for hosting the eruption associated with the event under study (discussed below). These are not visible as seen from any of the observers that carry instrumentation imaging the solar surface or lower corona. We therefore constrain the timing and location of the associated solar eruption based on the following observations. 

\begin{SCfigure*}[][th!]  
    \centering
        \includegraphics[width=0.65\textwidth]{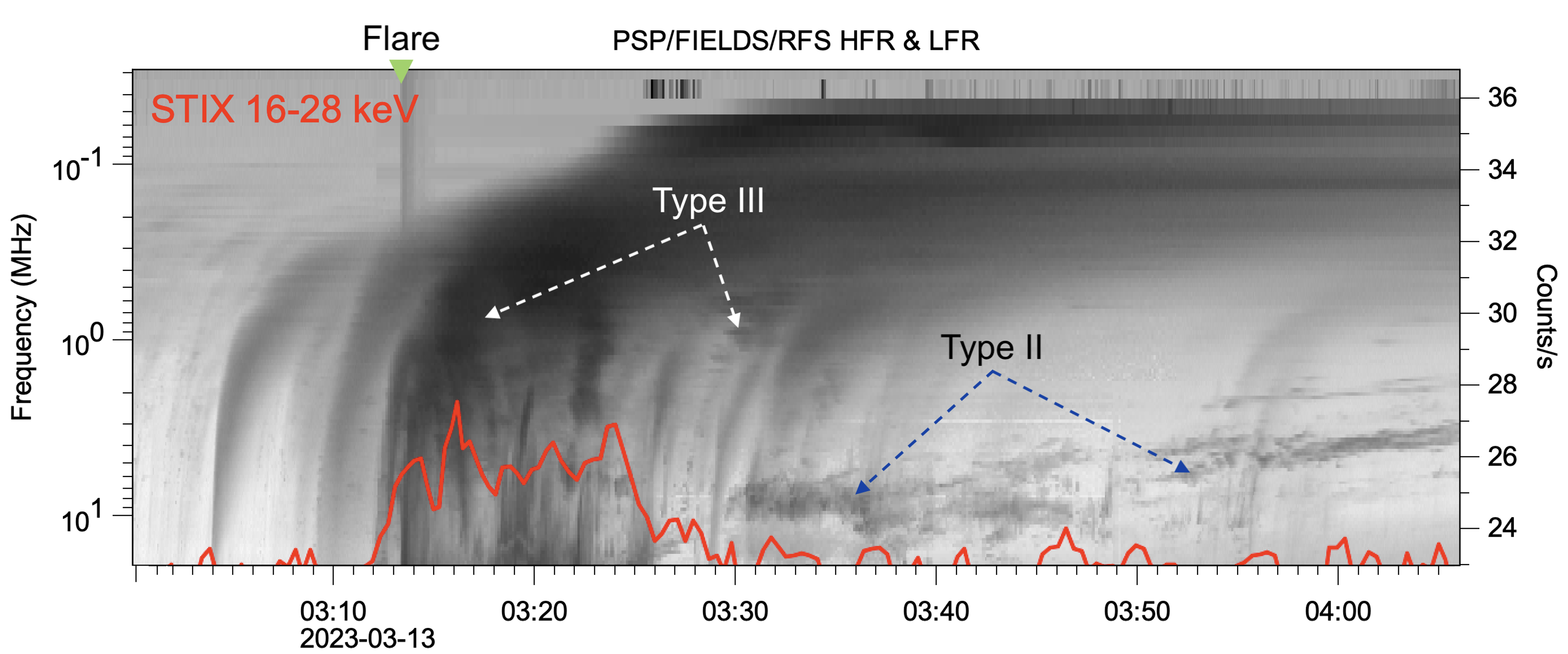}
        \caption{Radio and hard-X-ray radiation at the time of the flare. The radio frequency-time spectrogram constructed from FIELDS/RFS measurements on board Parker is shown in gray scale. The 30s averaged time series of the Solar Orbiter/STIX non-thermal (16--28 keV) X-ray photon counts (s\(^{-1}\)) is over plotted in red. Both FIELDS/RFS and Solar Orbiter/STIX measurements are time shifted to the Sun for a direct comparison.} 
        \label{fig:stix_rfs}
\end{SCfigure*}

Given the lack of solar imaging capabilities at Parker, we utilized the available radio observations. Figure~\ref{fig:stix_rfs} displays a radio spectrogram observed by Parker/FIELDS/RFS (time-shifted to the Sun), marking the onset of the solar eruption. A full description of the radio spectra and the associated emissions can be found in \cite{Jebaraj24b}. Here, we mark only the radio bursts that are associated with most of the powerful solar eruptions. This includes a series of fast-drifting type III radio bursts, a prominent slower-drifting type II radio burst, and multiple associated emission lanes. At 03:15~UT (03:13~UT shifted back to the Sun), concurrently with the first group of type III radio bursts, a dispersionless signal is observed, which has been marked using the small green arrow. This signal is attributed to photoelectrons excited by the fastest particles during a strong flare. Another indication of the association between the event at the solar surface and the radio emissions is the significantly higher intensity of the type III bursts observed at this time compared to earlier ones.  The type III before the dispersionless signal are either pre-flare releases of sub-relativistic electrons or unrelated to the flare. In the case of the former, type III storms are usually associated with strong active regions prior to major flaring activity \citep[e.g.,][]{norsham2019multiwavelength,Pulupa24}. These features, along with the temporal association with the type II burst (indicative of shock formation) which is also marked in Fig.~\ref{fig:stix_rfs}, allow us to infer that the main eruption occurred at 03:13~UT (at the Sun). 

The red line overlaid in Fig.~\ref{fig:stix_rfs} represents hard X-ray observations taken by Solar Orbiter/STIX, also time-shifted to the Sun. Although the temporal profile of the flare light curve agrees well with the time of the suggested solar eruption, we note that the main portion of the flare was likely behind the limb as seen from Solar Orbiter. The STIX data center\footnote{\url{https://datacenter.stix.i4ds.net/stix}} reports a flare right at the east limb, slightly in the northern hemisphere, with a description of `occulted flare?'. The estimated STIX flare class, which represents an equivalent to the GOES soft-X-ray flare class, \citep[see][for further details]{Xiao2023} is B2. This is unexpectedly low for such an energetic SEP event (cf.\ Sect.~\ref{subsec:seps}), which supports the scenario of a behind-the-limb flare as seen from Solar Orbiter.

Inspection of STEREO~A/EUVI, SDO/AIA, and Solar Orbiter/EUI images of the solar corona during periods before (22 Feb--5 Mar 2023) and after (14--22 Mar 2023) the solar event under study reveals at least two ARs potentially responsible for the studied eruption. One AR (AR1), which is the likely candidate producing the occulted flare as seen from Solar Orbiter, is the large AR 13258 (identified in Carrington rotation 2268) situated in the northern hemisphere, spanning a sector of [15, 39]$^{\circ}$ in Carrington longitude and [17, 33]$^{\circ}$ in latitude (see also Appendix~\ref{app:remotecme}). This region is closer to the east limb as seen from the front-sided observers. It is marked by a black arrow in Fig.~\ref{fig:solar-mach} (top), and its most eastern edge marks the eastern limit of the gray-shaded area.

A second AR (AR2), located in the southern hemisphere, which provides another potential candidate was AR 13256 (identified in Carrington rotation 2268), covering a longitudinal sector of [343, 12]$^{\circ}$ at Carrington longitude and a sector of [-12, -35]$^{\circ}$ in latitude (see Appendix~\ref{app:remotecme}). This region is further far-sided as seen from Earth and marked by the gray arrow in Fig.~\ref{fig:solar-mach} (top). Appendix~\ref{app:remotecme} shows in detail how the two ARs were identified. 
We used the most western edge of AR2 as the western limit of the eruption sector, which has in total a longitudinal extent of 56$^{\circ}$ Carrington longitude and 68$^{\circ}$ latitude (cf. Table~\ref{tab:coordinates}). It is clear that all S/C, except Parker, are situated at well-separated longitudes from either of these far-sided ARs. Each of their magnetic footpoint's longitudinal separation angles with this sector are well above 100$^{\circ}$, which clearly classifies this event as a widespread SEP event as defined by \citet[][]{Dresing2014}.

\sidecaptionvpos{figure}{c}


\subsection{Magnetic connectivity} \label{subsec:connectivity}
\begin{table*}
\caption{Coordinates of candidate ARs and S/C, and their and magnetic connectivity to the Sun.}
\label{tab:coordinates}      
\centering          
\begin{tabular}{l cccccccccc}     
\toprule
(1) & (2) & (3) & (4) & (5) & (6) & (7) & (8) & (9) & (10) \\
  & & & & \multicolumn{2}{c}{Parker spiral$^{(a)}$} & & \multicolumn{4}{c}{EUHFORIA$^{(c)}$} \\ \cline{5-6} \cline{8-11}
  Location & r & Lon$^{(b)}$ & Lat$^{(b)}$ & Lon$^{(b)}$ & Lon sep$^{(d)}$& V$_{obs}$  &  Lon$^{(b)}$ & Lat$^{(b)}$ & Lon sep$^{(d)}$\\
    &  (au) & ($^{\circ}$) & ($^{\circ}$) & ($^{\circ}$) & ($^{\circ}$) &  (km s$^{-1}$)  &  ($^{\circ}$) & ($^{\circ}$) & ($^{\circ}$)\\
  \hline
  AR1 (AR 13258)&  & 15--39 & 17--33 &  &  &  &  &  &  & \\
  AR2 (AR 13256)&  & 343--12 & --(12--35) &  &  &  &  &  &  & \\
  Eruption sector &  & 39--343 & --(35--33) &  &  &  &  &  & &  \\
  Parker  & 0.23 & 354.7 & 3.2 &  9.1 &   & 490   & 8.5 & 11.2&  \\
  BepiColombo & 0.4 & 119 & 0.2 &  143.9 & --104.9 & 400$^{(e)}$ & 140.2 & --20.4 & --101.2 \\
  Solar Orbiter  & 0.61 & 121.6 & $-$5.0 &  159.2 & --120.2 &400$^{(e)}$ &  149.9 & --21.8& --110.9\\
  STA  & 0.97 & 135.7 & $-$7.3 & 212.8 & 130.2  & 310 &   180.2 &  --27.4& --141.2\\
  near Earth  & 0.98 & 148.1 & $-$7.2 & 218.9 & 124.1 & 347 & 182.7 & --26.0 & --143.7\\
  MAVEN & 1.64 & 97.3 & --3.4 & 223.0$^{(f)}$ & 120 & 325 &  181.0 & --26.3& --142.0\\
\bottomrule
\end{tabular}
\tablefoot{Columns (2)--(4): coordinates of feature ior observer provided in Col.~(1). Columns (5) and (8)--(9): estimated magnetic footpoint coordinates of the observers.
\tablefoottext{a}{Magnetic footpoints at the solar surface based on ballistic backmapping assuming a nominal Parker spiral field line based on the solar wind speed provided in Col. (7). The backmapped latitudes (not shown) are the same as the spacecrafts' latitudes given in Col. (4);}
\tablefoottext{b}{Longitude and latitude values are given in the
Carrington coordinate system;}
\tablefoottext{c}{Magnetic footpoint at the solar surface based on EUHFORIA simulation\footnote{The magnetic footpoints in the EUHFORIA simulation are based on a potential field source surface (PFSS) extrapolation extended with the Schatten current sheet model up to 0.1~au, followed by the MHD simulation beyond that distance.};}
\tablefoottext{d}{Longitudinal separation angle between the observer's magnetic footpoint and the closer edge of the eruption sector. Positive (negative) values denote that the source sector edge is to the west (east) of the magnetic footpoint;}
\tablefoottext{e}{No solar wind measurements available, using therefore the nominal value of 400~km~s\(^{-1}\);}
\tablefoottext{f}{Solar wind speed is provied by the Mars Express mission.}
}
\end{table*}
Table~\ref{tab:coordinates} lists the coordinates of the two ARs (AR1 and AR2 described above), which provide potential source regions of the eruption, and which determine the potential eruption sector (cf. Sect.~\ref{subsec:constellation}), also provided in the table. The coordinates of all observers of the SEP event (columns 2--4), as well as those of their magnetic footpoints at the Sun at the time of the solar eruption (columns 5 and 8--9), are also listed. 

We determined the magnetic footpoints with two methods: 1) assuming a nominal Parker spiral magnetic field and backmapping the field line using measured solar wind speed values (illustrated in Fig.~\ref{fig:solar-mach}), where possible (Col.~5)\footnote{The latitude of the magnetic footpoint determined with this method is identical to the latitude of the S/C position.}, and 2) using the output of a simulation of the European Heliospheric Forecasting Information Asset \citep[EUHFORIA;][Sect.~\ref{subsec:euhforia_sim}]{Pomoell_Poedts2018} shown in Col.~8--9, which takes into account the presence of five pre-event CMEs and one post-event CME (Appendix~\ref{app:cme_reconstructions}) additional to the main eruption. 
As discussed in Sect.~\ref{subsec:constellation}, the only observer that is well-connected to the far-sided eruption is Parker, whose magnetic footpoint lies within the eruption sector (cf.\ Fig.~\ref{fig:solar-mach} (top) and Table~\ref{tab:coordinates}). The longitudinal separation angles of the magnetic footpoints of the other observer locations with their respective closest eruption sector edge are all larger than 100$^{\circ}$, whether using the magnetic footpoints determined by simple ballistic backmapping (as shown in Col.~6 of Table~\ref{tab:coordinates}) or the results from the EUHFORIA simulation
(as shown in Col.~10 in Table \ref{tab:coordinates}).

The EUHFORIA simulation suggests that the pairs STEREO~A and near-Earth observers, as well as BepiColombo and Solar Orbiter, have very similar magnetic footpoint longitudes. 
In the case of MAVEN at Mars, both methods suggest a similar magnetic footpoint to that of Earth and STEREO~A. When further comparing the magnetic footpoints of the nominal Parker spiral with those determined by EUHFORIA, all observers' footpoints are shifted eastward. The shift is not significant in the case of Solar Orbiter and BepiColombo given the usually assumed uncertainties of at least 10° \citep[e.g.,][]{Ippolito2005, Klein2008}. However, for STEREO~A and Earth, the shift leads to both S/C being closer connected to the eastern eruption sector instead of the western sector edge as suggested by the ballistic backmapping. This shift is primarily due to the presence of pre-event CMEs in the EUHFORIA simulation, which reduces the curvature of the field lines compared to the nominal Parker spiral, visible in Fig.~\ref{fig:euhforia_snap}, which is further discussed in Sect.~\ref{subsec:euhforia_sim}. However, it is important to note that this effect is largely a consequence of using the unmagnetized cone model for the pre-event CMEs in EUHFORIA. Had the pre-event CMEs contained a magnetic cloud, the interplanetary field lines would have likely been pushed aside rather than passing through the CME, leading rather to a westward shift of the footpoints and therefore a closer connection to the western edge of the eruption sector.

According to the ballistic backmapping both STEREO~A and near-Earth observers are more closely connected to the solar source sector via the Sun's western limb, with longitudinal separations of about 215$^{\circ}$ to the eruption sector edge (gray shade in Fig.~\ref{fig:solar-mach}). In contrast, BepiColombo and Solar Orbiter are more closely connected to the opposite sector edge via the eastern solar limb, with separations of about 105$^{\circ}$ and 120$^{\circ}$, respectively, from the sector edge (see Table~\ref{tab:coordinates}). This difference in magnetic connection to the source region may explain the varying SEP characteristics observed at STEREO~A and near Earth's perspective compared to those at Solar Orbiter and BepiColombo, suggesting different particle acceleration sites.
Based on the above, we consider the longitudinal separation angles based on ballistic backmapping (column 6 in Table~\ref{tab:coordinates}) more realistic than those of the EUHFORIA simulation. We note, however, that a reliable estimation is almost impossible for this event due to the far-sided eruption source and the very complex state of the interplanetary medium.

\subsection{The shock and CMEs of the main eruption}\label{subsec:coronal_shock}

As mentioned in Sect.~\ref{subsec:constellation}, the source region of the 13 March 2023 eruption was located on the far side of the Sun from the viewpoints of all the observing S/C equipped with solar disk imaging cameras. Hence, any available EUV telescope could have only observed the eruption off the solar limb---as was the case for STEREO A, Solar Orbiter, and near-Earth probes. Furthermore, the event produced a prominent halo CME as observed from Earth's and STEREO~A's perspectives. 

\begin{figure*}[!ht]
    \centering
    \includegraphics[width=0.89\textwidth]{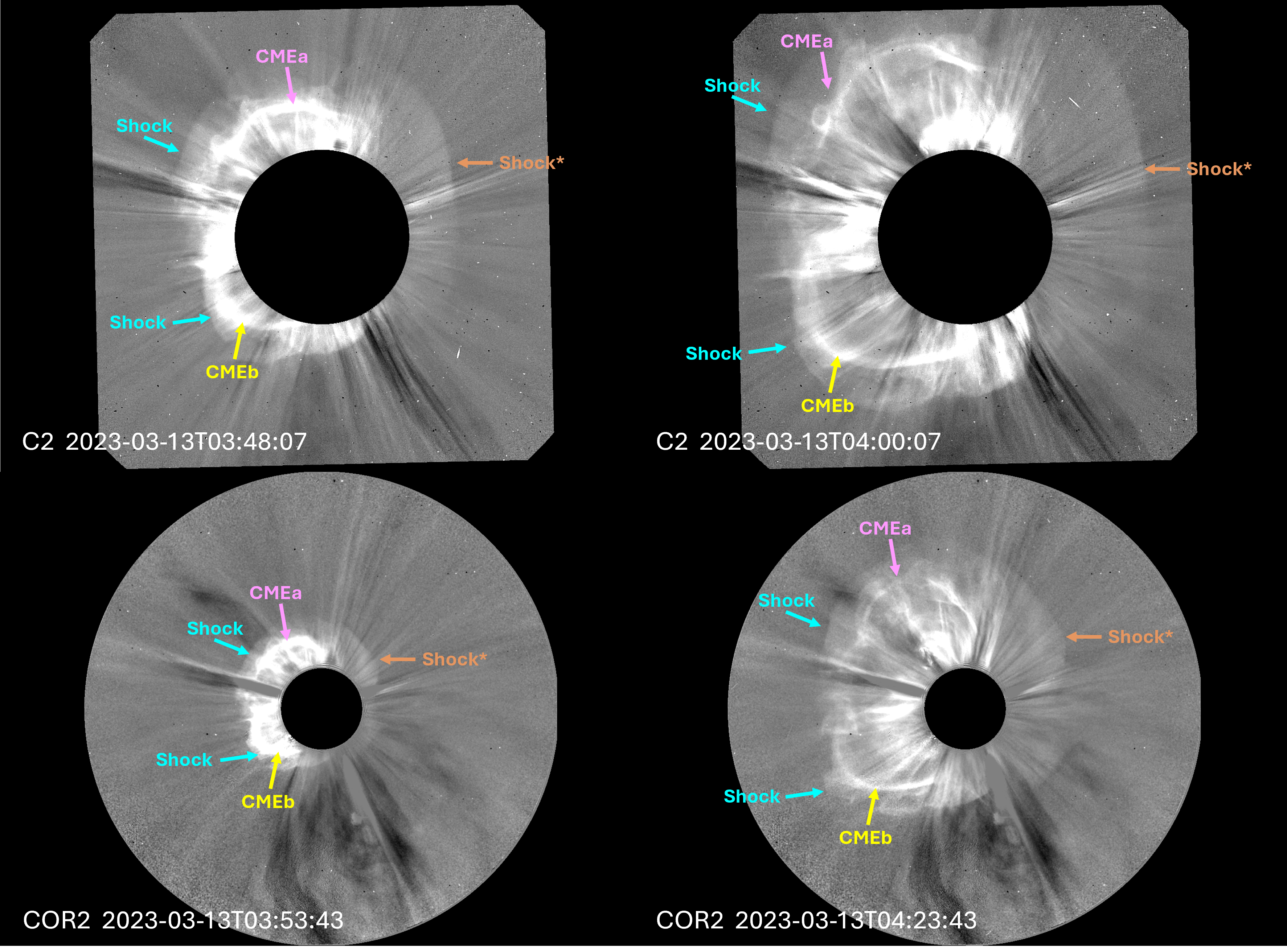}
    \caption{Base-difference images of the white-light signatures generated in the low corona by the two combined CMEs forming the eruption discussed in this manuscript, marked as CMEa (magenta arrow) and CMEb (yellow arrow), as seen by SOHO LASCO C2 and STEREO COR2 instruments. The cyan arrow indicates a shock front associated with the eruption. A freely propagating shock, marked as Shock*, is propagating above the western limb in the instruments' field of view, as indicated by the orange arrow.}
    \label{fig:coronagraph_images}
\end{figure*}

Figure~\ref{fig:coronagraph_images} shows coronagraph observations by SOHO/LASCO/C2 (top) and STEREO~A/COR2 (bottom) at two-time stamps depicting this halo CME. A more detailed look at the annotated difference images suggests, however, the presence of two distinct CMEs (marked as CMEa and CMEb), CMEa propagating northwards and CMEb south-eastwards, respectively. The CME-driven shock (marked by 'Shock') could be either a combination of two shocks or a single, potentially merged shock driven by the two CMEs. In contrast to the CMEs themselves, the shock front in the plane of sky extends clearly all around the Sun. The western part of the shock (marked 'Shock*') represents a driverless, freely-propagating wave. We note that an alternative explanation to the double-front appearance of the CME in white-light data is represented by a multi-stage and/or asymmetric filament eruption \citep[][]{Liu2009, Lynch2021}.

If two CMEs erupted consecutively, the fast-forward shocks they generate ahead of them are additive. In ideal MHD, this would lead to an increase in the amplitude of the fast-mode shock \citep[][]{Kennel88,Russell16}. This acts as additional energy which allows the wave to expand further into the heliosphere and could lead to the driverless part of the shock (marked 'Shock*') observed at the western solar limb (Fig.~\ref{fig:coronagraph_images}). 

Parker was not only situated within the eruption sector but also at a very close distance to the Sun (49 solar radii) allowing \citet{Jebaraj2024, Jebaraj24b} to study the shock in situ while it was still propagating through the upper solar corona. 
\revise{Theyat it} was near-parallel (\(\theta_{Bn} \approx 8\pm4^{\circ}\)), extremely fast (\(2800\pm300\)km s\(^{-1}\)), and exceptionally strong (\(M_\mathrm{A} \sim 9.1\pm1.35\)) for a shock evolving in a highly magnetized plasma. Their results obtained at a distance of 0.25~au also implicate that the CME may have been even faster in the lower corona.  

Furthermore, \cite{Jebaraj2024} found self-consistent injection and acceleration of electrons and ions in the presence of intense electromagnetic wave activity. These results reinforce the statistical findings by \cite{Dresing2022}, who found that the strongest near-sun shocks are capable of accelerating electrons and ions to high energies in a synchronized manner. The event culminated in one of the strongest particle events observed by Parker, saturating EPI-Lo and plasma instrumentation. While \citet{Jebaraj2024} studied the in-situ properties of the shock and its capabilities of accelerating particles in detail, they did not speculate on its global characteristics and evolution. 
Furthermore, the properties of the driving CME were not studied in situ.


\section{SEP observations and interplanetary context} \label{sec:in_situ_obs}

\begin{figure*}[!ht]  
    \centering
        \includegraphics[width=0.51\textwidth]{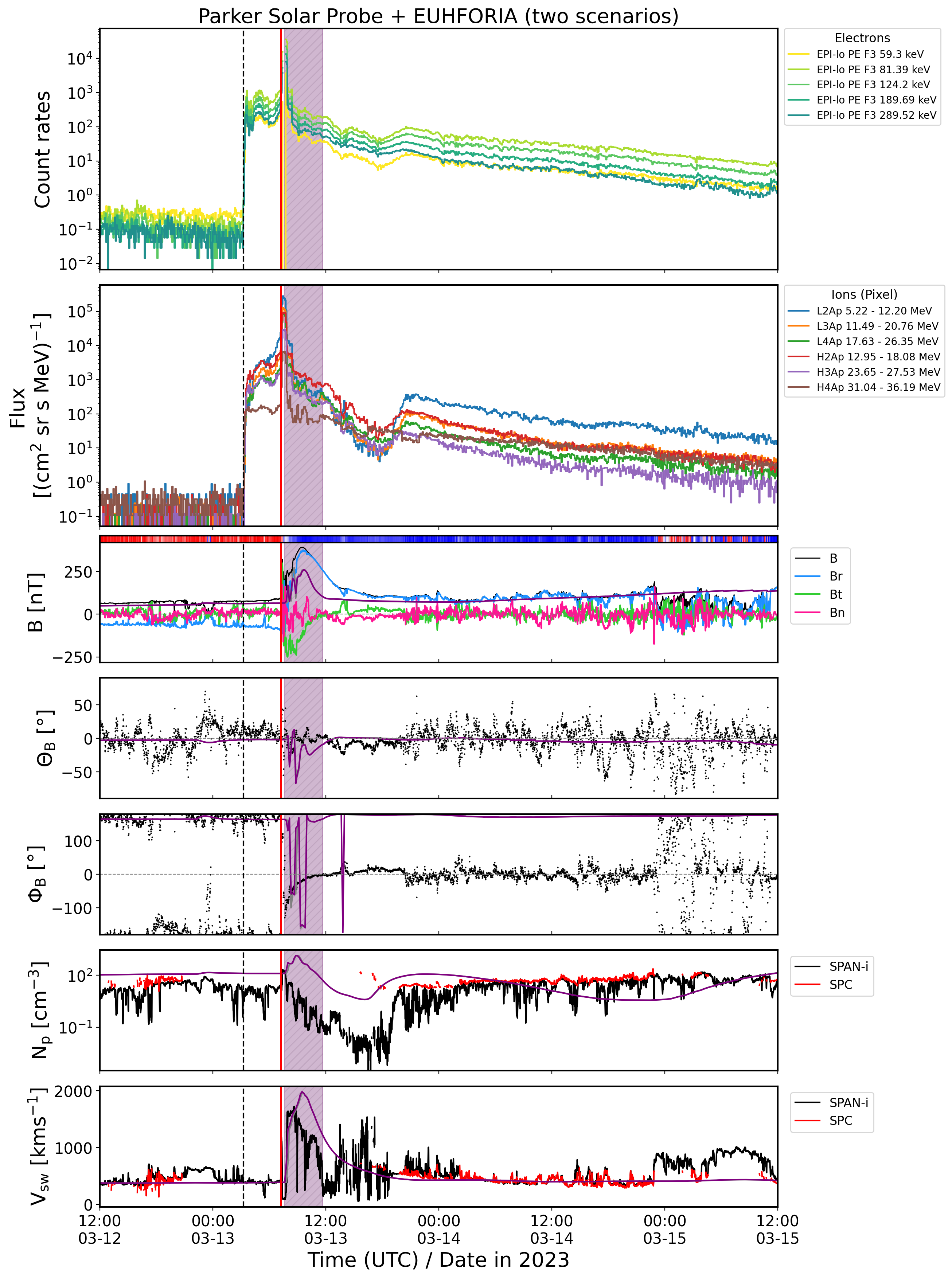}
        \includegraphics[width=0.454\textwidth]{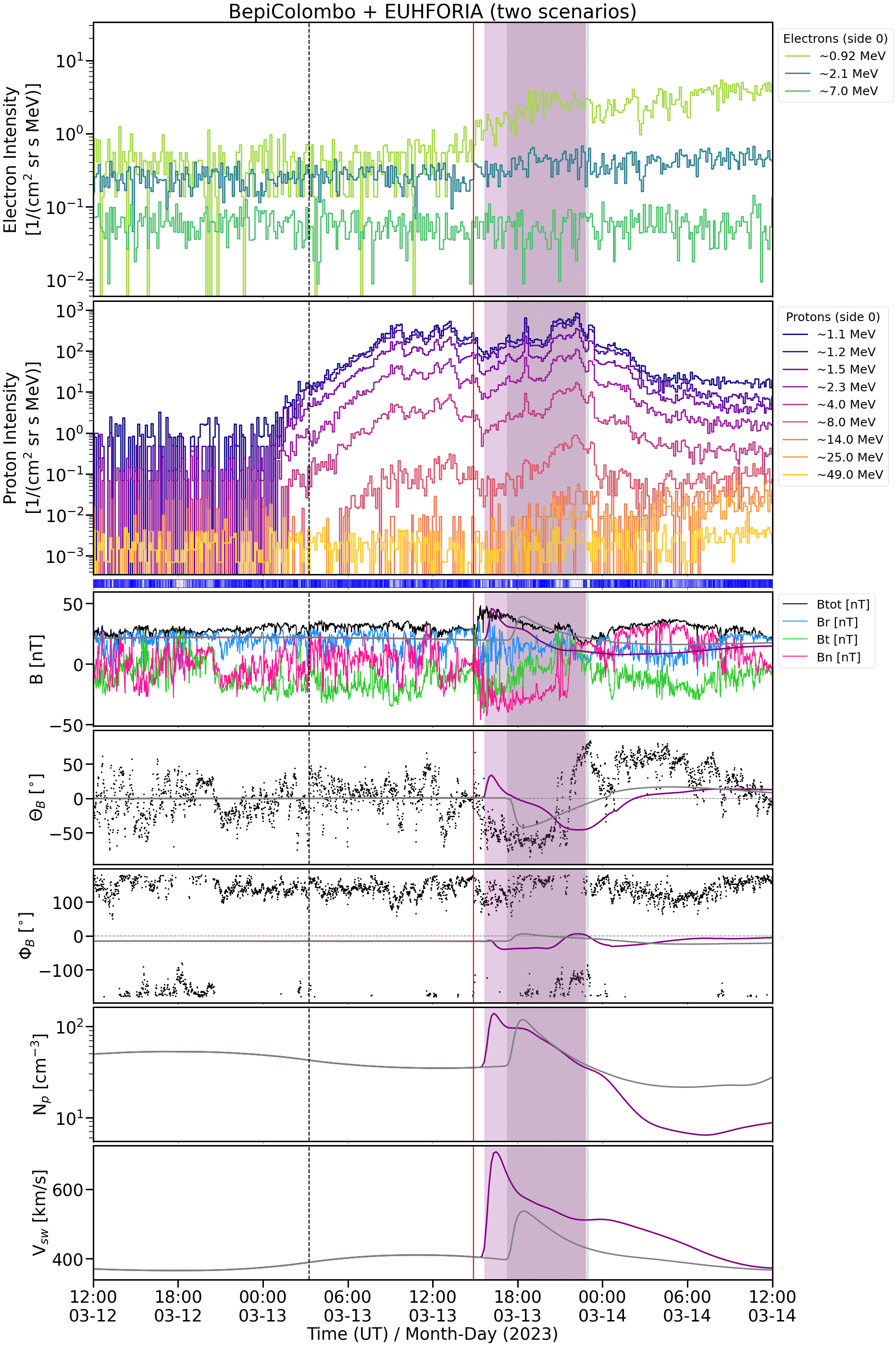} 
 
        \caption{In-situ observations in comparison with EUHFORIA simulation results for Parker (left) and BepiColombo (right). From top to bottom, we show energetic electron, and proton measurements, magnetic field (magnitude and RTN components, latitudinal, and azimuthal angle), followed by solar wind density and speed. The results of two different EUHFORIA simulations (purple: blast-wave scenario, gray: pre-CME scenario) are overlaid and show the simulated magnetic field magnitude, magnetic field angles, the solar wind density and speed in comparison with the observations (if available). The shaded regions correspond to the duration of CMEs passing the S/C according to the two different EUHFORIA simulations (same color coding as above). In case of the pre-event CME scenario, CME5 passes the S/C and in case of the blast-wave scenario it is the blast wave connected with the main eruption. The black dashed line marks the onset of the solar eruption, the red line marks the time of the observed IP shock arrivals at the S/C.}
        \label{fig:in-situ1}
\end{figure*}

\begin{figure*}[!ht]  
        \includegraphics[width=0.48\textwidth]{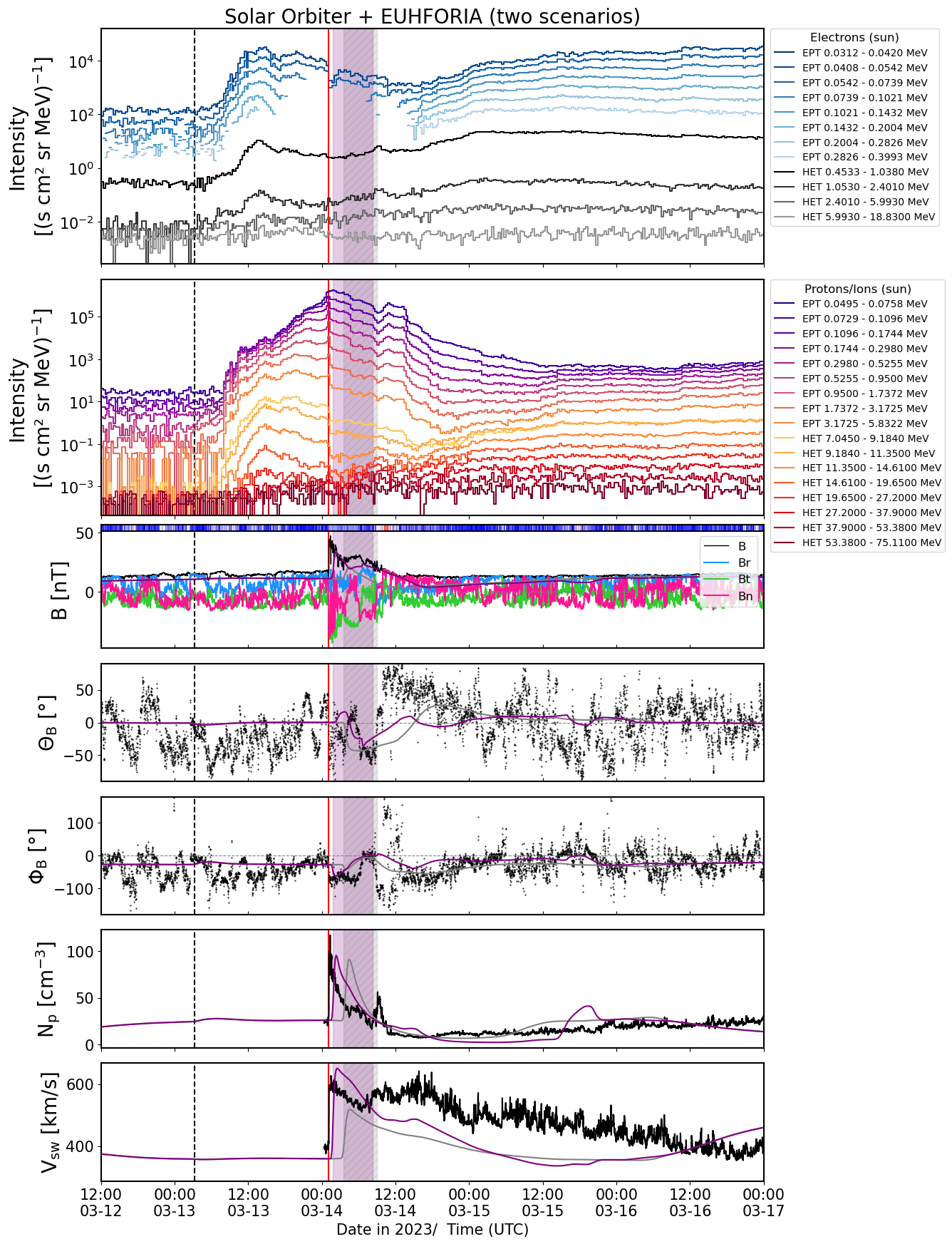} 
        \includegraphics[width=0.495\textwidth]{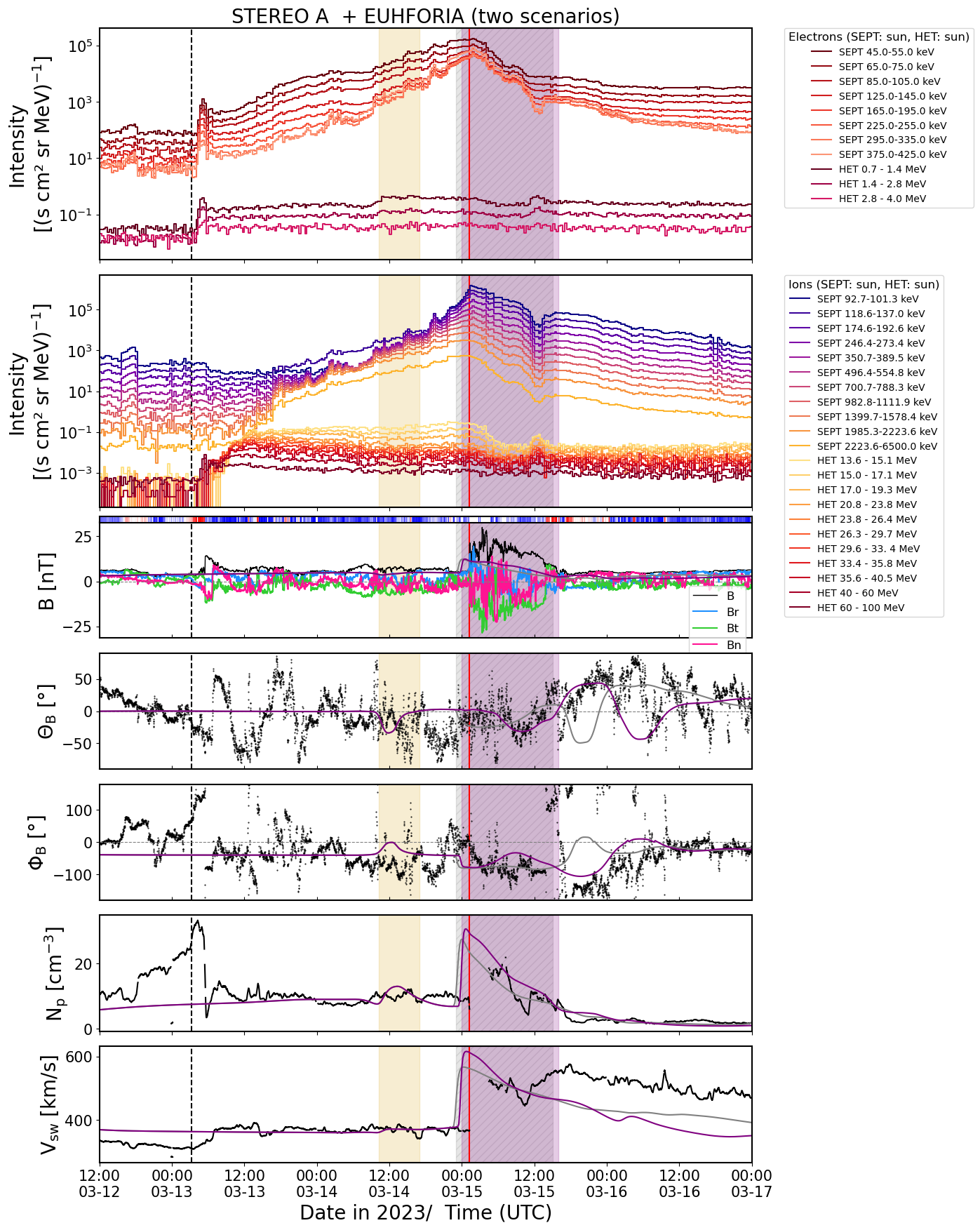} 
        \caption{In-situ observations in comparison with EUHFORIA simulation results for Solar Orbiter (left) and STEREO~A (right). Same format as in Fig.~\ref{fig:in-situ1}. The yellow shade marks arrival of CME2 (see Table~\ref{table:GCS_fit}), which is predicted similarly by both simulations. } 
        \label{fig:in-situ2}
\end{figure*}

Figures~\ref{fig:in-situ1}--\ref{fig:in-situ3} respectively show in-situ observations at Parker and BepiColombo (Fig.~\ref{fig:in-situ1}), Solar Orbiter and STEREO~A (Fig.~\ref{fig:in-situ2}), as well as near Earth and at MAVEN in orbit around Mars (Fig.~\ref{fig:in-situ3}). All plots show energetic proton and electron observations (top panels) in comparison with magnetic field observations (following panels) and, where available, solar wind plasma measurements (bottom panels; details given in the figure captions). The dashed black line in each plot marks the time of the 13 March 2023 eruption, which we define by the onset of the associated radio bursts (see Sect.~\ref{sec:overview}). Shaded regions mark the predicted arrival times of enhanced density according to the wake of the circumsolar wave or CME5, respectively according to two different EUHFORIA simulations (discussed in Sect.~\ref{subsec:euhforia_sim} below). 
In the following, we first discuss the S/C measurements (Sect.~\ref{subsec:seps}), as well as the IP shock crossings at all S/C (Sect.~\ref{subsec:IP_shock}). We refer the reader to Appendix~\ref{app:instrumentation} for an \revise {overview of the instrumentation used}. Then, we introduce two potential scenarios, which we propose to explain the widespread SEP and ESP event observations. EUHFORIA simulations of these two scenarios are presented in Sect.~\ref{subsec:euhforia_sim}, and compared with the in-situ observations in Sect.~\ref{subsec:insitu_model_comp}. While the following analysis focuses on the global SEP and ESP event, we present a more detailed analysis of the first-arriving SEPs in the Appendix, where we provide an analysis of the inferred SEP injection times at the Sun (Appendix~\ref{app:timing}) and of the particle anisotropies (Appendix~\ref{app:ani}) as observed by the different S/C.

\subsection{Multi-S/C SEP observations} \label{subsec:seps}

{\it Parker:} \\
Parker, which was situated close to the Sun and was well-connected to the eruption, observed an exceptionally strong SEP event, including a shock crossing only four hours after the solar eruption onset. \citet{Jebaraj2024, Jebaraj24b} closely investigated the particle acceleration by the shock while it passed the S/C and found that electrons were still in the progress of being accelerated reaching relativistic energies, namely in the MeV range. Protons reach energies up to at least 60~MeV.
Due to the very strong event EPI-Lo was saturated for a brief period of time and EPI-Hi transitioned into its higher ``dynamic threshold'' modes.  EPI-Lo data are only affected directly during the IP shock crossing at 07:13~UT, while in dynamic threshold modes, the EPI-Hi instruments LET and HET reduce their sensitivity, and energy range coverage, for electrons, protons, and helium (leaving heavy ion measurements unaffected).  At the highest dynamic threshold level, EPI-Hi cannot distinguish the species or energy of incoming electrons, protons, or helium.  During such periods the so-called pixel data (Appendix~\ref{app:instrumentation}), which are single-detector ion measurements within the detector stacks, can be used as a proxy for the protons (under the assumption that protons are the dominant foreground population). 
The second panel of Fig.~\ref{fig:in-situ1} (left) shows these pixel data, in various energy bands, which correspond to pixels of individual detectors within the LET and HET detector stacks. Higher-energy channels correspond to detectors that lie deeper in the stack. One of the caveats of the pixel data is that particles entering the telescopes from both sides are measured and not distinguished in any given pixel rate. Thus, especially for the deeper-lying detectors, measuring higher energies, the rate has contributions at similar energies from both sides of the telescope leading to higher rates (effectively `double counting'). The lower-energy channels are more reliable because the contribution of lower-energy ions in the outer detectors will generally be much more intense than that of the few high-energy particles reaching the detector from the other side.
Despite the instrumental issues, the data clearly show an outstandingly intense SEP event observed at Parker, including a prompt onset and a local intensity peak at the time of shock crossing with fluxes increasing more than an order of magnitude above the rest of the SEP event. \\\\

\begin{figure*}[!ht]  
        \includegraphics[height=0.485\textheight]{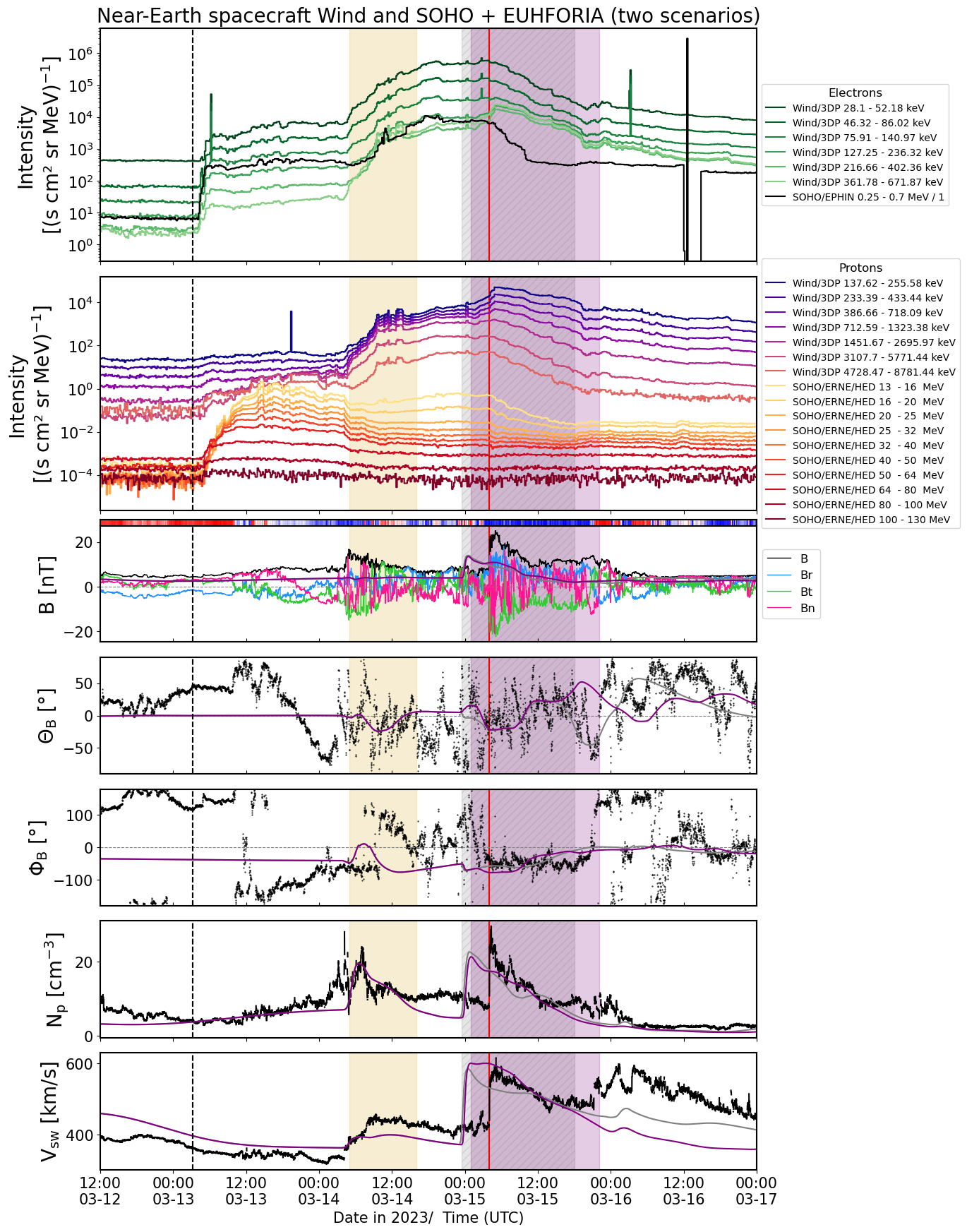} 
        \includegraphics[height=0.51\textheight]{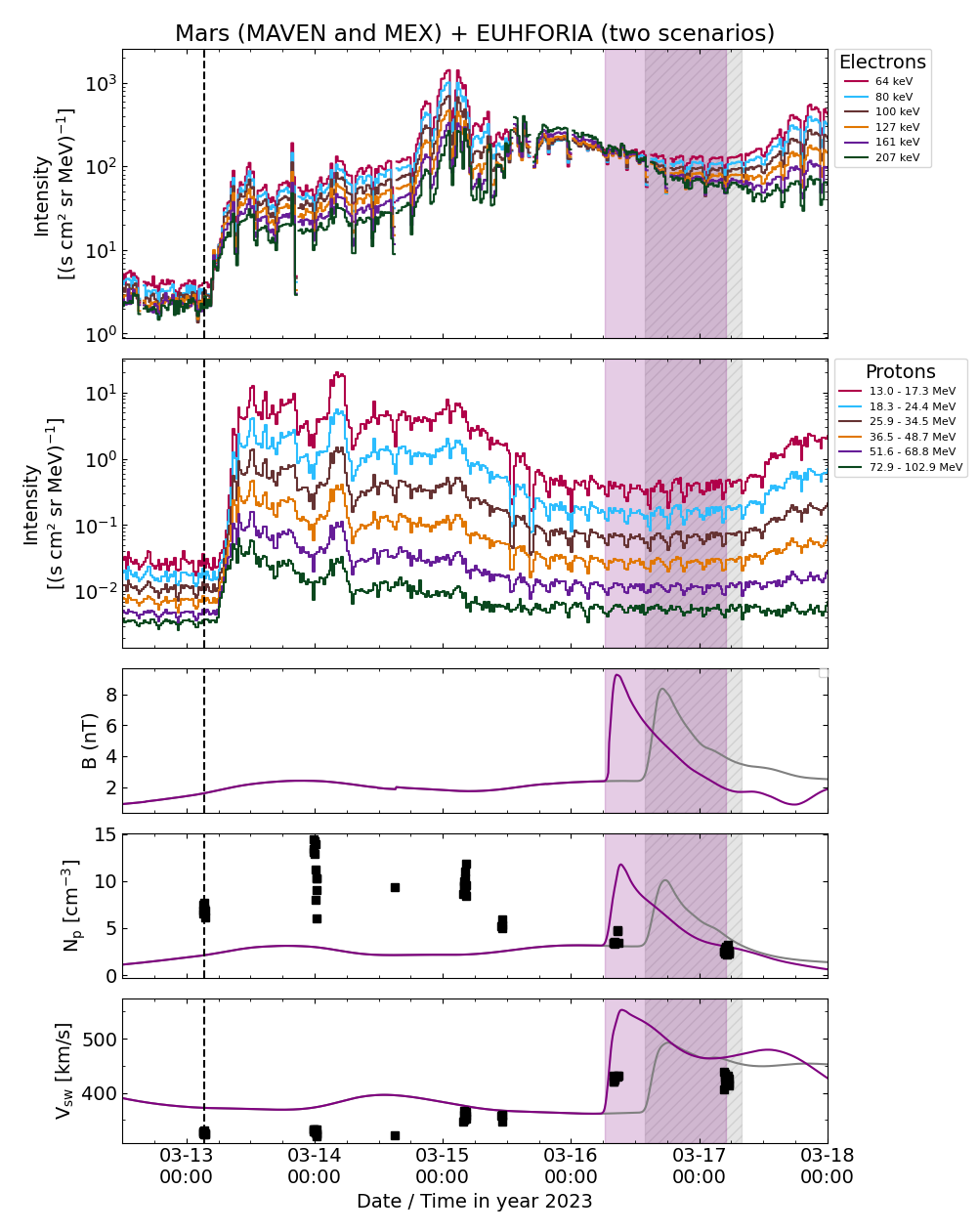} 
        \caption{In-situ observations in comparison with EUHFORIA simulation results for near-Earth S/C (left) and at Mars (right). Same format as in Fig.~\ref{fig:in-situ2}. At Mars energetic particle fluxes were measured by MAVEN and the solar wind speed and density were measured by the Mars Express mission (MEX). The gray shade in the Mars plot represents the CME arrival time based on the pre-event CME simulation, is the CME of the main eruption} 
        \label{fig:in-situ3}
\end{figure*}

{\it Earth-sided observers:}\\
Although far separated from the eruption sector, first-arriving energetic particles are detected at all Earth-sided S/C with rather short delays. A timing analysis presented in Appendix~\ref{app:timing}, allowed us to unambiguously associate these with the main eruption under study. 
In Sect.~\ref{sec:overview} (also Fig.~\ref{fig:solar-mach}, bottom) it was already mentioned that the energetic particle event observed by the front-sided observers differs significantly depending on their magnetic connection to the eastern versus the western edge of the eruption sector. While a rather prompt particle onset (see Figs.~\ref{fig:stereo_pad}, \ref{fig:wind_pad}) and a higher-intensity event is observed at Earth and STEREO~A (both presumably connected closer to the western eruption sector edge), BepiColombo and Solar Orbiter (likely connected closer to the eastern eruption sector edge) observe later and more gradual particle events, as well as lower particle intensities when comparing $\sim$25~MeV protons (Fig.~\ref{fig:solar-mach}, bottom). This difference is even more peculiar given BepiColombo's and Solar Orbiter's smaller radial distances to the Sun, which would typically be expected to facilitate earlier SEP arrival times and higher SEP intensities. Furthermore, Figs.~\ref{fig:sixs_pad_protons}--\ref{fig:ani_soho} show that near-Earth S/C observed SEP anisotropic pitch-angle distributions (PADs), suggesting a magnetic connection with the particle injection region. While the anisotropic PADs are less clear at STEREO~A, BepiColombo measured completely isotropic PADs. We note, however, that Solar Orbiter observed anisotropic PADs, however less during the event onset phase but rather during a later phase, which could suggest that the connection to the source improved with time.

Further differences are found for the peak energies of the observed event: Focusing on proton observations, BepiColombo and Solar Orbiter do not measure a significant intensity increase above 50~MeV (cf.\ Fig.~\ref{fig:in-situ1}, right and \ref{fig:in-situ2}, left). In contrast, the event is clearly observed in the highest proton energy channel of STEREO/HET (60--100 MeV) and in the 80--100 MeV channel of SOHO/ERNE (Fig.~\ref{fig:in-situ2}, right and Fig.~\ref{fig:in-situ3}, left)\footnote{The event was also observed by the Alpha Magnetic Spectrometer (AMS) on board the International Space Station reaching proton rigidities of $>$1 GV, corresponding to $>$400 MeV protons (based on a talk by C.~Consolandi at the European Cosmic Ray Symposium 2024).}. Furthermore, while STEREO~A and SOHO observe clear velocity dispersion in the proton arrival times (second panel from the top in Fig.~\ref{fig:in-situ2}, right, and Fig.~\ref{fig:in-situ3}, left), an inverse velocity dispersion \citep[e.g.,][]{Cohen2024}, that is delayed arrival times of high-energy particles, is observed by BepiColombo and Solar Orbiter (second panel in Fig.~\ref{fig:in-situ1}, right, and second panel in Fig.~\ref{fig:in-situ2}, left). This suggests that BepiColombo and Solar Orbiter are initially connected to regions with less efficient particle acceleration and only hours later accessed a more efficient acceleration region, most likely being the shock. A similar trend is observed for electrons at Solar Orbiter, while at BepiColombo, the electron event is strongly delayed. However, all energy channels at BepiColombo below ${\sim}1$~MeV are strongly contaminated by protons and therefore cannot be used reliably in the analysis.

A peculiar feature of the SEP event at BepiColombo is the presence of lower-energy protons, up to $\sim$4~MeV, which start to rise around 01:00~UT, well before the solar eruption and exclusively observed at BepiColombo. Their origin is ambiguous. Based on the timing association, only one candidate radio type III burst was observed at 23:40~UT on 12 March. The burst is almost exclusively observed by Parker, only a very faint, low-frequency trace is seen at Wind and STEREO~A (see Fig.~\ref{fig:time_line_full}). This suggests that its source region is situated far behind the limb as seen from Solar Orbiter, STEREO~A, and Wind. The lower-energy protons observed by BepiColombo could therefore have diffused to the S/C from this remote region, in agreement with their very gradual rise phase, however not reaching Solar Orbiter nor the other observers. We note that although the type III radio burst was well observed by Parker, no related SEPs were observed by the S/C (Fig.~\ref{fig:in-situ1}, left), suggesting a rather narrow particle injection. Based on the very similar time profiles of ${\leq}4$~MeV protons observed at BepiColombo and Solar Orbiter at times after the main eruption onset (cf.\ Fig.~\ref{fig:in-situ1}, right and Fig.~\ref{fig:in-situ2}, left), we are confident that this part of the event observed at BepiColombo is dominated by the 13 March eruption.

All front-sided, inner-heliospheric S/C observe an ESP event characterized by peaking particle intensities at the time of an in-situ shock crossing. The times of these in-situ shock arrivals are marked by the vertical red lines in each plot. The characteristics of the shocks and their drivers are discussed in detail in  Sect.~\ref{subsec:IP_shock}. 

At Mars (Fig.~\ref{fig:in-situ3}, right), increases in the fluxes of energetic electrons and protons were observed, indicating that the event reached also Mars. The top panel shows the ${\sim}64-207$ keV electron fluxes that are directly measured by the MAVEN/SEP instrument. The second panel shows the derived fluxes of ${\sim}13-103$ MeV energetic protons from the penetrating proton count rates measured by MAVEN/SEP \citep[][see also Appendix A for further details about the derived flux data product]{Lee2023}. We note that these fluxes were scaled with a factor of 0.01 to match the background levels of the other S/C. The prompt onset shows a good timing agreement with the associated solar eruption, and the extended time profiles, especially those of the electrons, resemble the shape of an ESP event. However, due to the S/C orbit geometry during this event, MAVEN was orbiting within the Martian magnetosphere and therefore was not sampling the upstream solar wind conditions. We can therefore not use the magnetic field measuremnets by MAVEN. Furhtermore, while MEX does provide solar wind speed and density observations, the scarce data sampling does not allow us to confirm a potential CME arrival nor the passage of an IP shock and a related ESP event at Mars.

%
\subsection{Characteristics of the interplanetary shock crossing and the interplanetary CME at various S/C}\label{subsec:IP_shock} 

We investigated the characteristics of each shock crossing at the various observers and present these results in Table~\ref{table:shock_params}. 
The shock normals (\(\mathbf{\hat{n}}_{\mathrm{RTN}}\)) indicate that the shock observed by all S/C was moving radially outward. Variations in the transversal components of \(\hat{\mathbf{n}}\) between BepiColombo, Solar Orbiter, STEREO~A, and Wind are, however, noteworthy to emphasize. These may arise from large-scale deformations and/or deflections of the shock front, which are likely to occur when shocks propagate in an inhomogeneous interplanetary medium \citep[][]{Rodriguez-Garcia2022CME,Wijsen23,palmerio2024mesoscale}.  

Based on the parameters shown in Table~\ref{table:shock_params}, we find two potential interpretations, explaining that a shock traversed all observers. First, a circumsolar shock could have formed as a result of the large eruption, potentially composed of two distinct eruptions, on the far side of the Sun, which is supported by the coronagraph observations discussed in Sec.~\ref{subsec:coronal_shock}. The second scenario involves another shock from a front-side eruption that traversed all S/C except for Parker \citep[discussed also by][]{Vlasova2024}. The basis of the first, circumsolar-shock scenario is the CME (likely composed of two simultaneously erupting CMEs as discussed in Sect.~\ref{subsec:coronal_shock}) associated with the main event on 13 March at 03:13~UT, which produced a strong shock. This shock was potentially circumsolar in nature, and part of it likely propagated as a free wave (see Sect.~\ref{subsec:coronal_shock} and Fig.~\ref{fig:coronagraph_images}). This together with the analysis presented by \citet{Wijsen24} aligns with the presence of a circumsolar shock. 

The observed in-situ shocks exhibit a consistent shock normal across all S/C and a shock speed that approximately matches the ballistic arrival times relative to the eruption time at the Sun. Ballistic arrival times assume that the shock speed remains constant from its onset until it arrives in situ. For example, if we assume a constant shock speed from the event onset to L1, we obtain \( V_\mathrm{sh, ballistic} \sim 830\, \mathrm{km\ s}^{-1} \), and to Solar Orbiter \( V_\mathrm{sh, ballistic} \sim 1150\, \mathrm{km\ s}^{-1} \). However, this assumption is flawed because the shock speed cannot remain constant over long distances. Maintaining a constant speed would require a positive rate of energy injection, which is unrealistic whether the shock is driven or freely propagating. Therefore, we adopt a different methodology. \cite{Wijsen24} developed a framework to estimate expected shock speeds for far-side observers by modifying the self-similar expansion solutions of \cite{Sedov46}. They found that the shock speed decreases over time following \( V_\mathrm{sh}(t) \propto t^{-1/3} \). Using a shock speed of approximately \(600\, \mathrm{km\,s}^{-1}\) based on observations at L1 and the arrival times, we estimated the shock speed at Solar Orbiter to be around \(640\, \mathrm{km\,s}^{-1}\), which is close to the observed value. We emphasize that this estimation is purely based on the scaling law obtained from the modified Sedov solutions presented in \cite{Wijsen24}. Since BepiColombo does not provide plasma measurements during its cruise phase, we used the same method to estimate the shock speed at BepiColombo, finding approximately \(655\, \mathrm{km\,s}^{-1}\). 


Table~\ref{table:shock_params} shows that as the shock propagates radially, the angle between the normal to the shock and the magnetic field lines changes from being quasi-parallel ($\theta_{Bn}\simeq0$) to oblique ($\theta_{Bn}\simeq45$) or eventually closer to quasi-perpendicular ($\theta_{Bn}\simeq90$) as the distance increases. This is due to the Parker spiral shape. Namely, the expected $\theta_{Bn}$ value at 1~au for a radially propagating spherical shock in a nominal magnetic field line configuration is close to 45$^\circ$. 
The configurations closer to quasi-perpendicular would mean that the magnetic field lines are distorted \citep[e.g., Figure 7 by][]{Rodriguez-Garcia2022CME}. 

\begin{figure*}[ht!]  
    \centering
        \includegraphics[width=1\textwidth]{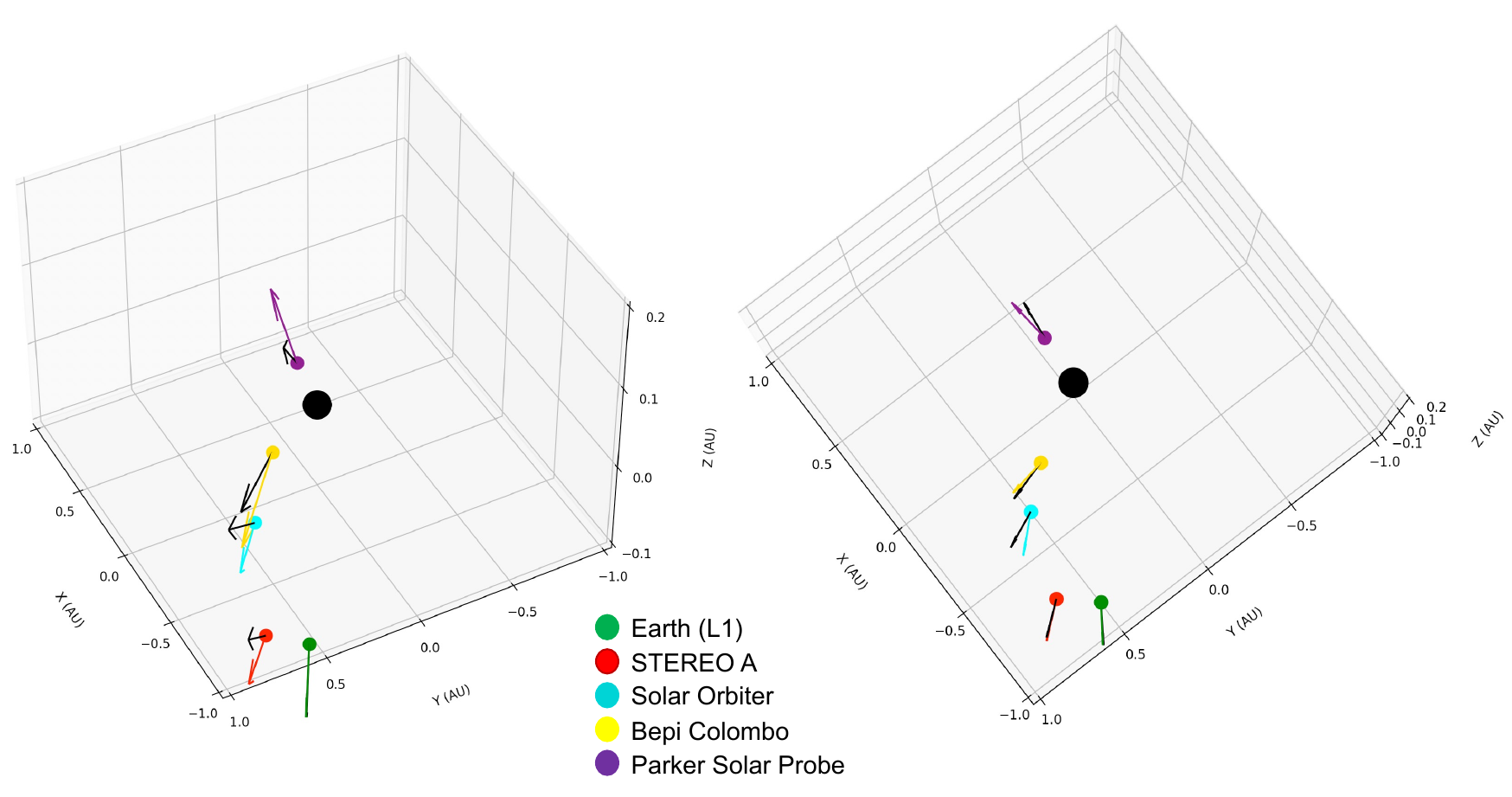}
        \caption{Three-dimensional Cartesian representation of the shock normal estimated at each observer from data and from the EUHFORIA simulation. The Sun (black sphere) is placed at the origin. The shock's normal vector estimated from data are shown with arrows in the same color as the observing S/C. The normal estimated from the EUHFORIA simulation of the circumsolar-shock scenario is shown with the black arrows. } 
        \label{fig:normals}
\end{figure*}

The second scenario for the circumsolar shock observation involves the contribution of different CMEs erupting in different directions around the Sun, each forming their own shock. In this case, the shock traversing Parker and the one passing the front-sided S/C would be driven by distinct CMEs. If this were the case, we would expect the Earth-sided S/C to detect a coherent magnetic driver following the shock. To explore this, we investigated the structures detected immediately after the shock arrivals at the different probes and found varying downstream solar wind characteristics depending on the observer.

At Parker (Fig.~\ref{fig:in-situ1}, left), a clear interplanetary CME (ICME) ejecta with flux rope signatures, that is clear rotations in the magnetic field components, was observed following a brief sheath region, indicating a direct encounter with the CME driver. However, the dominance of the \(B_{R}\) component over the other two components suggests that Parker may have encountered the CME closer to its flank rather than at the nose. At BepiColombo (Fig.~\ref{fig:in-situ1}, right), the shock was followed by highly fluctuating but slightly rotating fields, indicating a possible ejecta passage that could not be clearly distinguished from the preceding sheath. At Solar Orbiter (Fig.~\ref{fig:in-situ2}, left), both a sheath and an ICME ejecta were more discernible, with the ejecta displaying multiple rotations in the \(B_{N}\) component, suggesting a ``more complex'' ejecta than a classic flux rope configuration \citep[e.g.,][]{Rodriguez-Garcia2022CME}. At STEREO~A (Fig.~\ref{fig:in-situ2}, right), turbulent sheath magnetic fields were followed by a smoother region, possibly remnants of ejecta material. Near Earth (Fig.~\ref{fig:in-situ3}, left), no clear ejecta signatures were detected, implying the passage of a driverless shock. Finally, at Mars orbit (Fig.~\ref{fig:in-situ3}, right), magnetic field data cannot be used because MAVEN is situated inside the magnetosphere during the event, and only scarce solar wind speed and density observations near Mars are available from MEX observations. It is therefore not possible to confirm the arrival of a shock wave. It is worth noting that those downstream characteristics with missing signatures of a CME ejecta align with predictions for a freely propagating shock, where the heated downstream plasma acts as a pseudo-driver. Since much of the downstream material is composed of compressed upstream plasma, it decays after a characteristic scale, eventually meeting the colder solar wind and forming a termination wave (reverse shock), which we could, however, not identify in the available data.

Alternatively, if the magnetic structure observed by front-sided S/C were due to a CME encountered at the flank, we would expect similar ICME and shock characteristics, but the shock normals would not be as radially aligned as seen in Table~\ref{table:shock_params}. In particular, the shock normal (\(\mathbf{\hat{n}}_{\mathrm{RTN}}\)) shown in Fig.~\ref{fig:normals} should deviate significantly if the CME had a strong northward or southward propagation such as CME5 (cf. Table~\ref{table:GCS_fit}), particularly affecting the transverse components of \(\mathbf{\hat{n}}_{\mathrm{RTN}}\). Our estimations of \(\mathbf{\hat{n}}_{\mathrm{RTN}}\), shown in Table~\ref{table:shock_params}, do not indicate such deviations. In Fig.~\ref{fig:normals}, both the data-derived \(\mathbf{\hat{n}}_{\mathrm{RTN}}\) (in colors) and the modeled values of the circumsolar-shock scenario \citep[][and Sect.~\ref{subsec:euhforia_sim}]{Wijsen24} are shown. While some local variations in latitude are present, they are consistent with expectations for a circumsolar shock.

\begin{table*}
\caption{In-situ shock characteristics at different observers.}             
\label{table:shock_params}      
\centering          
\begin{tabular}{l l l l l l l l}     
\hline   \\
 \multicolumn{2}{c}{Arrival time}  &   Observer & Shock normal & Geometry & Speed (\(V_\mathrm
 {sh}\)) & Mach number & Gas \\  
 &&&(\(\mathbf{\hat{n}}_{\mathrm{RTN}}\))&(\(\theta_{Bn}\)) & (km s\(^{-1}\))& (\(M_\mathrm{A}\)) & compression\\
 \hline \\
& 13/03/2023 & Parker & \(0.984,-0.07, -0.128\)&\(8^{\circ}\pm 4^{\circ}\)& \(2800\pm 300\) & \(9.1\pm1.35\) & 4\\
& 07:14 UT& & \(\pm(0.01, 0.036, 0.096\)) & & & & \\
& 13/03/2023  & BepiColombo& \(0.878, -0.316, -0.275\)&\(41^{\circ}\pm 18^{\circ}\)& \(-\) & \(1.4\pm0.1\) & \(1.5\pm0.15\)\\
& 14:52 UT & & \(\pm(0.106, 0.148, 0.213\)) & & &  & \\
& 14/03/2023 & Solar Orbiter & \(0.946, -0.141, 0.189\) &\(35^{\circ}\pm 11^{\circ}\)& \(653\pm28 \) & \(3.8\pm0.7\) & 3.7
\\
& 01:04 UT & & \(\pm(0.041, 0.196, 0.081\)) & & & & \\
& 15/03/2023 & STEREO A & (\(0.947, -0.135, -0.091\))&\(51^{\circ}\pm 9^{\circ}\)& \(570\pm 26\) & \(4.2\pm1.04\) & 2.8 \\
& 01:16 UT & & \(\pm(0.087, 0.266, 0.198\)) & & & & \\

& 15/03/2023 & Wind & \(0.961, -0.192, 0.235\) &\(62^{\circ}\pm 12^{\circ}\)& \(607\pm 33\) & \(3.68\pm0.31\) & 2.7 \\
& 04:01 UT & & \(\pm(0.032, 0.183, 0.145\)) & & & & \\
\hline
\end{tabular}
\end{table*}
%
\subsection{EUHFORIA simulations of two different scenarios}
\label{subsec:euhforia_sim}
We conducted large-scale MHD simulations of the inner heliosphere using the EUHFORIA model to investigate the two scenarios discussed above that could explain the circumsolar ESP event. EUHFORIA is a 3D MHD code designed to model the large-scale structure of the solar wind beyond 0.1~au by solving ideal MHD equations with inner boundary conditions derived empirically from solar magnetograms \citep{Pomoell_Poedts2018}. To achieve this, EUHFORIA combines the potential field source surface \citep[PFSS;][]{Wang1992} model with the Schatten current sheet \citep[SCS;][]{Schatten1969} model to extend solar magnetograms to 0.1~au. In our work, we employed the Air Force Data Assimilative Photospheric Flux Transport (ADAPT) model in combination with the Global Oscillation Network Group (GONG) global photospheric magnetic field maps \citep{arge2010,hickmann2015}. ADAPT produces 12 synoptic maps of the Sun's surface magnetic field, accounting for uncertainties in photospheric flows through varying model parameters. These maps provide boundary conditions for the coronal magnetic field extrapolations, which, in turn, are used by the empirical Wang--Sheeley--Arge \citep[WSA;][]{arge2004} model to determine the plasma parameters at 0.1~au, necessary to initiate the heliospheric MHD model. For our simulations, we utilized the first realization of the GONG ADAPT map from 00:00~UT on 13 March 2023.

We then inserted multiple CMEs at the inner boundary (0.1~au) to explore different scenarios. Specifically, we included the five pre-event CMEs (CME1--CME5) and one post-event CME (CME6), detailed in Table~\ref{table:GCS_fit}. Columns (2)--(11) show the summary of the 3D-fitted CME parameters. The sensitivity (deviations) in the parameters of the GCS analysis is given in Table 2 of \cite{Thernisien2009}. 
Columns (2)--(3) show the longitude and latitude of the CME nose in Stonyhurst coordinates, respectively. Columns (4)--(7) respectively represent the angle with respect to the solar equator (tilt), the height from the Sun center of the final step of the reconstruction, the half-angle, and the aspect ratio of the CME. Column (8) shows the  ${R\textsubscript{maj}}=R\textsubscript{min}$+ half-angle. Column (9) represents $R\textsubscript{min}=\arcsin(ratio)$, following \cite{Thernisien2011}. Columns (10)--(11) give the 3D CME nose speed and the UT time in 2023 when the CME nose reaches a height of 21.5 R\textsubscript{$\odot$}. Details on the CME reconstructions are outlined in Appendix~\ref{app:cme_reconstructions}. 

\begin{table*}[htbp]
\caption{3D CME properties derived from the GCS fits}
\label{table:GCS_fit}.
\begin{tabularx}{1.0\textwidth}{ccccccccccc} 
\hline
\hline
 Date-Time  & {Lon}&{Lat}&{Tilt}&{Height}&{$\alpha$}&{Ratio}&{$R\textsubscript{maj}$} &{$R\textsubscript{min}$} & Speed & Date-Time\\
 (UT in 2023, C2)& {(deg)} & {(deg)} & {(deg)} & {(R\textsubscript{$\odot$})} & {(-)}& {(deg)}& {(deg)}& {(deg)}& (km s\textsuperscript{-1})& (UT in 2023, 21.5 (R\textsubscript{$\odot$}) ) \\
\hline
(1)&(2)&(3)&(4)&(5)&(6)&(7)&(8)&(9)&(10)&(11)\\
 \hline
 \hline
03-10 13:36  & 50 & -3  & -3  &  14.20  & 25 &0.31&43&18&502& CME1 03-10 19:29 \\
03-10 17:24  & 28 & -31  & -80  &  15.40& 32 & 0.32 &51 &19&580& CME2 03-10 23:45 \\
03-11 16:36  & 38 & -50  & 1  &  13.40& 20 & 0.26 &35 &15&638&CME3 03-11 22:11 \\
03-12 04:00  & -111 & 35  & 68  &  15.0&33  & 0.35 &54 &20&600&CME4 03-12 09:49 \\
03-12 19:12  & 2 & -42  & -82  &  20.20& 22 & 0.37 &44 &22&781&CME5 03-13 00:05 \\
03-13 10:36  & 31 & 33  & 55  &  14.20& 20 & 0.32 &39 &19&1179&CME6 03-13 13:59 \\
\hline
\\
\end{tabularx}
\footnotesize{ \textbf{Notes.} Column 1: Date and time UT in 2023 of the first appearance in SOHO LASCO C2 field of view. Columns (2)--(3): Stonyhurst coordinates of the CME nose. Columns (4)--(7): angle with respect to the solar equator, height from the Sun center of the final step of the reconstruction, half-angle and aspect ratio. Columns (8)--(9): face-on half-width ($R\textsubscript{min}$ + half-angle), and edge-on half-width ($\arcsin(ratio)$) according to \cite{Thernisien2011}. Column (10): 3D CME nose speed. Column (11): UT time in 2023 at a height of 21.5 (R\textsubscript{$\odot$}.)
}
\end{table*}

\begin{table}
\caption{EUHFORIA parameters for the main event CMEs.}
\label{tab:spheromaks}
\begin{tabular}{lll}
\hline
\hline
& \textbf{CME~a} & \textbf{CME~b} \\
\hline\hline
Date \& Time & 2023-03-13 05:15 & 2023-03-13 05:15 \\
Latitude [deg]         & 15               & -12 \\
Longitude [deg]        & -169             & -109 \\
Radius [R\(_s\)]       & 16               & 12 \\
Density [kg m$^{-1}$]  & 10$^{-18}$         & 10$^{-18}$ \\
Temperature [MK]       & 8.0              & 8.0 \\
Speed [km~s\(^{-1}\)]  & 1800.0           & 1800.0 \\
Tilt Angle [deg]       & 90.0             & 0.0 \\
Hel. Sign [\(\pm 1\)]  & 1.0              & 1.0 \\
Flux [Wb]              & $5 \times 10^{13}$ & $5 \times 10^{13}$ \\
\hline
\end{tabular}
\footnotesize{ \textbf{Notes.} The date and time correspond to the insertion time of the spheromaks at the inner boundary of EUHFORIA at 0.1~au. The latitudes and longitudes indicate the insertion locations of the CMEs' central axis. The kinematic and thermodynamic parameters are assumed to be constant. The magnetic parameters (tilt, helicity, and flux) are described in detail in \citet{verbeke2019}. }
\end{table}
 \begin{figure*}
    \centering
    \includegraphics[width=0.95\linewidth]{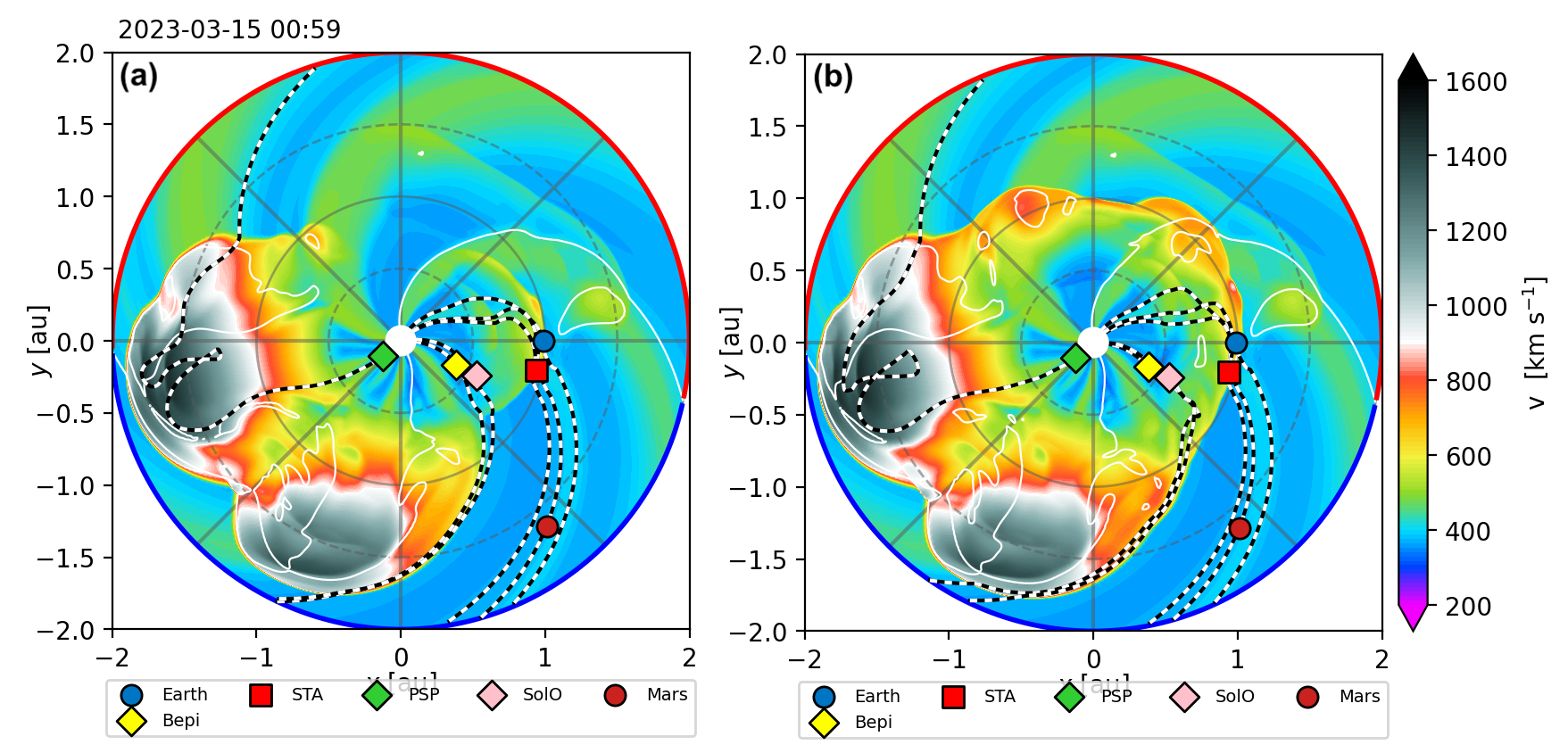}
    \caption{Snapshot from the EUHFORIA simulations showing the solar wind speed in the solar equatorial plane at March 15, 00:59 UT. Panel~(a) depicts the scenario where CME5 is inserted with an increased initial density (details given in the main text), while panel~(b) illustrates the circumsolar blast wave scenario. In both cases, a shock wave is visible arriving at STEREO~A and Earth. The regions with speeds exceeding 1000 km~s\(^{-1}\) identify the two spheromak CMEs. Dashed lines represent the magnetic field lines passing through the various S/C, projected onto the solar equatorial plane. An accompanying movie is available online.}
    \label{fig:euhforia_snap}
\end{figure*}

The pre- and \revise{post-event CMEs} in the EUHFORIA simulation are modeled as cone CMEs, which are ellipsoidal blobs of plasma with increased uniform density, temperature, and speed, while preserving the background solar wind magnetic field. In particular, we used EUHFORIA's default values for the CME's density 
$\rho=10^{-18}$~kg/m$^{-3}$
and temperature $T = 8$~MK \citep[see][]{Pomoell_Poedts2018}. 
The kinematic injection parameters for these CMEs were determined by fitting coronagraph images, as explained in  Appendix~\ref{app:cme_reconstructions}.  

The main event was modelled by injecting two spheromak CMEs \citep[][]{verbeke2019}, with their injection longitudes and latitudes determined based on the ARs identified in Sect.~\ref{sec:overview} as the likely source regions. The spheromaks were assigned injection speeds of 1800~km~s\(^{-1}\), producing a shock speed of approximately 3000~km~s\(^{-1}\), in agreement with the findings of \citet{Jebaraj2024}. The magnetic parameters are based on EUHFORIA’s default values \citep[e.g,][]{verbeke2019} and, as shown in Figure~\ref{fig:in-situ1}, yield reasonable results at Parker. A detailed derivation of the magnetic parameters for the associated magnetic cloud is beyond the scope of this study. An overview of the injection parameters for the spheromak CMEs is provided in Table~\ref{tab:spheromaks}.

As discussed, the observed shock wave near Earth, STEREO~A, and Solar Orbiter could potentially be attributed to a flank encounter with one of the pre-event CMEs. In the EUHFORIA simulation setup described above, however, no pre-event CME shock reaches the front-sided observers at the observed shock time. The only viable candidate is the CME first observed on March 12 at 19:12~UT, which was injected at a longitude of $2^{\circ}$ and a latitude of $-42^{\circ}$, resulting in a mostly southward propagation (CME5 in Table~\ref{table:GCS_fit}). This positioning means that STEREO~A and Earth encounter the central part of the CME's northern flank, while BepiColombo and Solar Orbiter observe its northeastern flank. In the simulation, however, this CME arrives at these S/C more than 12 hours after the observed shock wave.

To better match observations, we increased the CME's injection density  by an order of magnitude from the default value, to \(\rho = 10^{-17}\) kg m\(^{-3}\). This adjustment allowed the northern flank to arrive approximately on time (within 5 hours) at the various S/C without significantly overshooting the observed densities.  Importantly, the modified density remains consistent with the CME density range at 21.5 \(R_s\), as estimated by \citet{Temmer2021}.  The results of this simulation are represented by the gray lines in Figs.~\ref{fig:in-situ1},~\ref{fig:in-situ2}, and~\ref{fig:in-situ3}. 

Since the S/C encounter only the northern tip of the CME's flank, it is plausible that they would not detect any ejecta, as the main magnetic cloud is likely confined to lower latitudes. Interestingly, Earth and STEREO~A, being slightly farther south than Solar Orbiter and BepiColombo, are positioned closer to the injection direction and might therefore have a higher likelihood of observing driver signatures. However, this does not seem to match the in-situ observations, which suggest that BepiColombo and Solar Orbiter may have detected some indication of the CME driver, while STEREO~A and the L1 S/C did not.

Alternatively, and as suggested by the coronagraph images (Fig.~\ref{fig:coronagraph_images}) and the analysis of the in-situ shocks (Sect.~\ref{subsec:IP_shock}), the shock detected in situ by the front-sided observers may be attributed to a broad, potentially circumsolar shock generated by the main event eruption.
To model this scenario, we inserted such a circumsolar shock wave into the EUHFORIA simulation together with the pre-event CMEs CME1--CME5 and the post-event CME6. For simplicity, we introduced the shock wave simultaneously at the inner boundary at 0.1~au on March 13 at 05:00~UT, assuming a blast wave speed of 3000 km~s\(^{-1}\) upon injection, consistent with the findings of \citet{Jebaraj2024}. It is important to note that in EUHFORIA simulations, we typically do not insert actual shock waves, but rather the CME that drives the shock wave. As such, specifying a shock speed directly is not possible with the default EUHFORIA setup.
To address this, we solved the Rankine--Hugoniot jump conditions using EUHFORIA's solar wind at 0.1~au as the upstream plasma together with an assumed shock speed. This yields the downstream values, which we then inserted into the domain.  For more comprehensive details regarding the simulation setup and methodology, we refer the reader to \cite{Wijsen24}.
The results of this simulation are shown by the purple lines in Figs.~\ref{fig:in-situ1}--~\ref{fig:in-situ3}.  
Remarkably, it can be seen that the modelled shock arrives within 1.5~hours of the observed shock arrival at BepiColomobo, Solar Orbiter, and STEREO~A. At L1, the shock arrives approximately 3 hours too early, which can be attributed to the underestimation of the modeled solar wind density upstream of the shock. 
%
Figure~\ref{fig:euhforia_snap} provides a snapshot of the modelled solar wind speed in the solar equatorial plane on March 15, at 00:59~UT, around the time the shock wave was observed in situ at STEREO~A and Earth. Panel (a) illustrates the results of the simulation with the increased density for CME5 (pre-event CME scenario), showing that this adjustment indeed allows the CME flank to arrive at STEREO~A and L1 near the observed shock crossing times. Panel (b) depicts the circumsolar shock wave scenario, also showing a shock wave arriving at the appropriate times. A movie showing the evolution of the simulation from 2023-03-10 22:58~UT to 2023-03-18 02:58~UT is available online.

\subsection{Comparison of in situ observations and EUHFORIA results}
\label{subsec:insitu_model_comp}

Out of the in total six pre- and \revise{post-event CMEs} included in the EUHFORIA simulations, all except CME4 are associated with source locations on the visible side as seen from Earth (cf.\ Table~\ref{table:GCS_fit}). However, only two CMEs (CME2 and CME5) are found to arrive at STEREO~A and Earth, and only one (CME5) at Solar Orbiter and BepiColombo. Each of the two simulations (denser pre-event CME and circumsolar shock scenario) predicts that it is the same structure that passes at all Earth-sided, inner-heliospheric S/C (BepiColombo, Solar Orbiter, STEREO~A, near-Earth S/C). However, depending on the scenario, this is either the circumsolar blast-wave or CME5. In the case of Parker and MAVEN, both simulations predict the main CME (composed of CMEa and CMEb, cf. Fig.~\ref{fig:coronagraph_images}) to arrive at the S/C.

The panels in Fig.~\ref{fig:in-situ1}--\ref{fig:in-situ3} showing the magnetic field and solar wind plasma observations also include the results of the two EUHFORIA simulations in gray (pre-CME scenario) and purple (blast-wave scenario).
The shaded regions in Figs.~\ref{fig:in-situ1}--\ref{fig:in-situ3} mark the predicted passage of the disturbance at the different S/C, with the end-time defined as the point when the density drops below the background solar wind level.
This boundary approximately indicates the onset of the rarefaction region following the passage of the shocked solar wind plasma.

The purple-shaded region corresponds to the passage of the structure connected with the eruption of the main event in the circumsolar-wave scenario. 
The gray-hatched, shaded region corresponds to the passage of CME5 in the pre-event-CME scenario. The yellow-shaded region (only in the STEREO~A and near-Earth plots) marks the arrival of CME2, which is the same in both EUHFORIA simulations. As can be seen in Figs.~\ref{fig:in-situ1}--\ref{fig:in-situ3}, both simulations predict the ESP-event-related shock arrival well in agreement with the time of the observed ESP event for all front-sided, inner-heliospheric observers (BepiColombo, Solar Orbiter, STEREO~A, and Earth). 
While the differences in the predicted arrival times of the ESP-generating shock are small for the two simulated scenarios, the blast-wave scenario (purple shade) matches the observed shock arrival and following turbulent magnetic field enhancement at each of the front-sided observers (BepiColombo, Solar Orbiter, STEREO~A, and Earth) slightly better than the pre-event-CME scenario. In the case of Mars (Fig.~\ref{fig:in-situ3}, right) both simulations predict that the CME of the main eruption (the composite of CMEa and CMEb) passes Mars. However, the predicted arrival times are significantly later than the observed SEP peaks. The lack of available upstream magnetic field observations together with only scarce solar wind measurements at Mars does not permit \revise{inferring} whether any CMEs arrived at this planetary location during the time of interest.

\section{Discussion and Conclusions} \label{sec:discussion}
The 13 March 2023 widespread SEP event is exceptional in various aspects. It shows extremely strong SEP intensities, not only at the well-connected observer, Parker, but also at other inner-heliospheric S/C situated at the farside of the solar eruption. 
Although no imaging observations of the eruption region on the solar surface or lower corona were available during the event, we identified two potential far-sided ARs, one in the northern and one in the southern hemisphere. Based on the area spanned by these two ARs we define a potential solar source region sector spanning 56° in longitude and 68° in latitude, which is separated from all Earth-sided observers by at least 100° in longitude. 

Exceptional are also the prompt SEP arrival times at the far separated observers near-Earth and at STEREO~A, as well as particle anisotropies (Appendix~\ref{app:ani}) observed near Earth. Considering a magnetic connection via a nominal Parker spiral, the longitudinal separation angles between the closest associated source sector and the S/C footpoints would be 124° and 130° for Earth and STEREO~A, respectively (cf.\ Table~\ref{tab:coordinates}). However, the observations rather suggest an early connection to the SEP source region, either by a different magnetic connection or a fast expanding source that rapidly connects to field lines leading to Earth and STEREO~A. 
A magnetic connection according to a nominal Parker-like field suggests that BepiColombo and Solar Orbiter were closer connected to the eastern-limb side of the sector, while STEREO~A and Earth were better connected to the other edge of the sector via the western limb. While the real magnetic field line connections were likely modified by the presence of various pre-event CMEs, the different SEP event characteristics observed at the various S/C still support the scenario of STEREO~A and Earth being connected to a different part of the source than BepiColombo and Solar Orbiter.
Considering a magnetic connection via a nominal Parker spiral, the longitudinal separation angles between the closest associated source sector and the S/C footpoints would be 124° and 130° for Earth and STEREO~A, respectively (cf.\ Table~\ref{tab:coordinates}). However, the prompt SEP arrival times at Earth and STEREO~A, as well as particle anisotropies (Appendix~\ref{app:ani}) observed near Earth suggest a direct connection to the particle source region was rapidly established, most likely by a quickly-expanding shock. In contrast, the SEP event at BepiColombo and Solar Orbiter shows later particle arrival times and lower SEP intensities and peak energies, suggesting a less favorable magnetic connection to the source as compared to STEREO~A and Earth. The inverse velocity dispersion, observed by BepiColombo and Solar Orbiter, might either be caused by the S/C being connected only later to more efficient particle acceleration regions along the shock front, where also higher-energy SEPs are injected, or by the shock itself taking longer to be able to accelerate particles to higher energies.

An important ingredient in forming this exceptional event could have been the sympathetic eruption of two CMEs (Fig.~\ref{fig:coronagraph_images}), which may have formed a commonly-driven and partly freely-propagating shock (cf.\ Sect.~\ref{subsec:coronal_shock}). 
The two ARs, which define our potential source sector, are likely candidates to host these two simultaneously erupting CMEs. 
However, the most surprising feature of this event is the observation of an in-situ shock and a corresponding ESP event at all inner-heliospheric observers all around the Sun. We present two scenarios that could explain this observation, 1) a circumsolar, partly freely-propagating shock that is propagating driverless in the direction towards the Earth-sided S/C, and 2) the involvement of two oppositely directed CMEs, both driving shocks, one passing Parker (associated with the eruption under study potentially composed of two CMEs) and the other one driven by a pre-event CME (CME5), potentially creating the ESP events observed by Solar Orbiter, BepiColombo, STEREO~A, and near Earth \citep[see also][]{Vlasova2024}. We modeled both scenarios using EUHFORIA taking into account the presence of five pre-event CMEs (CME1--CME5, Table~\ref{table:GCS_fit})) and one \revise{post-event CME} (CME6) launched in the period from 10--13 March, and the main eruption, composed of two simultaneously erupting CMEs (CMEa and CMEb) in the simulation. 

We note, however, that in the case of the second scenario, a default EUHFORIA setup using standard CME densities was not able to reproduce the observations at all and only when increasing the density of CME5 by an order of magnitude the simulation produced a shock wave reaching the front-sided observers in time.

Furthermore, CME5, which would be responsible for the ESP event in the pre-CME scenario, was launched at the Sun without an associated SEP event. The CME was first visible in the field of view of LASCO at 19:12~UT on 12 March, but no SEPs, nor a prominent type~III radio burst were observed at any of the inner-heliospheric S/C in temporal coincidence (see Fig.~\ref{fig:in-situ1}--\ref{fig:in-situ3}). Therefore, if CME5 did indeed drive a shock responsible for the multi-S/C ESP event, it most likely formed only further out in the IP medium and must have re(or further)-accelerated SEPs associated with the main eruption of 13 March 2023. 

In the circumsolar shock scenario, the shock propagates freely on the far side of the eruption center toward Earth. This means that shocks observed by S/C other than Parker do not have a driving piston \citep[][]{Howard2016}. This occurs because the magnetic piston had an extremely high internal pressure during its near-Sun expansion, allowing the shock to expand rapidly and spherically. Unlike the piston-driven shock wave detected by Parker, the circumsolar shock exhibits self-similar expansion. Evidence for this self-similarity is observed in the estimated shock parameters across different S/C, as shown in Table~\ref{table:shock_params}. \cite{Wijsen24} demonstrates that while the shock toward Parker is non-spherical and non-self-similar due to its piston-driven nature, the shock remains spherical for observers on the opposite side.


While inhomogeneities in the solar wind may locally deform the shock, overall self-similarity is generally maintained. Furthermore, shock waves can gain energy from the solar wind through interactions with other fast wave modes, which retain their additive properties under ideal MHD. For instance, large-amplitude fast waves from previous CMEs could interact with the shock, increasing its amplitude and allowing it to persist over longer distances. 

A similar scenario was discussed by \cite{Liu17}, who studied the extreme CME event on 23 July 2012. They observed that although the shock exhibited spherical expansion, the shock in the wake region (180$^{\circ}$ from the primary propagation direction) decayed rapidly before reaching 0.3~au (Mercury's orbit). In contrast, in our case, the shock was able to propagate in directions close to the wake (with L1 separated by approximately $130^{\circ}$ on the front side). This extended propagation may be due to the characteristics of the medium and the additional expansion provided by the potential presence of two CMEs, CMEa and CMEb (cf. Fig.~\ref{fig:coronagraph_images}).

To our knowledge, an IP circumsolar shock remaining intact up to 1~au has not been previously reported. However, \citet{Gomez-Herrero2015} suggested a circumsolar shock in the solar corona, and \citet[][]{Lario2016} investigated a very wide CME-driven shock reaching 1~au and extending at least $190^{\circ}$ in longitude. Although the circumsolar shock wave scenario may seem extreme, the EUHFORIA simulation matches the observations exceptionally well, even outperforming the denser-pre-event-CME scenario.

Additionally, several other observations support the circumsolar-shock scenario. The shock normals of the IP shocks observed by all S/C point roughly radially outward, and no clear shock drivers were identified at Earth and STEREO~A, suggesting a freely propagating rather than a CME-driven shock (see Sect.~\ref{subsec:IP_shock}). Furthermore, the scaling of the shock speed is consistent with what is expected from a freely propagating wave. It should be noted that possible ejecta material contributing to a complex ICME may have been present in BepiColombo and Solar Orbiter observations, respectively. While these could be candidates for driving the shock, it is also possible that they are remnants of CME5, arriving coincidentally with the blast wave shock.

Our study reveals that an expansive shock driven by a powerful and wide-ranging solar eruption can have significant implications for SEP production and observations across a wide range of longitudes. This was made possible by the exceptional constellation of S/C available for observations, allowing for a comprehensive analysis of the shock's propagation and effects.

\begin{acknowledgements}
      We acknowledge funding by the European Union’s Horizon 2020 / Horizon Europe research and innovation program under grant agreement No.\ 101004159 (SERPENTINE) and No.\ 101134999 (SOLER). The paper reflects only the authors' view and the European Commission is not responsible for any use that may be made of the information it contains.
      %
      Work in the University of Turku was performed under the umbrella of Finnish Centre of Excellence in Research of Sustainable Space (FORESAIL) funded by the Research Council of Finland (grant No.\ 352847). N.D.\ and I.C.J.\ are grateful for support by the Research Council of Finland (SHOCKSEE, grant No.\ 346902).
      We thank the members of the data analysis working group at the Space Research Laboratory of the University of Turku, Finland for useful discussions.
      N.W.\ acknowledges funding from the Research Foundation -- Flanders (FWO -- Vlaanderen, fellowship no.\ 1184319N) and the KU Leuven research project 3E241013. Computational resources used for the EUHFORIA simulations presented in this work were provided by the VSC (Flemish Supercomputer Center), funded by FWO -- Vlaanderen and the Flemish Government – department EWI.
      E.P.\ acknowledges support from NASA's PSP-GI (grant No.\ 80NSSC22K0349), HGI (grant No.\ 80NSSC23K0447), LWS (grant No.\ 80NSSC19K0067), and LWS-SC (grant No.\ 80NSSC22K0893) programmes, as well as NSF's PREEVENTS (grant No.\ ICER-1854790) programme.
      L.R.-G.\ acknowledges support through the European Space Agency (ESA) research fellowship programme.
      C.O.L.\ acknowledges support from the MAVEN project funded through the NASA Mars Exploration Program, NASA LWS and MDAP (Grants 80NSSC21K1325, 80NSSC21K0119, 80NSSC19K1224) and the IMPACT Investigation funded by the NASA Heliophysics Division through the STEREO Project Office at NASA GSFC.
      W.W.\ acknowledges support from NASA LWS (Grants 80NSSC21K1325, 80NSSC21K0119).
      A.K. acknowledges financial support from NASA’s NNN06AA01C (SO-SIS Phase-E and Parker Solar Probe) contract.
      B.E.\ and P.D.\ acknowledge support from NASA MDAP (Grant 80NSSC19K1224).
      We acknowledge funding by the “Long-term program of support of the Ukrainian research teams at the Polish Academy of Sciences carried out in collaboration with the U.S. National Academy of Sciences with the financial support of external partners” (Grant PAN.BFB.S.BWZ.363.022.2023).
      
      We thank the Solar Orbiter instrument teams for providing the data. Solar Orbiter is a mission of international cooperation between ESA and NASA, operated by ESA.
      BepiColombo is a joint ESA -- JAXA science mission with instruments and contributions directly funded by ESA Member States and JAXA.
      Parker Solar Probe was designed, built, and is now operated by the Johns Hopkins Applied Physics Laboratory as part of NASA’s Living with a Star (LWS) program (contract NNN06AA01C). Support from the LWS management and technical team has played a critical role in the success of the Parker Solar Probe mission.

      B.S.-C. acknowledges support through STFC Ernest Rutherford Fellowship ST/V004115/1
      %
      E.A. acknowledges support from the Research Council of Finland (Academy Research Fellow project SOFTCAT, grant No. 355659). 

\end{acknowledgements}

\bibliographystyle{aa} 
\bibliography{references_imported,referencev2}

\begin{thebibliography}{125}
\expandafter\ifx\csname natexlab\endcsname\relax\def\natexlab#1{#1}\fi

\bibitem[{{Abraham-Shrauner}(1972)}]{Abraham72}
{Abraham-Shrauner}, B. 1972, Journal of Geophysical Research, 77, 736

\bibitem[{{Acu{\~n}a} {et~al.}(2008){Acu{\~n}a}, {Curtis}, {Scheifele}, {Russell}, {Schroeder}, {Szabo}, \& {Luhmann}}]{Acuna2008}
{Acu{\~n}a}, M.~H., {Curtis}, D., {Scheifele}, J.~L., {et~al.} 2008, \ssr, 136, 203

\bibitem[{{Arge} {et~al.}(2010){Arge}, {Henney}, {Koller}, {Compeau}, {Young}, {MacKenzie}, {Fay}, \& {Harvey}}]{arge2010}
{Arge}, C.~N., {Henney}, C.~J., {Koller}, J., {et~al.} 2010, in American Institute of Physics Conference Series, Vol. 1216, Twelfth International Solar Wind Conference, ed. M.~{Maksimovic}, K.~{Issautier}, N.~{Meyer-Vernet}, M.~{Moncuquet}, \& F.~{Pantellini} (AIP), 343--346

\bibitem[{{Arge} {et~al.}(2004){Arge}, {Luhmann}, {Odstrcil}, {Schrijver}, \& {Li}}]{arge2004}
{Arge}, C.~N., {Luhmann}, J.~G., {Odstrcil}, D., {Schrijver}, C.~J., \& {Li}, Y. 2004, Journal of Atmospheric and Solar-Terrestrial Physics, 66, 1295

\bibitem[{Bale {et~al.}(2005)Bale, Balikhin, Horbury, Krasnoselskikh, Kucharek, M{\"o}bius, Walker, Balogh, Burgess, {et~al.}}]{Bale05}
Bale, S., Balikhin, M., Horbury, T., {et~al.} 2005, Space Science Reviews, 118, 161

\bibitem[{{Bale} {et~al.}(2016){Bale}, {Goetz}, {Harvey}, {Turin}, {Bonnell}, {Dudok de Wit}, {Ergun}, {MacDowall}, {Pulupa}, {Andre}, {Bolton}, {Bougeret}, {Bowen}, {Burgess}, {Cattell}, {Chandran}, {Chaston}, {Chen}, {Choi}, {Connerney}, {Cranmer}, {Diaz-Aguado}, {Donakowski}, {Drake}, {Farrell}, {Fergeau}, {Fermin}, {Fischer}, {Fox}, {Glaser}, {Goldstein}, {Gordon}, {Hanson}, {Harris}, {Hayes}, {Hinze}, {Hollweg}, {Horbury}, {Howard}, {Hoxie}, {Jannet}, {Karlsson}, {Kasper}, {Kellogg}, {Kien}, {Klimchuk}, {Krasnoselskikh}, {Krucker}, {Lynch}, {Maksimovic}, {Malaspina}, {Marker}, {Martin}, {Martinez-Oliveros}, {McCauley}, {McComas}, {McDonald}, {Meyer-Vernet}, {Moncuquet}, {Monson}, {Mozer}, {Murphy}, {Odom}, {Oliverson}, {Olson}, {Parker}, {Pankow}, {Phan}, {Quataert}, {Quinn}, {Ruplin}, {Salem}, {Seitz}, {Sheppard}, {Siy}, {Stevens}, {Summers}, {Szabo}, {Timofeeva}, {Vaivads}, {Velli}, {Yehle}, {Werthimer}, \& {Wygant}}]{Bale2016}
{Bale}, S.~D., {Goetz}, K., {Harvey}, P.~R., {et~al.} 2016, \ssr, 204, 49

\bibitem[{{Barabash} {et~al.}(2006){Barabash}, {Lundin}, {Andersson}, {Brinkfeldt}, {Grigoriev}, {Gunell}, {Holmstr{\"o}m}, {Yamauchi}, {Asamura}, {Bochsler}, {Wurz}, {Cerulli-Irelli}, {Mura}, {Milillo}, {Maggi}, {Orsini}, {Coates}, {Linder}, {Kataria}, {Curtis}, {Hsieh}, {Sandel}, {Frahm}, {Sharber}, {Winningham}, {Grande}, {Kallio}, {Koskinen}, {Riihel{\"a}}, {Schmidt}, {S{\"a}les}, {Kozyra}, {Krupp}, {Woch}, {Livi}, {Luhmann}, {McKenna-Lawlor}, {Roelof}, {Williams}, {Sauvaud}, {Fedorov}, \& {Thocaven}}]{Barabash2006}
{Barabash}, S., {Lundin}, R., {Andersson}, H., {et~al.} 2006, \ssr, 126, 113

\bibitem[{{Benkhoff} {et~al.}(2021){Benkhoff}, {Murakami}, {Baumjohann}, {Besse}, {Bunce}, {Casale}, {Cremosese}, {Glassmeier}, {Hayakawa}, {Heyner}, {Hiesinger}, {Huovelin}, {Hussmann}, {Iafolla}, {Iess}, {Kasaba}, {Kobayashi}, {Milillo}, {Mitrofanov}, {Montagnon}, {Novara}, {Orsini}, {Quemerais}, {Reininghaus}, {Saito}, {Santoli}, {Stramaccioni}, {Sutherland}, {Thomas}, {Yoshikawa}, \& {Zender}}]{Benkhoff2021}
{Benkhoff}, J., {Murakami}, G., {Baumjohann}, W., {et~al.} 2021, \ssr, 217, 90

\bibitem[{{Br{\"u}dern} {et~al.}(2022){Br{\"u}dern}, {Berger}, {Heber}, {Heidrich-Meisner}, {Klassen}, {Kollhoff}, {K{\"u}hl}, {Strauss}, {Wimmer-Schweingruber}, \& {Dresing}}]{Bruedern2022}
{Br{\"u}dern}, M., {Berger}, L., {Heber}, B., {et~al.} 2022, \aap, 663, A89

\bibitem[{Brueckner {et~al.}(1995)Brueckner, Howard, Koomen, Korendyke, Michels, Moses, Socker, Dere, Lamy, Llebaria, Bout, Schwenn, Simnett, Bedford, \& Eyles}]{Brueckner1995}
Brueckner, G.~E., Howard, R.~A., Koomen, M.~J., {et~al.} 1995, \solphys, 162, 357

\bibitem[{{Cohen} {et~al.}(2024){Cohen}, {Leske}, {Christian}, {Cummings}, {de Nolfo}, {Desai}, {Giacalone}, {Hill}, {Labrador}, {McComas}, {McNutt}, {Mewaldt}, {Mitchell}, {Mitchell}, {Muro}, {Rankin}, {Schwadron}, {Sharma}, {Shen}, {Szalay}, {Wiedenbeck}, {Xu}, {Romeo}, {Vourlidas}, {Bale}, {Pulupa}, {Kasper}, {Larson}, {Livi}, \& {Whittlesey}}]{Cohen2024}
{Cohen}, C.~M.~S., {Leske}, R.~A., {Christian}, E.~R., {et~al.} 2024, \apj, 966, 148

\bibitem[{{Colburn} \& {Sonett}(1966)}]{Colburn66}
{Colburn}, D.~S. \& {Sonett}, C.~P. 1966, Space Science Reviews, 5, 439

\bibitem[{{Dalla} {et~al.}(2024){Dalla}, {Hutchinson}, {Hyndman}, {Kihara}, {Nitta}, {Rodriguez-Garcia}, {Laitinen}, {Waterfall}, \& {Brown}}]{Dalla2024}
{Dalla}, S., {Hutchinson}, A., {Hyndman}, R.~A., {et~al.} 2024, arXiv e-prints, arXiv:2411.08211

\bibitem[{{Ding} {et~al.}(2024){Ding}, {Li}, {Mason}, {Poedts}, {Kouloumvakos}, {Ho}, {Wijsen}, {Wimmer-Schweingruber}, \& {Rodr{\'\i}guez-Pacheco}}]{Ding2024}
{Ding}, Z., {Li}, G., {Mason}, G., {et~al.} 2024, \aap, 681, A92

\bibitem[{{Domingo} {et~al.}(1995){Domingo}, {Fleck}, \& {Poland}}]{Domingo1995}
{Domingo}, V., {Fleck}, B., \& {Poland}, A.~I. 1995, \solphys, 162, 1

\bibitem[{{Dresing} {et~al.}(2014){Dresing}, {G{\'o}mez-Herrero}, {Heber}, {Klassen}, {Malandraki}, {Dr{\"o}ge}, \& {Kartavykh}}]{Dresing2014}
{Dresing}, N., {G{\'o}mez-Herrero}, R., {Heber}, B., {et~al.} 2014, \aap, 567, A27

\bibitem[{{Dresing} {et~al.}(2018){Dresing}, {G{\'o}mez-Herrero}, {Heber}, {Klassen}, {Temmer}, \& {Veronig}}]{Dresing2018}
{Dresing}, N., {G{\'o}mez-Herrero}, R., {Heber}, B., {et~al.} 2018, \aap, 613, A21

\bibitem[{{Dresing} {et~al.}(2012){Dresing}, {G{\'o}mez-Herrero}, {Klassen}, {Heber}, {Kartavykh}, \& {Dr{\"o}ge}}]{Dresing2012}
{Dresing}, N., {G{\'o}mez-Herrero}, R., {Klassen}, A., {et~al.} 2012, \solphys, 281, 281

\bibitem[{{Dresing} {et~al.}(2022){Dresing}, {Kouloumvakos}, {Vainio}, \& {Rouillard}}]{Dresing2022}
{Dresing}, N., {Kouloumvakos}, A., {Vainio}, R., \& {Rouillard}, A. 2022, \apjl, 925, L21

\bibitem[{{Dresing} {et~al.}(2023){Dresing}, {Rodr{\'\i}guez-Garc{\'\i}a}, {Jebaraj}, {Warmuth}, {Wallace}, {Balmaceda}, {Podladchikova}, {Strauss}, {Kouloumvakos}, {Palmroos}, {Krupar}, {Gieseler}, {Xu}, {Mitchell}, {Cohen}, {de Nolfo}, {Palmerio}, {Carcaboso}, {Kilpua}, {Trotta}, {Auster}, {Asvestari}, {da Silva}, {Dr{\"o}ge}, {Getachew}, {G{\'o}mez-Herrero}, {Grande}, {Heyner}, {Holmstr{\"o}m}, {Huovelin}, {Kartavykh}, {Laurenza}, {Lee}, {Mason}, {Maksimovic}, {Mieth}, {Murakami}, {Oleynik}, {Pinto}, {Pulupa}, {Richter}, {Rodr{\'\i}guez-Pacheco}, {S{\'a}nchez-Cano}, {Schuller}, {Ueno}, {Vainio}, {Vecchio}, {Veronig}, \& {Wijsen}}]{Dresing2023}
{Dresing}, N., {Rodr{\'\i}guez-Garc{\'\i}a}, L., {Jebaraj}, I.~C., {et~al.} 2023, \aap, 674, A105

\bibitem[{{Dresing} {et~al.}(2024){Dresing}, {Yli-Laurila}, {Valkila}, {Gieseler}, {Morosan}, {Farwa}, {Kartavykh}, {Palmroos}, {Jebaraj}, {Jensen}, {K{\"u}hl}, {Heber}, {Espinosa}, {G{\'o}mez-Herrero}, {Kilpua}, {Linho}, {Oleynik}, {Hayes}, {Warmuth}, {Schuller}, {Collier}, {Xiao}, {Asvestari}, {Trotta}, {Mitchell}, {Cohen}, {Labrador}, {Hill}, \& {Vainio}}]{Dresing2024}
{Dresing}, N., {Yli-Laurila}, A., {Valkila}, S., {et~al.} 2024, \aap, 687, A72

\bibitem[{{Dr{\"o}ge} {et~al.}(2014){Dr{\"o}ge}, {Kartavykh}, {Dresing}, {Heber}, \& {Klassen}}]{Droege2014}
{Dr{\"o}ge}, W., {Kartavykh}, Y.~Y., {Dresing}, N., {Heber}, B., \& {Klassen}, A. 2014, Journal of Geophysical Research (Space Physics), 119, 6074

\bibitem[{{Dr{\"o}ge} {et~al.}(2016){Dr{\"o}ge}, {Kartavykh}, {Dresing}, \& {Klassen}}]{Droege2016}
{Dr{\"o}ge}, W., {Kartavykh}, Y.~Y., {Dresing}, N., \& {Klassen}, A. 2016, \apj, 826, 134

\bibitem[{{Fox} {et~al.}(2016){Fox}, {Velli}, {Bale}, {Decker}, {Driesman}, {Howard}, {Kasper}, {Kinnison}, {Kusterer}, {Lario}, {Lockwood}, {McComas}, {Raouafi}, \& {Szabo}}]{Fox2016}
{Fox}, N.~J., {Velli}, M.~C., {Bale}, S.~D., {et~al.} 2016, \ssr, 204, 7

\bibitem[{{Freeland} \& {Handy}(1998)}]{Freeland1998}
{Freeland}, S.~L. \& {Handy}, B.~N. 1998, \solphys, 182, 497

\bibitem[{Galvin {et~al.}(2008)Galvin, Kistler, Popecki, Farrugia, Simunac, Ellis, M\"{o}bius, Lee, Boehm, Carroll, Crawshaw, Conti, Demaine, Ellis, Gaidos, Googins, Granoff, Gustafson, Heirtzler, King, Knauss, Levasseur, Longworth, Singer, Turco, Vachon, Vosbury, Widholm, Blush, Karrer, Bochsler, Daoudi, Etter, Fischer, Jost, Opitz, Sigrist, Wurz, Klecker, Ertl, Seidenschwang, Wimmer-Schweingruber, Koeten, Thompson, \& Steinfeld}]{Galvin2008}
Galvin, A.~B., Kistler, L.~M., Popecki, M.~A., {et~al.} 2008, \ssr, 136, 437

\bibitem[{{Gedalin} \& {Ganushkina}(2022)}]{Gedalin22JPP}
{Gedalin}, M. \& {Ganushkina}, N. 2022, Journal of Plasma Physics, 88, 905880301

\bibitem[{Gedalin {et~al.}(2022)Gedalin, Golan, Pogorelov, \& Roytershteyn}]{Gedalin22}
Gedalin, M., Golan, M., Pogorelov, N.~V., \& Roytershteyn, V. 2022, The Astrophysical Journal, 940, 21

\bibitem[{Golub \& Van~Loan(2013)}]{Golub13}
Golub, G.~H. \& Van~Loan, C.~F. 2013, Matrix computations (JHU press)

\bibitem[{{G{\'o}mez-Herrero} {et~al.}(2015){G{\'o}mez-Herrero}, {Dresing}, {Klassen}, {Heber}, {Lario}, {Agueda}, {Malandraki}, {Blanco}, {Rodr{\'{\i}}guez-Pacheco}, \& {Banjac}}]{Gomez-Herrero2015}
{G{\'o}mez-Herrero}, R., {Dresing}, N., {Klassen}, A., {et~al.} 2015, \apj, 799, 55

\bibitem[{{G{\'o}mez-Herrero} {et~al.}(2021){G{\'o}mez-Herrero}, {Pacheco}, {Kollhoff}, {Espinosa Lara}, {Freiherr von Forstner}, {Dresing}, {Lario}, {Balmaceda}, {Krupar}, {Malandraki}, {Aran}, {Bu{\v{c}}{\'\i}k}, {Klassen}, {Klein}, {Cernuda}, {Eldrum}, {Reid}, {Mitchell}, {Mason}, {Ho}, {Rodr{\'\i}guez-Pacheco}, {Wimmer-Schweingruber}, {Heber}, {Berger}, {Allen}, {Janitzek}, {Laurenza}, {De Marco}, {Wijsen}, {Kartavykh}, {Dr{\"o}ge}, {Horbury}, {Maksimovic}, {Owen}, {Vecchio}, {Bonnin}, {Kruparova}, {P{\'\i}{\v{s}}a}, {Sou{\v{c}}ek}, {Louarn}, {Fedorov}, {O'Brien}, {Evans}, {Angelini}, {Zucca}, {Prieto}, {S{\'a}nchez-Prieto}, {Carrasco}, {Blanco}, {Parra}, {Rodr{\'\i}guez-Polo}, {Mart{\'\i}n}, {Terasa}, {Boden}, {Kulkarni}, {Ravanbakhsh}, {Yedla}, {Xu}, {Andrews}, {Schlemm}, {Seifert}, {Tyagi}, {Lees}, \& {Hayes}}]{Gomez-Herrero2021}
{G{\'o}mez-Herrero}, R., {Pacheco}, D., {Kollhoff}, A., {et~al.} 2021, \aap, 656, L3

\bibitem[{Heyner {et~al.}(2021)Heyner, Auster, Forna{\c{c}}on, Carr, Richter, Mieth, Kolhey, Exner, Motschmann, Baumjohann, Matsuoka, Magnes, Berghofer, Fischer, Plaschke, Nakamura, Narita, Delva, Volwerk, Balogh, Dougherty, Horbury, Langlais, Mandea, Masters, Oliveira, S{\'{a}}nchez-Cano, Slavin, Vennerstr{\o}m, Vogt, Wicht, \& Glassmeier}]{Heyner2021}
Heyner, D., Auster, H.-U., Forna{\c{c}}on, K.-H., {et~al.} 2021, Space Science Reviews, 217, 52

\bibitem[{{Hickmann} {et~al.}(2015){Hickmann}, {Godinez}, {Henney}, \& {Arge}}]{hickmann2015}
{Hickmann}, K.~S., {Godinez}, H.~C., {Henney}, C.~J., \& {Arge}, C.~N. 2015, \solphys, 290, 1105

\bibitem[{{Hill} {et~al.}(2017){Hill}, {Mitchell}, {Andrews}, {Cooper}, {Gurnee}, {Hayes}, {Layman}, {McNutt}, {Nelson}, {Parker}, {Schlemm}, {Stokes}, {Begley}, {Boyle}, {Burgum}, {Do}, {Dupont}, {Gold}, {Haggerty}, {Hoffer}, {Hutcheson}, {Jaskulek}, {Krimigis}, {Liang}, {London}, {Noble}, {Roelof}, {Seifert}, {Strohbehn}, {Vandegriff}, \& {Westlake}}]{Hill2017}
{Hill}, M.~E., {Mitchell}, D.~G., {Andrews}, G.~B., {et~al.} 2017, J. Geophys. Res. Space Phys., 122, 1513

\bibitem[{{Horbury} {et~al.}(2020){Horbury}, {O'Brien}, {Carrasco Blazquez}, {Bendyk}, {Brown}, {Hudson}, {Evans}, {Oddy}, {Carr}, {Beek}, {Cupido}, {Bhattacharya}, {Dominguez}, {Matthews}, {Myklebust}, {Whiteside}, {Bale}, {Baumjohann}, {Burgess}, {Carbone}, {Cargill}, {Eastwood}, {Erd{\"o}s}, {Fletcher}, {Forsyth}, {Giacalone}, {Glassmeier}, {Goldstein}, {Hoeksema}, {Lockwood}, {Magnes}, {Maksimovic}, {Marsch}, {Matthaeus}, {Murphy}, {Nakariakov}, {Owen}, {Owens}, {Rodriguez-Pacheco}, {Richter}, {Riley}, {Russell}, {Schwartz}, {Vainio}, {Velli}, {Vennerstrom}, {Walsh}, {Wimmer-Schweingruber}, {Zank}, {M{\"u}ller}, {Zouganelis}, \& {Walsh}}]{Horbury2020}
{Horbury}, T.~S., {O'Brien}, H., {Carrasco Blazquez}, I., {et~al.} 2020, \aap, 642, A9

\bibitem[{Howard {et~al.}(2008)Howard, Moses, Vourlidas, Newmark, Socker, Plunkett, Korendyke, Cook, Hurley, Davila, Thompson, {St Cyr}, Mentzell, Mehalick, Lemen, Wuelser, Duncan, Tarbell, Wolfson, Moore, Harrison, Waltham, Lang, Davis, Eyles, Mapson-Menard, Simnett, Halain, Defise, Mazy, Rochus, Mercier, Ravet, Delmotte, Auchere, Delaboudiniere, Bothmer, Deutsch, Wang, Rich, Cooper, Stephens, Maahs, Baugh, McMullin, \& Carter}]{Howard2008}
Howard, R.~A., Moses, J.~D., Vourlidas, A., {et~al.} 2008, \ssr, 136, 67

\bibitem[{Howard \& Pizzo(2016)}]{Howard2016}
Howard, T.~A. \& Pizzo, V.~J. 2016, The Astrophysical Journal, 824, 92

\bibitem[{Huovelin {et~al.}(2020)Huovelin, Vainio, Kilpua, Lehtolainen, Korpela, Esko, Muinonen, Bunce, Martindale, Grande, Andersson, Nenonen, Lehti, Schmidt, Genzer, Vihavainen, Saari, Peltonen, Valtonen, Talvioja, Portin, Narendranath, Jarvinen, Okada, Milillo, Laurenza, Heino, \& Oleynik}]{Huovelin2020}
Huovelin, J., Vainio, R., Kilpua, E., {et~al.} 2020, Space Science Reviews, 216

\bibitem[{{Ippolito} {et~al.}(2005){Ippolito}, {Pommois}, {Zimbardo}, \& {Veltri}}]{Ippolito2005}
{Ippolito}, A., {Pommois}, P., {Zimbardo}, G., \& {Veltri}, P. 2005, \aap, 438, 705

\bibitem[{{Jakosky} {et~al.}(2015){Jakosky}, {Lin}, {Grebowsky}, {Luhmann}, {Mitchell}, {Beutelschies}, {Priser}, {Acuna}, {Andersson}, {Baird}, {Baker}, {Bartlett}, {Benna}, {Bougher}, {Brain}, {Carson}, {Cauffman}, {Chamberlin}, {Chaufray}, {Cheatom}, {Clarke}, {Connerney}, {Cravens}, {Curtis}, {Delory}, {Demcak}, {DeWolfe}, {Eparvier}, {Ergun}, {Eriksson}, {Espley}, {Fang}, {Folta}, {Fox}, {Gomez-Rosa}, {Habenicht}, {Halekas}, {Holsclaw}, {Houghton}, {Howard}, {Jarosz}, {Jedrich}, {Johnson}, {Kasprzak}, {Kelley}, {King}, {Lankton}, {Larson}, {Leblanc}, {Lefevre}, {Lillis}, {Mahaffy}, {Mazelle}, {McClintock}, {McFadden}, {Mitchell}, {Montmessin}, {Morrissey}, {Peterson}, {Possel}, {Sauvaud}, {Schneider}, {Sidney}, {Sparacino}, {Stewart}, {Tolson}, {Toublanc}, {Waters}, {Woods}, {Yelle}, \& {Zurek}}]{Jakosky2015}
{Jakosky}, B.~M., {Lin}, R.~P., {Grebowsky}, J.~M., {et~al.} 2015, \ssr, 195, 3

\bibitem[{{Jebaraj} {et~al.}(2024{\natexlab{a}}){Jebaraj}, {Agapitov}, {Krasnoselskikh}, {Vuorinen}, {Gedalin}, {Choi}, {Palmerio}, {Wijsen}, {Dresing}, {Cohen}, {Kouloumvakos}, {Balikhin}, {Vainio}, {Kilpua}, {Afanasiev}, {Verniero}, {Mitchell}, {Trotta}, {Hill}, {Raouafi}, \& {Bale}}]{Jebaraj2024}
{Jebaraj}, I.~C., {Agapitov}, O., {Krasnoselskikh}, V., {et~al.} 2024{\natexlab{a}}, \apjl, 968, L8

\bibitem[{{Jebaraj} {et~al.}(2024{\natexlab{b}}){Jebaraj}, {Agapitov}, {Gedalin}, {Vuorinen}, {Miceli}, {Cohen}, {Voshchepynets}, {Kouloumvakos}, {Dresing}, {Marmyleva}, {Krasnoselskikh}, {Balikhin}, {Mitchell}, {Labrador}, {Wijsen}, {Palmerio}, {Colomban}, {Pomoell}, {Kilpua}, {Pulupa}, {Mozer}, {Raouafi}, {McComas}, {Bale}, \& {Vainio}}]{Jebaraj24b}
{Jebaraj}, I.~C., {Agapitov}, O.~V., {Gedalin}, M., {et~al.} 2024{\natexlab{b}}, \apjl, 976, L7

\bibitem[{{Jebaraj} {et~al.}(2023{\natexlab{a}}){Jebaraj}, {Dresing}, {Krasnoselskikh}, {Agapitov}, {Gieseler}, {Trotta}, {Wijsen}, {Larosa}, {Kouloumvakos}, {Palmroos}, {Dimmock}, {Kolhoff}, {K{\"u}hl}, {Fleth}, {Fedeli}, {Valkila}, {Lario}, {Khotyaintsev}, \& {Vainio}}]{Jebaraj2023b}
{Jebaraj}, I.~C., {Dresing}, N., {Krasnoselskikh}, V., {et~al.} 2023{\natexlab{a}}, \aap, 680, L7

\bibitem[{{Jebaraj} {et~al.}(2023{\natexlab{b}}){Jebaraj}, {Kouloumvakos}, {Dresing}, {Warmuth}, {Wijsen}, {Palmroos}, {Gieseler}, {Marmyleva}, {Vainio}, {Krupar}, {Wiegelmann}, {Magdalenic}, {Schuller}, {Battaglia}, \& {Fedeli}}]{Jebaraj2023a}
{Jebaraj}, I.~C., {Kouloumvakos}, A., {Dresing}, N., {et~al.} 2023{\natexlab{b}}, \aap, 675, A27

\bibitem[{{Kaiser} {et~al.}(2008){Kaiser}, {Kucera}, {Davila}, {St. Cyr}, {Guhathakurta}, \& {Christian}}]{Kaiser2008}
{Kaiser}, M.~L., {Kucera}, T.~A., {Davila}, J.~M., {et~al.} 2008, \ssr, 136, 5

\bibitem[{{Kasper} {et~al.}(2016){Kasper}, {Abiad}, {Austin}, {Balat-Pichelin}, {Bale}, {Belcher}, {Berg}, {Bergner}, {Berthomier}, {Bookbinder}, {Brodu}, {Caldwell}, {Case}, {Chandran}, {Cheimets}, {Cirtain}, {Cranmer}, {Curtis}, {Daigneau}, {Dalton}, {Dasgupta}, {DeTomaso}, {Diaz-Aguado}, {Djordjevic}, {Donaskowski}, {Effinger}, {Florinski}, {Fox}, {Freeman}, {Gallagher}, {Gary}, {Gauron}, {Gates}, {Goldstein}, {Golub}, {Gordon}, {Gurnee}, {Guth}, {Halekas}, {Hatch}, {Heerikuisen}, {Ho}, {Hu}, {Johnson}, {Jordan}, {Korreck}, {Larson}, {Lazarus}, {Li}, {Livi}, {Ludlam}, {Maksimovic}, {McFadden}, {Marchant}, {Maruca}, {McComas}, {Messina}, {Mercer}, {Park}, {Peddie}, {Pogorelov}, {Reinhart}, {Richardson}, {Robinson}, {Rosen}, {Skoug}, {Slagle}, {Steinberg}, {Stevens}, {Szabo}, {Taylor}, {Tiu}, {Turin}, {Velli}, {Webb}, {Whittlesey}, {Wright}, {Wu}, \& {Zank}}]{Kasper2016}
{Kasper}, J.~C., {Abiad}, R., {Austin}, G., {et~al.} 2016, \ssr, 204, 131

\bibitem[{{Kennel}(1988)}]{Kennel88}
{Kennel}, C.~F. 1988, \jgr, 93, 8545

\bibitem[{{Khoo} {et~al.}(2024){Khoo}, {S{\'a}nchez-Cano}, {Lee}, {Rodr{\'\i}guez-Garc{\'\i}a}, {Kouloumvakos}, {Palmerio}, {Carcaboso}, {Lario}, {Dresing}, {Cohen}, {McComas}, {Lynch}, {Fraschetti}, {Jebaraj}, {Mitchell}, {Nieves-Chinchilla}, {Krupar}, {Pacheco}, {Giacalone}, {Auster}, {Benkhoff}, {Bonnin}, {Christian}, {Ehresmann}, {Fedeli}, {Fischer}, {Heyner}, {Holmstr{\"o}m}, {Leske}, {Maksimovic}, {Mieth}, {Oleynik}, {Pinto}, {Richter}, {Rodr{\'\i}guez-Pacheco}, {Schwadron}, {Schmid}, {Telloni}, {Vecchio}, \& {Wiedenbeck}}]{Khoo2024}
{Khoo}, L.~Y., {S{\'a}nchez-Cano}, B., {Lee}, C.~O., {et~al.} 2024, \apj, 963, 107

\bibitem[{{Klassen} {et~al.}(2016){Klassen}, {Dresing}, {G{\'o}mez-Herrero}, {Heber}, \& {M{\"u}ller-Mellin}}]{Klassen2016}
{Klassen}, A., {Dresing}, N., {G{\'o}mez-Herrero}, R., {Heber}, B., \& {M{\"u}ller-Mellin}, R. 2016, \aap, 593, A31

\bibitem[{Klein {et~al.}(2008)Klein, Krucker, Lointier, \& Kerdraon}]{Klein2008}
Klein, K.-L., Krucker, S., Lointier, G., \& Kerdraon, A. 2008, \aap, 486, 589

\bibitem[{{Klein} {et~al.}(2022){Klein}, {Musset}, {Vilmer}, {Briand}, {Krucker}, {Francesco Battaglia}, {Dresing}, {Palmroos}, \& {Gary}}]{Klein2022}
{Klein}, K.-L., {Musset}, S., {Vilmer}, N., {et~al.} 2022, \aap, 663, A173

\bibitem[{{Klein} {et~al.}(2024){Klein}, {Salas Matamoros}, {Hamini}, \& {Kollhoff}}]{Klein2024}
{Klein}, K.-L., {Salas Matamoros}, C., {Hamini}, A., \& {Kollhoff}, A. 2024, \aap, 690, A382

\bibitem[{{Kollhoff} {et~al.}(2021){Kollhoff}, {Kouloumvakos}, {Lario}, {Dresing}, {G{\'o}mez-Herrero}, {Rodr{\'\i}guez-Garc{\'\i}a}, {Malandraki}, {Richardson}, {Posner}, {Klein}, {Pacheco}, {Klassen}, {Heber}, {Cohen}, {Laitinen}, {Cernuda}, {Dalla}, {Espinosa Lara}, {Vainio}, {K{\"o}berle}, {K{\"u}hl}, {Xu}, {Berger}, {Eldrum}, {Br{\"u}dern}, {Laurenza}, {Kilpua}, {Aran}, {Rouillard}, {Bu{\v{c}}{\'\i}k}, {Wijsen}, {Pomoell}, {Wimmer-Schweingruber}, {Martin}, {B{\"o}ttcher}, {Freiherr von Forstner}, {Terasa}, {Boden}, {Kulkarni}, {Ravanbakhsh}, {Yedla}, {Janitzek}, {Rodr{\'\i}guez-Pacheco}, {Prieto Mateo}, {S{\'a}nchez Prieto}, {Parra Espada}, {Rodr{\'\i}guez Polo}, {Mart{\'\i}nez Hell{\'\i}n}, {Carcaboso}, {Mason}, {Ho}, {Allen}, {Bruce Andrews}, {Schlemm}, {Seifert}, {Tyagi}, {Lees}, {Hayes}, {Bale}, {Krupar}, {Horbury}, {Angelini}, {Evans}, {O'Brien}, {Maksimovic}, {Khotyaintsev}, {Vecchio}, {Steinvall}, \& {Asvestari}}]{Kollhoff2021}
{Kollhoff}, A., {Kouloumvakos}, A., {Lario}, D., {et~al.} 2021, \aap, 656, A20

\bibitem[{{Kouloumvakos} {et~al.}(2022){Kouloumvakos}, {Kwon}, {Rodr{\'\i}guez-Garc{\'\i}a}, {Lario}, {Dresing}, {Kilpua}, {Vainio}, {T{\"o}r{\"o}k}, {Plotnikov}, {Rouillard}, {Downs}, {Linker}, {Malandraki}, {Pinto}, {Riley}, \& {Allen}}]{Kouloumvakos2022}
{Kouloumvakos}, A., {Kwon}, R.~Y., {Rodr{\'\i}guez-Garc{\'\i}a}, L., {et~al.} 2022, \aap, 660, A84

\bibitem[{{Kouloumvakos} {et~al.}(2024){Kouloumvakos}, {Papaioannou}, {Waterfall}, {Dalla}, {Vainio}, {Mason}, {Heber}, {K{\"u}hl}, {Allen}, {Cohen}, {Ho}, {Anastasiadis}, {Rouillard}, {Rodr{\'\i}guez-Pacheco}, {Guo}, {Li}, {H{\"o}rl{\"o}ck}, \& {Wimmer-Schweingruber}}]{Kouloumvakos2024}
{Kouloumvakos}, A., {Papaioannou}, A., {Waterfall}, C.~O.~G., {et~al.} 2024, \aap, 682, A106

\bibitem[{{Kouloumvakos} {et~al.}(2023){Kouloumvakos}, {Vainio}, {Gieseler}, \& {Price}}]{Kouloumvakos2023}
{Kouloumvakos}, A., {Vainio}, R., {Gieseler}, J., \& {Price}, D.~J. 2023, \aap, 669, A58

\bibitem[{{Krucker} {et~al.}(2020){Krucker}, {Hurford}, {Grimm}, {K{\"o}gl}, {Gr{\"o}belbauer}, {Etesi}, {Casadei}, {Csillaghy}, {Benz}, {Arnold}, {Molendini}, {Orleanski}, {Schori}, {Xiao}, {Kuhar}, {Hochmuth}, {Felix}, {Schramka}, {Marcin}, {Kobler}, {Iseli}, {Dreier}, {Wiehl}, {Kleint}, {Battaglia}, {Lastufka}, {Sathiapal}, {Lapadula}, {Bednarzik}, {Birrer}, {Stutz}, {Wild}, {Marone}, {Skup}, {Cichocki}, {Ber}, {Rutkowski}, {Bujwan}, {Juchnikowski}, {Winkler}, {Darmetko}, {Michalska}, {Seweryn}, {Bia{\l}ek}, {Osica}, {Sylwester}, {Kowalinski}, {{\'S}cis{\l}owski}, {Siarkowski}, {St{\k{e}}{\'s}licki}, {Mrozek}, {Podg{\'o}rski}, {Meuris}, {Limousin}, {Gevin}, {Le Mer}, {Brun}, {Strugarek}, {Vilmer}, {Musset}, {Maksimovi{\'c}}, {F{\'a}rn{\'\i}k}, {Koz{\'a}{\v{c}}ek}, {Ka{\v{s}}parov{\'a}}, {Mann}, {{\"O}nel}, {Warmuth}, {Rendtel}, {Anderson}, {Bauer}, {Dionies}, {Paschke}, {Pl{\"u}schke}, {Woche}, {Schuller}, {Veronig}, {Dickson}, {Gallagher}, {Maloney}, {Bloomfield}, {Piana}, {Massone}, {Benvenuto}, {Massa},
  {Schwartz}, {Dennis}, {van Beek}, {Rodr{\'\i}guez-Pacheco}, \& {Lin}}]{Krucker2020}
{Krucker}, S., {Hurford}, G.~J., {Grimm}, O., {et~al.} 2020, \aap, 642, A15

\bibitem[{{Laitinen} {et~al.}(2016){Laitinen}, {Kopp}, {Effenberger}, {Dalla}, \& {Marsh}}]{Laitinen2016}
{Laitinen}, T., {Kopp}, A., {Effenberger}, F., {Dalla}, S., \& {Marsh}, M.~S. 2016, \aap, 591, A18

\bibitem[{Lario {et~al.}(2013)Lario, Aran, G{\'o}mez-Herrero, Dresing, Heber, Ho, Decker, \& Roelof}]{Lario2013}
Lario, D., Aran, A., G{\'o}mez-Herrero, R., {et~al.} 2013, \apj, 767, 41

\bibitem[{{Lario} {et~al.}(2016){Lario}, {Kwon}, {Vourlidas}, {Raouafi}, {Haggerty}, {Ho}, {Anderson}, {Papaioannou}, {G{\'o}mez-Herrero}, {Dresing}, \& {Riley}}]{Lario2016}
{Lario}, D., {Kwon}, R.-Y., {Vourlidas}, A., {et~al.} 2016, \apj, 819, 72

\bibitem[{{Lario} {et~al.}(2014){Lario}, {Raouafi}, {Kwon}, {Zhang}, {G{\'o}mez-Herrero}, {Dresing}, \& {Riley}}]{Lario2014}
{Lario}, D., {Raouafi}, N.~E., {Kwon}, R.-Y., {et~al.} 2014, \apj, 797, 8

\bibitem[{{Lario} {et~al.}(2022){Lario}, {Wijsen}, {Kwon}, {S{\'a}nchez-Cano}, {Richardson}, {Pacheco}, {Palmerio}, {Stevens}, {Szabo}, {Heyner}, {Dresing}, {G{\'o}mez-Herrero}, {Carcaboso}, {Aran}, {Afanasiev}, {Vainio}, {Riihonen}, {Poedts}, {Br{\"u}den}, {Xu}, \& {Kollhoff}}]{Lario2022}
{Lario}, D., {Wijsen}, N., {Kwon}, R.~Y., {et~al.} 2022, \apj, 934, 55

\bibitem[{{Larson} {et~al.}(2015){Larson}, {Lillis}, {Lee}, {Dunn}, {Hatch}, {Robinson}, {Glaser}, {Chen}, {Curtis}, {Tiu}, {Lin}, {Luhmann}, \& {Jakosky}}]{Larson2015}
{Larson}, D.~E., {Lillis}, R.~J., {Lee}, C.~O., {et~al.} 2015, \ssr, 195, 153

\bibitem[{{Lee} {et~al.}(2023){Lee}, {Dunn}, \& {Ehresmann}}]{Lee2023}
{Lee}, C.~O., {Dunn}, P., \& {Ehresmann}, B. 2023, in AGU Fall Meeting Abstracts, Vol. 2023, P43H--3366

\bibitem[{{Lee} {et~al.}(2018){Lee}, {Jakosky}, {Luhmann}, {Brain}, {Mays}, {Hassler}, {Holmstr{\"o}m}, {Larson}, {Mitchell}, {Mazelle}, \& {Halekas}}]{Lee2018}
{Lee}, C.~O., {Jakosky}, B.~M., {Luhmann}, J.~G., {et~al.} 2018, \grl, 45, 8871

\bibitem[{{Lemen} {et~al.}(2012){Lemen}, {Title}, {Akin}, {Boerner}, {Chou}, {Drake}, {Duncan}, {Edwards}, {Friedlaender}, {Heyman}, {Hurlburt}, {Katz}, {Kushner}, {Levay}, {Lindgren}, {Mathur}, {McFeaters}, {Mitchell}, {Rehse}, {Schrijver}, {Springer}, {Stern}, {Tarbell}, {Wuelser}, {Wolfson}, {Yanari}, {Bookbinder}, {Cheimets}, {Caldwell}, {Deluca}, {Gates}, {Golub}, {Park}, {Podgorski}, {Bush}, {Scherrer}, {Gummin}, {Smith}, {Auker}, {Jerram}, {Pool}, {Soufli}, {Windt}, {Beardsley}, {Clapp}, {Lang}, \& {Waltham}}]{Lemen2012}
{Lemen}, J.~R., {Title}, A.~M., {Akin}, D.~J., {et~al.} 2012, \solphys, 275, 17

\bibitem[{{Lepping} {et~al.}(1995){Lepping}, {Ac{\~{u}}na}, {Burlaga}, {Farrell}, {Slavin}, {Schatten}, {Mariani}, {Ness}, {Neubauer}, {Whang}, {Byrnes}, {Kennon}, {Panetta}, {Scheifele}, \& {Worley}}]{Lepping1995}
{Lepping}, R.~P., {Ac{\~{u}}na}, M.~H., {Burlaga}, L.~F., {et~al.} 1995, \ssr, 71, 207

\bibitem[{{Lin} {et~al.}(1995){Lin}, {Anderson}, {Ashford}, {Carlson}, {Curtis}, {Ergun}, {Larson}, {McFadden}, {McCarthy}, {Parks}, {R{\`e}me}, {Bosqued}, {Coutelier}, {Cotin}, {D'Uston}, {Wenzel}, {Sanderson}, {Henrion}, {Ronnet}, \& {Paschmann}}]{Lin1995}
{Lin}, R.~P., {Anderson}, K.~A., {Ashford}, S., {et~al.} 1995, \ssr, 71, 125

\bibitem[{{Liu} {et~al.}(2009){Liu}, {Alexander}, \& {Gilbert}}]{Liu2009}
{Liu}, R., {Alexander}, D., \& {Gilbert}, H.~R. 2009, \apj, 691, 1079

\bibitem[{{Liu} {et~al.}(2017){Liu}, {Hu}, {Zhu}, {Luhmann}, \& {Vourlidas}}]{Liu17}
{Liu}, Y.~D., {Hu}, H., {Zhu}, B., {Luhmann}, J.~G., \& {Vourlidas}, A. 2017, \apj, 834, 158

\bibitem[{{Livi} {et~al.}(2022){Livi}, {Larson}, {Kasper}, {Abiad}, {Case}, {Klein}, {Curtis}, {Dalton}, {Stevens}, {Korreck}, {Ho}, {Robinson}, {Tiu}, {Whittlesey}, {Verniero}, {Halekas}, {McFadden}, {Marckwordt}, {Slagle}, {Abatcha}, {Rahmati}, \& {McManus}}]{Livi2022}
{Livi}, R., {Larson}, D.~E., {Kasper}, J.~C., {et~al.} 2022, \apj, 938, 138

\bibitem[{{Luhmann} {et~al.}(2008){Luhmann}, {Curtis}, {Schroeder}, {McCauley}, {Lin}, {Larson}, {Bale}, {Sauvaud}, {Aoustin}, {Mewaldt}, {Cummings}, {Stone}, {Davis}, {Cook}, {Kecman}, {Wiedenbeck}, {von Rosenvinge}, {Acuna}, {Reichenthal}, {Shuman}, {Wortman}, {Reames}, {Mueller-Mellin}, {Kunow}, {Mason}, {Walpole}, {Korth}, {Sanderson}, {Russell}, \& {Gosling}}]{Luhmann2008}
{Luhmann}, J.~G., {Curtis}, D.~W., {Schroeder}, P., {et~al.} 2008, \ssr, 136, 117

\bibitem[{{Lynch} {et~al.}(2021){Lynch}, {Palmerio}, {DeVore}, {Kazachenko}, {Dahlin}, {Pomoell}, \& {Kilpua}}]{Lynch2021}
{Lynch}, B.~J., {Palmerio}, E., {DeVore}, C.~R., {et~al.} 2021, \apj, 914, 39

\bibitem[{{Masson} {et~al.}(2013){Masson}, {Antiochos}, \& {DeVore}}]{Masson2013}
{Masson}, S., {Antiochos}, S.~K., \& {DeVore}, C.~R. 2013, \apj, 771, 82

\bibitem[{Mazur {et~al.}(2000)Mazur, Mason, Dwyer, Giacalone, Jokipii, \& Stone}]{Mazur2000}
Mazur, J.~E., Mason, G.~M., Dwyer, J.~R., {et~al.} 2000, \apj, 532, L79

\bibitem[{{McComas} {et~al.}(2016){McComas}, {Alexander}, {Angold}, {Bale}, {Beebe}, {Birdwell}, {Boyle}, {Burgum}, {Burnham}, {Christian}, {Cook}, {Cooper}, {Cummings}, {Davis}, {Desai}, {Dickinson}, {Dirks}, {Do}, {Fox}, {Giacalone}, {Gold}, {Gurnee}, {Hayes}, {Hill}, {Kasper}, {Kecman}, {Klemic}, {Krimigis}, {Labrador}, {Layman}, {Leske}, {Livi}, {Matthaeus}, {McNutt}, {Mewaldt}, {Mitchell}, {Nelson}, {Parker}, {Rankin}, {Roelof}, {Schwadron}, {Seifert}, {Shuman}, {Stokes}, {Stone}, {Vandegriff}, {Velli}, {von Rosenvinge}, {Weidner}, {Wiedenbeck}, \& {Wilson}}]{McComas2016}
{McComas}, D.~J., {Alexander}, N., {Angold}, N., {et~al.} 2016, \ssr, 204, 187

\bibitem[{{Miteva} {et~al.}(2014){Miteva}, {Klein}, {Kienreich}, {Temmer}, {Veronig}, \& {Malandraki}}]{Miteva2014}
{Miteva}, R., {Klein}, K.-L., {Kienreich}, I., {et~al.} 2014, \solphys, accepted

\bibitem[{{M{\"u}ller} {et~al.}(2017){M{\"u}ller}, {Nicula}, {Felix}, {Verstringe}, {Bourgoignie}, {Csillaghy}, {Berghmans}, {Jiggens}, {Garc{\'\i}a-Ortiz}, {Ireland}, {Zahniy}, \& {Fleck}}]{Muller2017}
{M{\"u}ller}, D., {Nicula}, B., {Felix}, S., {et~al.} 2017, \aap, 606, A10

\bibitem[{{M{\"u}ller} {et~al.}(2020){M{\"u}ller}, {St. Cyr}, {Zouganelis}, {Gilbert}, {Marsden}, {Nieves-Chinchilla}, {Antonucci}, {Auch{\`e}re}, {Berghmans}, {Horbury}, {Howard}, {Krucker}, {Maksimovic}, {Owen}, {Rochus}, {Rodriguez-Pacheco}, {Romoli}, {Solanki}, {Bruno}, {Carlsson}, {Fludra}, {Harra}, {Hassler}, {Livi}, {Louarn}, {Peter}, {Sch{\"u}hle}, {Teriaca}, {del Toro Iniesta}, {Wimmer-Schweingruber}, {Marsch}, {Velli}, {De Groof}, {Walsh}, \& {Williams}}]{Muller2020}
{M{\"u}ller}, D., {St. Cyr}, O.~C., {Zouganelis}, I., {et~al.} 2020, \aap, 642, A1

\bibitem[{{M{\"u}ller-Mellin} {et~al.}(2008){M{\"u}ller-Mellin}, {Gomez-Herrero}, {B{\"o}ttcher}, {Klassen}, {Wimmer-Schweingruber}, {Duvet}, \& {Sanderson}}]{Muller-Mellin2008}
{M{\"u}ller-Mellin}, R., {Gomez-Herrero}, R., {B{\"o}ttcher}, S., {et~al.} 2008, International Cosmic Ray Conference, 1, 371

\bibitem[{M\"{u}ller-Mellin {et~al.}(1995)M\"{u}ller-Mellin, Kunow, Fleissner, Pehlke, Rode, R\"{o}schmann, Scharmberg, Sierks, Rusznyak, Mckenna-Lawlor, Elendt, Sequeiros, Meziat, Sanchez, Medina, Peral, Witte, Marsden, \& Henrion}]{Muller-Mellin1995}
M\"{u}ller-Mellin, R., Kunow, H., Fleissner, V., {et~al.} 1995, \solphys, 162, 483

\bibitem[{{Norsham} \& {Sharizat Hamidi}(2019)}]{norsham2019multiwavelength}
{Norsham}, N. A.~M. \& {Sharizat Hamidi}, Z. 2019, in Journal of Physics Conference Series, Vol. 1411, Journal of Physics Conference Series (IOP), 012011

\bibitem[{{Ogilvie} {et~al.}(1995){Ogilvie}, {Chornay}, {Fritzenreiter}, {Hunsaker}, {Keller}, {Lobell}, {Miller}, {Scudder}, {Sittler}, {Torbert}, {Bodet}, {Needell}, {Lazarus}, {Steinberg}, {Tappan}, {Mavretic}, \& {Gergin}}]{Ogilvie1995}
{Ogilvie}, K.~W., {Chornay}, D.~J., {Fritzenreiter}, R.~J., {et~al.} 1995, \ssr, 71, 55

\bibitem[{{Ogilvie} \& {Desch}(1997)}]{Ogilvie1997}
{Ogilvie}, K.~W. \& {Desch}, M.~D. 1997, Advances in Space Research, 20, 559

\bibitem[{{Owen} {et~al.}(2020){Owen}, {Bruno}, {Livi}, {Louarn}, {Al Janabi}, {Allegrini}, {Amoros}, {Baruah}, {Barthe}, {Berthomier}, {Bordon}, {Brockley-Blatt}, {Brysbaert}, {Capuano}, {Collier}, {DeMarco}, {Fedorov}, {Ford}, {Fortunato}, {Fratter}, {Galvin}, {Hancock}, {Heirtzler}, {Kataria}, {Kistler}, {Lepri}, {Lewis}, {Loeffler}, {Marty}, {Mathon}, {Mayall}, {Mele}, {Ogasawara}, {Orlandi}, {Pacros}, {Penou}, {Persyn}, {Petiot}, {Phillips}, {P{\v{r}}ech}, {Raines}, {Reden}, {Rouillard}, {Rousseau}, {Rubiella}, {Seran}, {Spencer}, {Thomas}, {Trevino}, {Verscharen}, {Wurz}, {Alapide}, {Amoruso}, {Andr{\'e}}, {Anekallu}, {Arciuli}, {Arnett}, {Ascolese}, {Bancroft}, {Bland}, {Brysch}, {Calvanese}, {Castronuovo}, {{\v{C}}erm{\'a}k}, {Chornay}, {Clemens}, {Coker}, {Collinson}, {D'Amicis}, {Dandouras}, {Darnley}, {Davies}, {Davison}, {De Los Santos}, {Devoto}, {Dirks}, {Edlund}, {Fazakerley}, {Ferris}, {Frost}, {Fruit}, {Garat}, {G{\'e}not}, {Gibson}, {Gilbert}, {de Giosa}, {Gradone}, {Hailey}, {Horbury},
  {Hunt}, {Jacquey}, {Johnson}, {Lavraud}, {Lawrenson}, {Leblanc}, {Lockhart}, {Maksimovic}, {Malpus}, {Marcucci}, {Mazelle}, {Monti}, {Myers}, {Nguyen}, {Rodriguez-Pacheco}, {Phillips}, {Popecki}, {Rees}, {Rogacki}, {Ruane}, {Rust}, {Salatti}, {Sauvaud}, {Stakhiv}, {Stange}, {Stubbs}, {Taylor}, {Techer}, {Terrier}, {Thibodeaux}, {Urdiales}, {Varsani}, {Walsh}, {Watson}, {Wheeler}, {Willis}, {Wimmer-Schweingruber}, {Winter}, {Yardley}, \& {Zouganelis}}]{Owen2020}
{Owen}, C.~J., {Bruno}, R., {Livi}, S., {et~al.} 2020, \aap, 642, A16

\bibitem[{{Palmerio} {et~al.}(2024){Palmerio}, {Carcaboso}, {Khoo}, {Salman}, {S{\'a}nchez-Cano}, {Lynch}, {Rivera}, {Pal}, {Nieves-Chinchilla}, {Weiss}, {Lario}, {Mieth}, {Heyner}, {Stevens}, {Romeo}, {Zhukov}, {Rodriguez}, {Lee}, {Cohen}, {Rodr{\'\i}guez-Garc{\'\i}a}, {Whittlesey}, {Dresing}, {Oleynik}, {Jebaraj}, {Fischer}, {Schmid}, {Richter}, {Auster}, {Fraschetti}, \& {Mierla}}]{palmerio2024mesoscale}
{Palmerio}, E., {Carcaboso}, F., {Khoo}, L.~Y., {et~al.} 2024, \apj, 963, 108

\bibitem[{{Palmerio} {et~al.}(2021){Palmerio}, {Kilpua}, {Witasse}, {Barnes}, {S{\'a}nchez-Cano}, {Weiss}, {Nieves-Chinchilla}, {M{\"o}stl}, {Jian}, {Mierla}, {Zhukov}, {Guo}, {Rodriguez}, {Lowrance}, {Isavnin}, {Turc}, {Futaana}, \& {Holmstr{\"o}m}}]{Palmerio2021}
{Palmerio}, E., {Kilpua}, E. K.~J., {Witasse}, O., {et~al.} 2021, Space Weather, 19, e2020SW002654

\bibitem[{{Palmroos} {et~al.}(2024){Palmroos}, {Dresing}, {Gieseler}, {Guti\'errez}, \& {Vainio}}]{Palmroos2024}
{Palmroos}, C., {Dresing}, N., {Gieseler}, J., {Guti\'errez}, C.~P., \& {Vainio}, R. 2024, A\&A, under revision

\bibitem[{{Paschmann} \& {Schwartz}(2000)}]{Paschmann00}
{Paschmann}, G. \& {Schwartz}, S.~J. 2000, in ESA Special Publication, Vol. 449, Cluster-II Workshop Multiscale / Multipoint Plasma Measurements, ed. R.~A. {Harris}, 99

\bibitem[{{Pesnell} {et~al.}(2012){Pesnell}, {Thompson}, \& {Chamberlin}}]{Pesnell2012}
{Pesnell}, W.~D., {Thompson}, B.~J., \& {Chamberlin}, P.~C. 2012, \solphys, 275, 3

\bibitem[{{Pomoell} \& {Poedts}(2018)}]{Pomoell_Poedts2018}
{Pomoell}, J. \& {Poedts}, S. 2018, Journal of Space Weather and Space Climate, 8, A35

\bibitem[{{Prise} {et~al.}(2014){Prise}, {Harra}, {Matthews}, {Long}, \& {Aylward}}]{Prise2014}
{Prise}, A.~J., {Harra}, L.~K., {Matthews}, S.~A., {Long}, D.~M., \& {Aylward}, A.~D. 2014, \solphys, 289, 1731

\bibitem[{{Pulupa} {et~al.}(2017){Pulupa}, {Bale}, {Bonnell}, {Bowen}, {Carruth}, {Goetz}, {Gordon}, {Harvey}, {Maksimovic}, {Mart{\'\i}nez-Oliveros}, {Moncuquet}, {Saint-Hilaire}, {Seitz}, \& {Sundkvist}}]{Pulupa2017}
{Pulupa}, M., {Bale}, S.~D., {Bonnell}, J.~W., {et~al.} 2017, J. Geophys. Res. Space Phys., 122, 2836

\bibitem[{{Pulupa} {et~al.}(2024){Pulupa}, {Bale}, {Jebaraj}, {Romeo}, \& {Krucker}}]{Pulupa24}
{Pulupa}, M., {Bale}, S.~D., {Jebaraj}, I.~C., {Romeo}, O., \& {Krucker}, S. 2024, arXiv e-prints, arXiv:2412.05464

\bibitem[{{Richardson} {et~al.}(2014){Richardson}, {von Rosenvinge}, {Cane}, {Christian}, {Cohen}, {Labrador}, {Leske}, {Mewaldt}, {Wiedenbeck}, \& {Stone}}]{Richardson2014}
{Richardson}, I.~G., {von Rosenvinge}, T.~T., {Cane}, H.~V., {et~al.} 2014, \solphys, 289, 3059

\bibitem[{{Rodr{\'\i}guez-Garc{\'\i}a} {et~al.}(2024){Rodr{\'\i}guez-Garc{\'\i}a}, {G{\'o}mez-Herrero}, {Dresing}, {Balmaceda}, {Palmerio}, {Kouloumvakos}, {Jebaraj}, {Espinosa Lara}, {Roco}, {Palmroos}, {Warmuth}, {Nicolaou}, {Mason}, {Guo}, {Laitinen}, {Cernuda}, {Nieves-Chinchilla}, {Fedeli}, {Lee}, {Cohen}, {Owen}, {Ho}, {Malandraki}, {Vainio}, \& {Rodr{\'\i}guez-Pacheco}}]{Rodriguez-Garcia2024}
{Rodr{\'\i}guez-Garc{\'\i}a}, L., {G{\'o}mez-Herrero}, R., {Dresing}, N., {et~al.} 2024, arXiv e-prints, arXiv:2409.04564

\bibitem[{{Rodr{\'\i}guez-Garc{\'\i}a} {et~al.}(2021){Rodr{\'\i}guez-Garc{\'\i}a}, {G{\'o}mez-Herrero}, {Zouganelis}, {Balmaceda}, {Nieves-Chinchilla}, {Dresing}, {Dumbovi{\'c}}, {Nitta}, {Carcaboso}, {dos Santos}, {Jian}, {Mays}, {Williams}, \& {Rodr{\'\i}guez-Pacheco}}]{Rodriguez-Garcia2021}
{Rodr{\'\i}guez-Garc{\'\i}a}, L., {G{\'o}mez-Herrero}, R., {Zouganelis}, I., {et~al.} 2021, \aap, 653, A137

\bibitem[{{Rodr\'{\i}guez-Garc\'{\i}a} {et~al.}(2022){Rodr\'{\i}guez-Garc\'{\i}a}, {Nieves-Chinchilla, T.}, {G\'omez-Herrero, R.}, {Zouganelis, I.}, {Vourlidas, A.}, {Balmaceda, L. A.}, {Dumbovi\'{}c, M.}, {Jian, L. K.}, {Mays, L.}, {Carcaboso, F.}, {dos Santos, L. F. G.}, \& {Rodr\'{\i}guez-Pacheco, J.}}]{Rodriguez-Garcia2022CME}
{Rodr\'{\i}guez-Garc\'{\i}a}, L., {Nieves-Chinchilla, T.}, {G\'omez-Herrero, R.}, {et~al.} 2022, A\&A, 662, A45

\bibitem[{{Rodr{\'\i}guez-Pacheco} {et~al.}(2020){Rodr{\'\i}guez-Pacheco}, {Wimmer-Schweingruber}, {Mason}, {Ho}, {S{\'a}nchez-Prieto}, {Prieto}, {Mart{\'\i}n}, {Seifert}, {Andrews}, {Kulkarni}, {Panitzsch}, {Boden}, {B{\"o}ttcher}, {Cernuda}, {Elftmann}, {Espinosa Lara}, {G{\'o}mez-Herrero}, {Terasa}, {Almena}, {Begley}, {B{\"o}hm}, {Blanco}, {Boogaerts}, {Carrasco}, {Castillo}, {da Silva Fari{\~n}a}, {de Manuel Gonz{\'a}lez}, {Drews}, {Dupont}, {Eldrum}, {Gordillo}, {Guti{\'e}rrez}, {Haggerty}, {Hayes}, {Heber}, {Hill}, {J{\"u}ngling}, {Kerem}, {Knierim}, {K{\"o}hler}, {Kolbe}, {Kulemzin}, {Lario}, {Lees}, {Liang}, {Mart{\'\i}nez Hell{\'\i}n}, {Meziat}, {Montalvo}, {Nelson}, {Parra}, {Paspirgilis}, {Ravanbakhsh}, {Richards}, {Rodr{\'\i}guez-Polo}, {Russu}, {S{\'a}nchez}, {Schlemm}, {Schuster}, {Seimetz}, {Steinhagen}, {Tammen}, {Tyagi}, {Varela}, {Yedla}, {Yu}, {Agueda}, {Aran}, {Horbury}, {Klecker}, {Klein}, {Kontar}, {Krucker}, {Maksimovic}, {Malandraki}, {Owen}, {Pacheco}, {Sanahuja}, {Vainio},
  {Connell}, {Dalla}, {Dr{\"o}ge}, {Gevin}, {Gopalswamy}, {Kartavykh}, {Kudela}, {Limousin}, {Makela}, {Mann}, {{\"O}nel}, {Posner}, {Ryan}, {Soucek}, {Hofmeister}, {Vilmer}, {Walsh}, {Wang}, {Wiedenbeck}, {Wirth}, \& {Zong}}]{Rodriguez-Pacheco2020}
{Rodr{\'\i}guez-Pacheco}, J., {Wimmer-Schweingruber}, R.~F., {Mason}, G.~M., {et~al.} 2020, \aap, 642, A7

\bibitem[{Russell {et~al.}(2016)Russell, Luhmann, \& Strangeway}]{Russell16}
Russell, C.~T., Luhmann, J.~G., \& Strangeway, R.~J. 2016, Space physics: An introduction (Cambridge University Press)

\bibitem[{{Schatten} {et~al.}(1969){Schatten}, {Wilcox}, \& {Ness}}]{Schatten1969}
{Schatten}, K.~H., {Wilcox}, J.~M., \& {Ness}, N.~F. 1969, \solphys, 6, 442

\bibitem[{Sedov(1946)}]{Sedov46}
Sedov, L.~I. 1946, Journal of Applied Mathematics and Mechanics, 10, 241

\bibitem[{{Sonnerup} \& {Scheible}(1998)}]{Sonnerup98}
{Sonnerup}, B. U.~{\"O}. \& {Scheible}, M. 1998, ISSI Scientific Reports Series, 1, 185

\bibitem[{{Strauss} {et~al.}(2017){Strauss}, {Dresing}, \& {Engelbrecht}}]{Strauss2017}
{Strauss}, R.~D.~T., {Dresing}, N., \& {Engelbrecht}, N.~E. 2017, \apj, 837, 43

\bibitem[{{Temmer} {et~al.}(2021){Temmer}, {Holzknecht}, {Dumbovi{\'c}}, {Vr{\v{s}}nak}, {Sachdeva}, {Heinemann}, {Dissauer}, {Scolini}, {Asvestari}, {Veronig}, \& {Hofmeister}}]{Temmer2021}
{Temmer}, M., {Holzknecht}, L., {Dumbovi{\'c}}, M., {et~al.} 2021, Journal of Geophysical Research (Space Physics), 126, e28380

\bibitem[{{Thernisien}(2011)}]{Thernisien2011}
{Thernisien}, A. 2011, \apjs, 194, 33

\bibitem[{{Thernisien} {et~al.}(2009){Thernisien}, {Vourlidas}, \& {Howard}}]{Thernisien2009}
{Thernisien}, A., {Vourlidas}, A., \& {Howard}, R.~A. 2009, \solphys, 256, 111

\bibitem[{{Thernisien} {et~al.}(2006){Thernisien}, {Howard}, \& {Vourlidas}}]{Thernisien2006GCS}
{Thernisien}, A.~F.~R., {Howard}, R.~A., \& {Vourlidas}, A. 2006, \apj, 652, 763

\bibitem[{{Torsti} {et~al.}(1995){Torsti}, {Valtonen}, {Lumme}, {Peltonen}, {Eronen}, {Louhola}, {Riihonen}, {Schultz}, {Teittinen}, {Ahola}, {Holmlund}, {Kelh{\"a}}, {Lepp{\"a}l{\"a}}, {Ruuska}, \& {Str{\"o}mmer}}]{Torsti1995}
{Torsti}, J., {Valtonen}, E., {Lumme}, M., {et~al.} 1995, \solphys, 162, 505

\bibitem[{{Verbeke} {et~al.}(2019){Verbeke}, {Pomoell}, \& {Poedts}}]{verbeke2019}
{Verbeke}, C., {Pomoell}, J., \& {Poedts}, S. 2019, \aap, 627, A111

\bibitem[{{Vlasova} {et~al.}(2024){Vlasova}, {Bazilevskaya}, {Ginzburg}, {Daibog}, {Kalegaev}, {Kaportseva}, {Logachev}, \& {Myagkova}}]{Vlasova2024}
{Vlasova}, N.~A., {Bazilevskaya}, G.~A., {Ginzburg}, E.~A., {et~al.} 2024, Cosmic Research, 62, 197

\bibitem[{{von Rosenvinge} {et~al.}(2008){von Rosenvinge}, {Reames}, {Baker}, {Hawk}, {Nolan}, {Ryan}, {Shuman}, {Wortman}, {Mewaldt}, {Cummings}, {Cook}, {Labrador}, {Leske}, \& {Wiedenbeck}}]{vonRosenvinge2008}
{von Rosenvinge}, T.~T., {Reames}, D.~V., {Baker}, R., {et~al.} 2008, \ssr, 136, 391

\bibitem[{{Wang} \& {Sheeley}(1992)}]{Wang1992}
{Wang}, Y.~M. \& {Sheeley}, N.~R., J. 1992, \apj, 392, 310

\bibitem[{{Whittlesey} {et~al.}(2020){Whittlesey}, {Larson}, {Kasper}, {Halekas}, {Abatcha}, {Abiad}, {Berthomier}, {Case}, {Chen}, {Curtis}, {Dalton}, {Klein}, {Korreck}, {Livi}, {Ludlam}, {Marckwordt}, {Rahmati}, {Robinson}, {Slagle}, {Stevens}, {Tiu}, \& {Verniero}}]{Whittlesey2020}
{Whittlesey}, P.~L., {Larson}, D.~E., {Kasper}, J.~C., {et~al.} 2020, \apjs, 246, 74

\bibitem[{{Wiedenbeck} {et~al.}(2017){Wiedenbeck}, {Angold}, {Birdwell}, {Burnham}, {Christian}, {Cohen}, {Cook}, {Cummings}, {Davis}, {Dirks}, {Do}, {Everett}, {Goodwin}, {Hanley}, {Hernandez}, {Kecman}, {Klemic}, {Labrador}, {Leske}, {Lopez}, {Link}, {McComas}, {Mewaldt}, {Miyasaka}, {Nahory}, {Rankin}, {Riggans}, {Rodriguez}, {Rusert}, {Shuman}, {Simms}, {Stone}, {von Rosenvinge}, {Weidner}, \& {White}}]{Wiedenbeck2017}
{Wiedenbeck}, M.~E., {Angold}, N.~G., {Birdwell}, B., {et~al.} 2017, in International Cosmic Ray Conference, Vol. 301, 35th International Cosmic Ray Conference (ICRC2017), 16

\bibitem[{{Wijsen} {et~al.}(2019){Wijsen}, {Aran}, {Pomoell}, \& {Poedts}}]{Wijsen2019}
{Wijsen}, N., {Aran}, A., {Pomoell}, J., \& {Poedts}, S. 2019, \aap, 624, A47

\bibitem[{{Wijsen} {et~al.}(2025){Wijsen}, {Jebaraj}, {Dresing}, {Kouloumvakos}, {Palmerio}, {Rodr\'iguez-Garc\'ia}, \& {Lario}}]{Wijsen24}
{Wijsen}, N., {Jebaraj}, I., {Dresing}, N., {et~al.} 2025, \aap, submitted

\bibitem[{{Wijsen} {et~al.}(2023){Wijsen}, {Lario}, {S{\'a}nchez-Cano}, {Jebaraj}, {Dresing}, {Richardson}, {Aran}, {Kouloumvakos}, {Ding}, {Niemela}, {Palmerio}, {Carcaboso}, {Vainio}, {Afanasiev}, {Pinto}, {Pacheco}, {Poedts}, \& {Heyner}}]{Wijsen23}
{Wijsen}, N., {Lario}, D., {S{\'a}nchez-Cano}, B., {et~al.} 2023, \apj, 950, 172

\bibitem[{{Wimmer-Schweingruber} {et~al.}(2023){Wimmer-Schweingruber}, {Berger}, {Kollhoff}, {K{\"u}hl}, {Heber}, {Yang}, {Heidrich-Meisner}, {Klassen}, {Gomez-Herrero}, {Rodriguez-Pacheco}, {Ho}, {Mason}, {Janitzek}, {Kouloumvakos}, {Wang}, {Warmuth}, {Lario}, {Carcaboso}, {Owen}, {Bu{\v{c}}{\'\i}k}, {Pacheco}, {Malandraki}, {Allen}, {Rodriguez}, {Shukhobodskaia}, {Espinosa Lara}, {Cernuda}, {B{\"o}ttcher}, {Eldrum}, {Fleth}, \& {Xu}}]{Wimmer2023}
{Wimmer-Schweingruber}, R.~F., {Berger}, L., {Kollhoff}, A., {et~al.} 2023, \aap, 678, A98

\bibitem[{Wuelser(2004)}]{Wuelser2004}
Wuelser, J.-P. 2004, in Proceedings of SPIE, Vol. 5171 (SPIE), 111--122

\bibitem[{{Xiao} {et~al.}(2023){Xiao}, {Maloney, Shane}, {Krucker, S\"am}, {Dickson, Ewan}, {Massa, Paolo}, {Lastufka, Erica}, {Francesco Battaglia, Andrea}, {Etesi, L\'aszl\'o}, {Hochmuth, Nicky}, {Schuller, Fr\'ed\'eric}, {Ryan, Daniel F.}, {Limousin, Olivier}, {Collier, Hannah}, {Warmuth, Alexander}, \& {Piana, Michele}}]{Xiao2023}
{Xiao}, H., {Maloney, Shane}, {Krucker, S\"am}, {et~al.} 2023, A\&A, 673, A142

\bibitem[{{Xie} {et~al.}(2019){Xie}, {St. Cyr}, {M{\"a}kel{\"a}}, \& {Gopalswamy}}]{Xie2019}
{Xie}, H., {St. Cyr}, O.~C., {M{\"a}kel{\"a}}, P., \& {Gopalswamy}, N. 2019, Journal of Geophysical Research (Space Physics), 124, 6384

\bibitem[{{Yashiro} {et~al.}(2004){Yashiro}, {Gopalswamy}, {Michalek}, {St. Cyr}, {Plunkett}, {Rich}, \& {Howard}}]{Yashiro2004}
{Yashiro}, S., {Gopalswamy}, N., {Michalek}, G., {et~al.} 2004, Journal of Geophysical Research (Space Physics), 109, A07105

\bibitem[{{Zhu} {et~al.}(2018){Zhu}, {Liu}, {Kwon}, \& {Wang}}]{Zhu2018}
{Zhu}, B., {Liu}, Y.~D., {Kwon}, R.-Y., \& {Wang}, R. 2018, \apj, 865, 138

\bibitem[{{Zhuang} {et~al.}(2024){Zhuang}, {Lugaz}, {Lario}, {Kwon}, {Chrysaphi}, {Niehof}, {Gou}, \& {Zhao}}]{Zhuang2024}
{Zhuang}, B., {Lugaz}, N., {Lario}, D., {et~al.} 2024, \apj, 963, 119

\end{thebibliography}

\begin{appendix}

\section{Instrumentation used in the analysis} \label{app:instrumentation}

{\it Parker Solar Probe}

Energetic particles measurements are provided by Parker's Integrated Science Investigation of the Sun \citep[IS$\odot$IS;][]{McComas2016} suite. We make use of low-energy electrons detected by the Energetic Particle Instrument-Low \citep[EPI-Lo;][]{Hill2017} and high-energy protons are provided by the Energetic Particle Instrument-High \citep[EPI-Hi;][]{Wiedenbeck2017} consisting of the Low Energy Telescopes (LETs) and High Energy Telescope (HET). Under normal conditions LET and HET, which consist of stacked solid-state detectors, use the standard $dE/dx$ versus residual energy technique to measure ions from $\sim$1 to $>$100 MeV/nuc and electrons in the range $\sim$ 0.5--6~MeV. However, due to very high particle fluxes during the IP shock crossing, EPI-Hi sensors went into dynamic threshold mode 3 (DT3) \citep[see also][]{Cohen2024}. In this mode the energy threshold is raised so much that electrons, protons, and helium are no longer distinguished by species and energy, but instead the instrument focuses on heavy ion measurements. As an alternative to proton measurements, we use the so-called pixel data. The pixels are small areas on several detectors of LET and HET, where the thresholds are not raised. However, as single-detector measurements, these pixels respond to all ions/electrons. Assuming, however, that the energetic ion environment is dominated by protons, allows us to employ these as proton proxies. 
 
Magnetic field measurements at Parker are obtained from the fluxgate magnetometer part of the FIELDS \citep{Bale2016} suite. FIELDS also hosts the Radio Frequency Spectrometer \citep[RFS;][]{Pulupa2017}, which provides radio observations.
Solar wind measurements are provided by the Solar Probe ANalyzer for Ions \citep[SPAN-I;][]{Livi2022} and the Solar Probe Cup (SPC) instrument, part of the Solar Wind Electrons Alphas and Protons \citep[SWEAP;][]{Kasper2016} investigation. We also inspect electron pitch-angle distributions from the Solar Probe ANalyzer for Electrons \citep[SPAN-E;][]{Whittlesey2020}, also part of the SWEAP suite.
\\
\\
{\it BepiColombo}

The BepiColombo S/C was still in its cruise phase en route to Mercury. We use energetic electron and proton measurements from the Solar Intensity X-Ray and Particle Spectrometer \citep[SIXS;][]{Huovelin2020} on board the Mercury Planetary Orbiter (MPO; the European S/C involved in the BepiColombo mission). The SIXS-P particle detector provides measurements of high-energy electrons and protons. The instrument consists of a CsI(Tl) scintillator bar surrounded by five orthogonal detectors, called ``Sides,'', which we use to determine SEP pitch-angle distributions  (see Sect.~\ref{app:ani}). However, Sides 0 and 4 are partially and totally obstructed by the S/C cruise shield, respectively and the detector of Side 3 has noise issues so that only three sides could be used. 
Magnetic field measurements are provided by the MPO magnetometer \citep[MPO-MAG;][]{Heyner2021}. 
\\
\\
{\it Solar Orbiter}

Energetic particles as measured by Solar Orbiter are studied using the Electron Proton Telescope (EPT) and the High Energy Telescope (HET) of the Energetic Particle Detector \citep[EPD;][]{Rodriguez-Pacheco2020}  suite. Both instruments provide four different viewing directions, used to determine SEP pitch-angle distributions (see Sect.~\ref{app:ani}). EPT measures ions and electrons in the energy ranges 20~keV -- 15~MeV and 20--400~keV, respectively, and HET relativistic electrons between 0.3 and 30 MeV and protons between 7 and 107 MeV.
Hard X-ray observations are provided by the Spectrometer/Telescope for Imaging X-rays \citep[STIX;][]{Krucker2020}. 
Magnetic fields at Solar Orbiter are measured with \citep[MAG;][]{Horbury2020} and the Solar Wind Analyzer \citep[SWA;][]{Owen2020} suite provides plasma parameters of the solar wind. 
\\
\\
{\it STEREO~A}



Energetic particles observed by STEREO~A are provided by the In situ Measurements of Particles And CME Transients \citep[IMPACT;][]{Luhmann2008}. The Solar Electron Proton Telescope \citep[SEPT;][]{Muller-Mellin2008}, which provides four viewing directions, is used to study pitch-angle distributions of near-relativistic electrons and low-energy ions. 
We note that since July 2015, after the solar conjunction, the S/C was rolled $180^{\circ}$ about the S/C--Sun line leading to a change in the nominal pointing directions changed, with the Sun and Asun telescopes, which were previously pointing along the nomainal Parker spiral towards and way from the Sun, now pointing perpendicular to it, North pointing southward and South pointing northward. 
Relativistic electron and high-energy proton observations are provided by the High Energy Telescope (HET) \citep{vonRosenvinge2008}.

The interplanetary magnetic field is measured by the Magnetic Field Experiment \citep[MFE;][]{Acuna2008}, part of the IMPACT suite and parameters of the solar wind plasma are obtained by the Plasma and Suprathermal Ion Composition \citep[PLASTIC;][]{Galvin2008} instrument. 
Remote-sensing observations from STEREO~A as used in this study are provided by the Sun Earth Connection Coronal and Heliospheric Investigation \citep[SECCHI;][]{Howard2008} instrument suite, including the Extreme UltraViolet Imager \citep[EUVI;][]{Wuelser2004}, and two coronagraphs (COR1 and COR2) imaging the corona from 1.4 up to 15\,$R_{\odot}$.
\\
\\
{\it Near-Earth S/C}

We use high-energy SEP measurements provided by the SOHO S/C. Protons are detected by the Energetic and Relativistic Nuclei and Electron \citep[ERNE;][]{Torsti1995} in the energy range of a few to a hundred MeV. Electron measurements in the relativistic energy range are provided by the Electron Proton Helium Instrument (EPHIN), part of the Comprehensive Suprathermal and Energetic Particle Analyser \citep[COSTEP;][]{Muller-Mellin1995} suite. Lower-energy SEPs are provided by Wind's Three-Dimensional Plasma and Energetic Particle Investigation \citep[3DP;][]{Lin1995}, measuring energetic electrons up to about 500~keV and protons up to a few MeV over a complete pitch-angle space. Pitch-angle distributions of are studied using 3DP pitch-angle data, which are provided as pre-binned data product into eight pitch-angle sectors. provides energetic particle measurements. Magnetic field and solar wind plasma observations are taken from the Magnetic Field Investigation \citep[MFI;][]{Lepping1995} and the Solar Wind Experiment \citep[SWE;][]{Ogilvie1995}, respectively.

 We also employ coronagraph observations by the Large Angle and Spectrometric Coronagraph \citep[LASCO;][]{Brueckner1995} and Extreme ultraviolet images of the solar corona provided by the Atmospheric Imaging Assembly \citep[AIA;][]{Lemen2012} on board the Solar Dynamics Observatory \citep[SDO;][]{Pesnell2012}.
\\
\\
{\it Mars}

Energetic particle measurements near Mars are obtained from the NASA Mars Atmosphere and Volatile Evolution \citep[MAVEN;][]{Jakosky2015} S/C. The SEP instrument \citep{Larson2015} measures fluxes of electrons with energies up to 200 keV and ions (mostly protons) up to 6 MeV. For higher energy protons, previous studies utilized the MAVEN SEP count rate data to characterize the arrival times of SEP protons at Mars \citep[see, e.g., ][]{Khoo2024, Lee2018}. Recently, work has been done to derive fluxes of $\geq$ 13 MeV protons from measured count rate data. Details of the derivation methodology, which involves running theoretical spectra through the SEP instrument response matrix and comparing the theoretical counts with measured counts, as well as performing statistical fittings (assume broken power laws) to determine the best-fitting proton spectra, is forthcoming \citep[Lee et al., manuscript in preparation; see also][]{Lee2023}.

The Mars Express (MEX) Analyzer of Space Plasmas and Energetic Atoms \citep[ASPERA-3;][]{Barabash2006} Ion Mass Analyzer (IMA) solar wind moments are calculated using a further development of the algorithm behind the official MEX solar wind moments available from the PSA\footnote{\url{ftp://psa.esac.esa.int/pub/mirror/MA
RS-EXPRESS/ASPERA-3/MEX-SUN-ASPERA3-4-SWM-V1.0/}}. Specifically, IMA's sector energy-geometric factors for $\mathrm{H^+}$ and $\mathrm{He^{2+}}$, as well as IMA's energy table, have been recalibrated to minimize systematic discrepancies with simultaneous MAVEN measurements of solar wind density and velocity. Detection of undisturbed upstream solar wind has been greatly improved to exclude measurements in the Martian magnetosheath and foreshock. And crucially for SEP-events, noise reduction is now based on a Poisson-likelihood signal-detection routine, greatly improving sensitivity during events such as the one studied here.

\section{Remote-sensing observations of the solar source regions} \label{app:remotecme}
Because no imaging observations of the flaring AR of the eruption on 13 March 2023 were available the source location is ambiguous. In order to identify the most likely AR candidates we analyzed imaging observations before and after the eruption when the potential source regions were in the field of view of various instruments.
The JHelioviewer \citep{Muller2017} application allows users to view different images from different instruments and S/C simultaneously. It can also be used to select a range of dates to display images as a movie. This helps to understand the dynamics of solar activity and monitor its evolution. Using this tool, we observed solar activity from the various coronagraphs and heliospheric imagers on board STEREO~A, SOHO, and Parker in the days before and after 13 March 2023. By combining the previous observations with the full disk images from SDO/AIA, STEREO~A/EUVI, and Solar Orbiter/EUI, it was then possible to trace the possible origins of the eruption. 
The activity and evolution of the active regions visible on the Sun's surface between 24 February and 22 March 2023 have been studied in detail, and we have been able to identify two active regions likely responsible for the studied eruption: one in the northern hemisphere, AR1, and another in the southern hemisphere, AR2.

It was clear that the AR at the origin of the eruption was not on the Earth-facing disc, so we focused our research on ARs on the far side of the Sun, i.e.\ ARs behind the limb, as seen from the front-sided observers at the time of the eruption (SDO/AIA, STA/EUVI, and Solar Orbiter/EUI). Among them, we eliminated those on the west far side, as they are unlikely to be responsible since the eruption is first seen on the east limb side. This corresponds to a longitude range of [330$^\circ$, 50$^\circ$] in Carrington coordinates.
During Carrington rotation 2267, ARs in this range of longitudes are AR13229, AR13233, AR13234, and AR13235 for the northern hemisphere, AR13230, AR13236, and AR13237 for the southern hemisphere. AR13229 is connected to AR13233 with a filament. AR13230 and AR13236 are very close to each other and will interact as we find them on rotation 2268 under the names AR13256, AR13257 and AR13259, forming a group of ARs that we will call the southern group. Three ARs disappear during solar rotation: AR13233, AR13235, AR13237. AR13234 will become AR13260 but seems to be too far east to be responsible for the March 13 eruption. It would have produced an outward eruption. AR13229 will become AR13258 and is AR1.

AR1 is clearly visible from 23 February to 26 February 2023 with STA/EUVI, where it is very active and shows several eruptions. During the corresponding Carrington rotation 2267 it was named AR13229. This active region disappeared from the view of STA/EUVI on 27 February, and reappeared on 15 March 2023, two days after the main event studied. It is then named AR13258 for this new Carrington rotation (number 2268). We determined the extent of this active region on 20 March 2023, when its contours were well-defined and its shape had changed little since the 13 March eruption, as it had not shown any other activity in the meantime. The Carrington latitude and longitude extensions (as provided in Table~\ref{tab:coordinates}) were measured directly in JHelioviewer using the extreme values of the active region. The center of the active region was determined as the center of the measured latitude and longitude extensions.

\begin{figure}[ht!]
    \centering
    \includegraphics[width=0.45\textwidth]{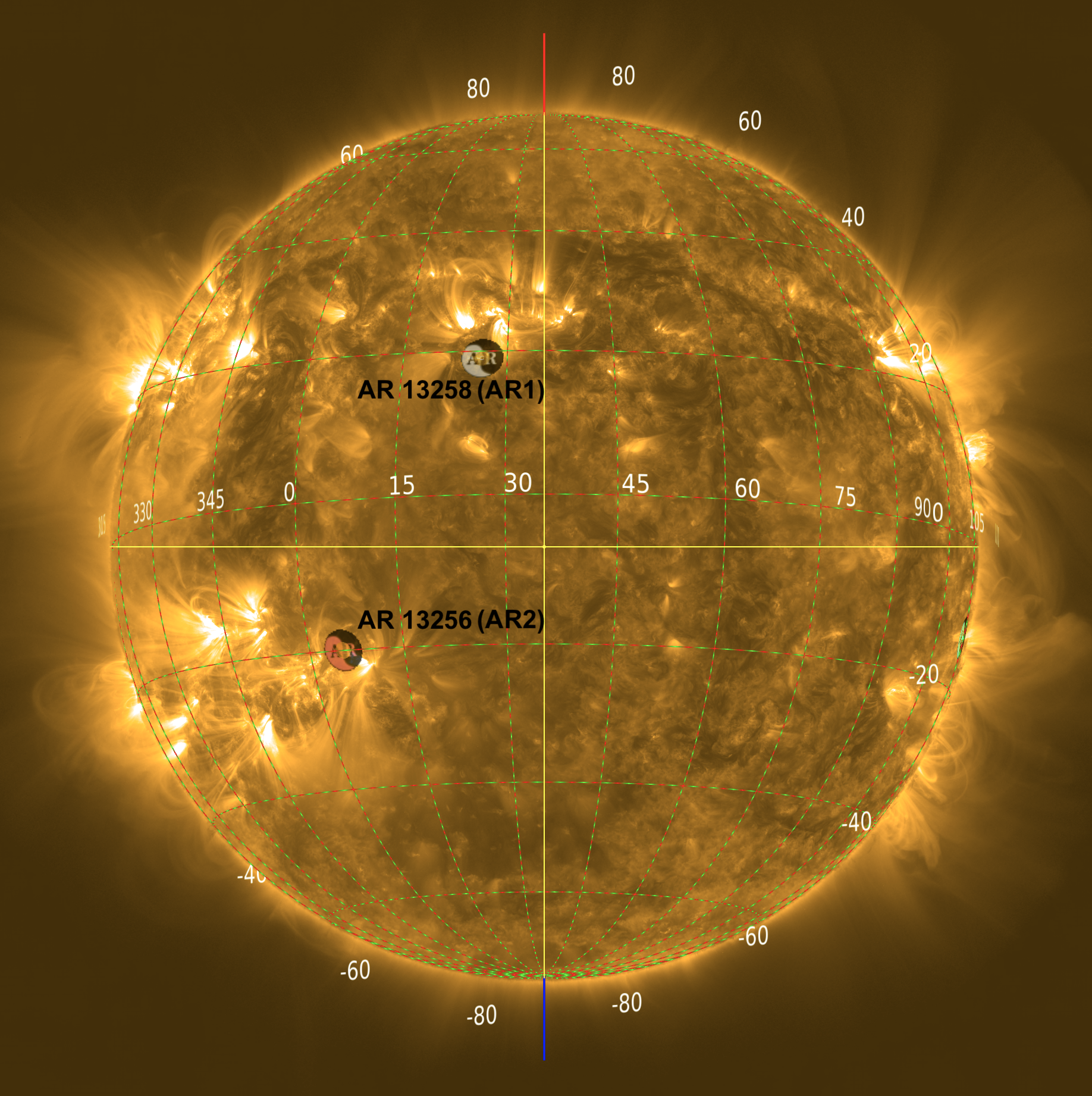}
    \includegraphics[width=0.45\textwidth]{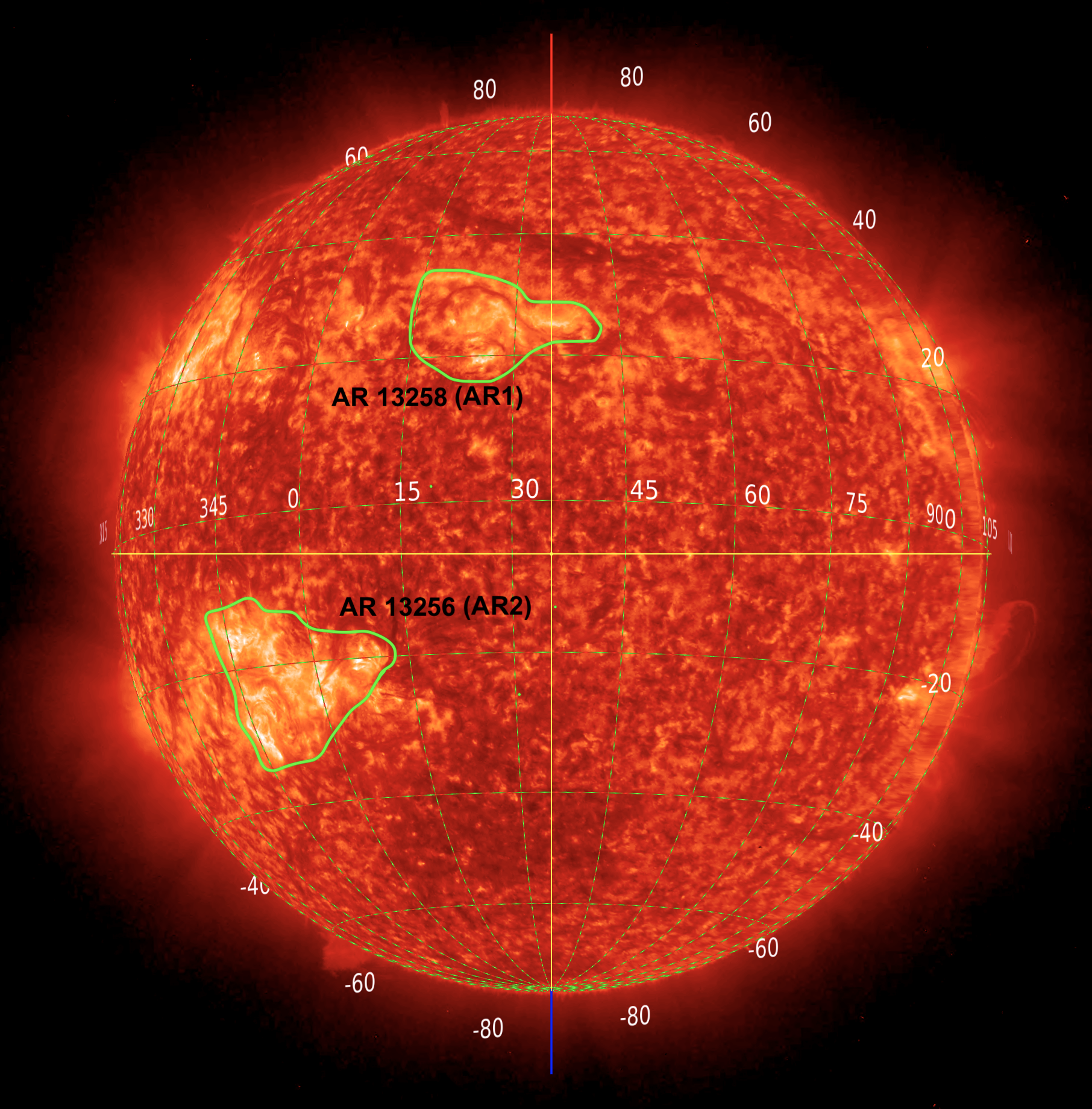}
    \caption{View of the Sun's disk on the 2023-03-21 at 16:50 UT with a grid in Carrington coordinates, from JHelioviewer. Top: SDO/AIA 171\AA. Active regions are identified by NOAA. Bottom: STEREO~A/EUVI 304\AA. The extension of these active regions is circled in green.}
    \label{fig:AR_identification}
\end{figure}

The same procedure was used to identify AR2. Named AR13236 during Carrington rotation 2267, it disappeared from the view of SDO/AIA on 5 March 2023 and reappeared on Solar Orbiter/EUI on 19 March 2023. It is then named AR13256 for Carrington rotation 2268 and difficult to distinguish with two others ARs: AR13257 and AR13259. The shape of this southern group of ARs varied greatly from one rotation to the next, but never exceeded 343$^\circ$ Carrington longitude on the west side. We determined its extension on 21 March 2023 in the same way as AR1.
These two active regions are shown in Fig.~\ref{fig:AR_identification} with their NOAA identification and shape highlighted in green.

\section{SEP anisotropies} \label{app:ani}
Pitch-angle distributions (PADs) of SEPs are an important tool to infer if a direct connection to the source region was present, which is marked by strongly anisotropic PADs. If particle diffusion dominates, isotropic PADs are observed. Pitch-angle scattering, the process that constitutes particle diffusion in magnetized plasmas, can occur either parallel to the mean magnetic field (parallel diffusion) or perpendicular to it (perpendicular diffusion) including also the possibility for particles to jump field lines and spread in longitude or latitude perpendicular to the magnetic field. Although a missing anisotropy seems to suggest a poor source connection, it can also be caused by the presence of parallel diffusion along the field line, and it is therefore not a proof for a poor magnetic connection. On the other hand, if significant anisotropies are observed, diffusive processes did not dominate and a magnetic connection to or near the particle injection region, is usually concluded \cite[e.g.,][]{Dresing2014, Gomez-Herrero2015, Lario2016}.

\begin{figure}[ht!]
    \centering
    \includegraphics[width=0.49\textwidth]{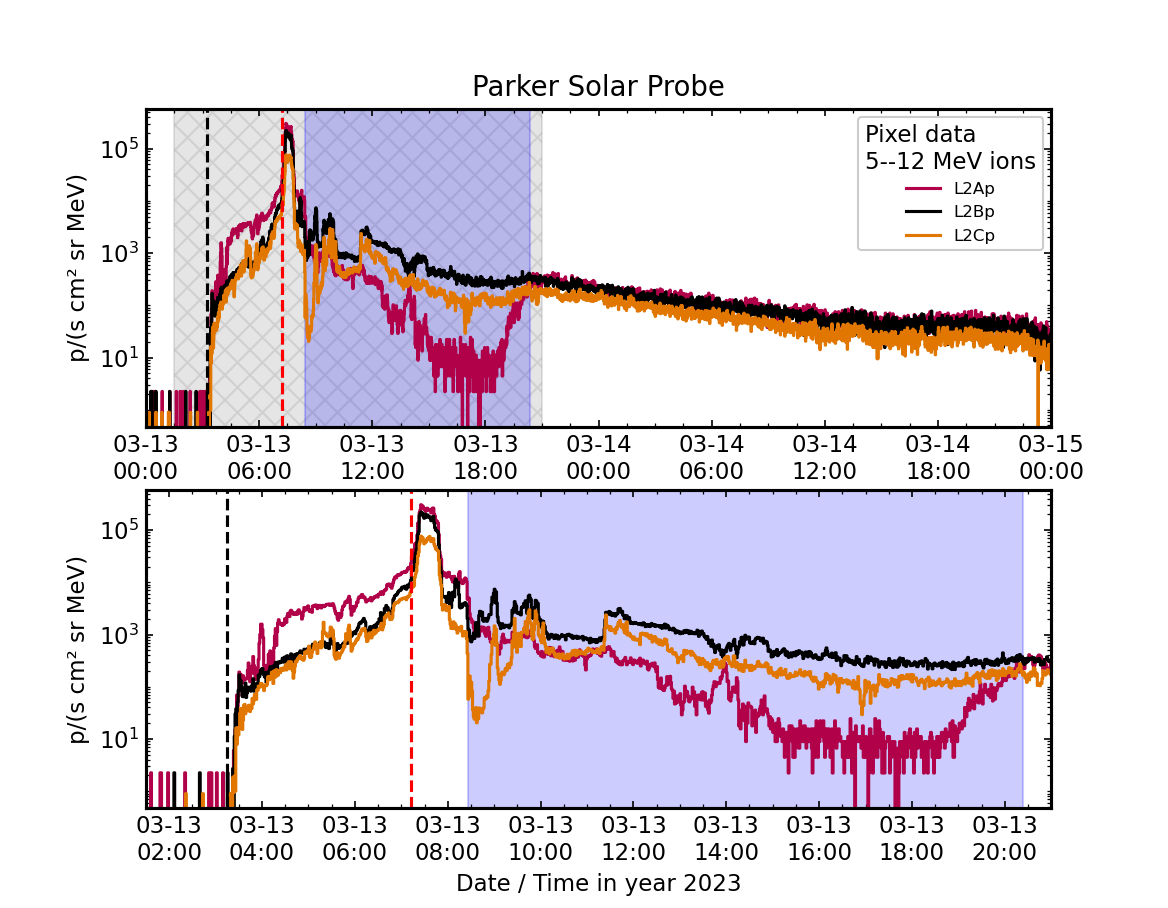}
    \caption{Pixel measurements of ${\sim}8$ MeV protons as detected by the three viewing directions of EPI-Hi/LET. The top figure shows the whole event as seen by Parker, the lower panel zooms in around the anisotropic period during the early phase of the event (gray shaded area in the top panel). The blue shaded region marks the CME ejecta passage as determined from solar wind magnetic field and plasma observations as well as electron PADs (shown in Fig.~\ref{fig:in-situ1}, left). The two dashed lines mark the onset time of the solar eruption (black) and the time of the in-situ shock arrival (red), respectively.}
    \label{fig:PSP_ani}
\end{figure}

\begin{figure*}[ht!]
    \centering
    \includegraphics[width=0.49\textwidth]{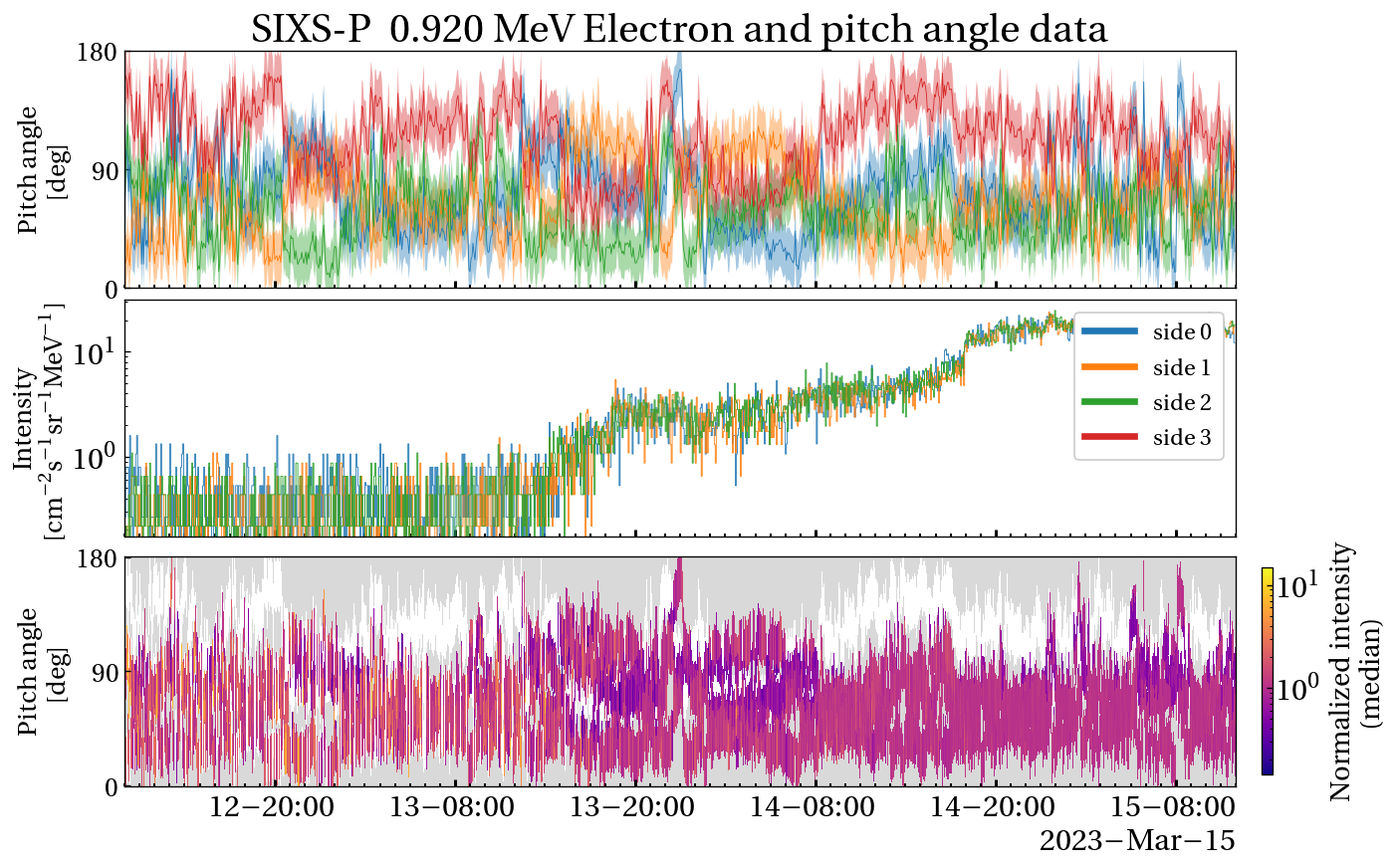}
    \includegraphics[width=0.49\textwidth]
    {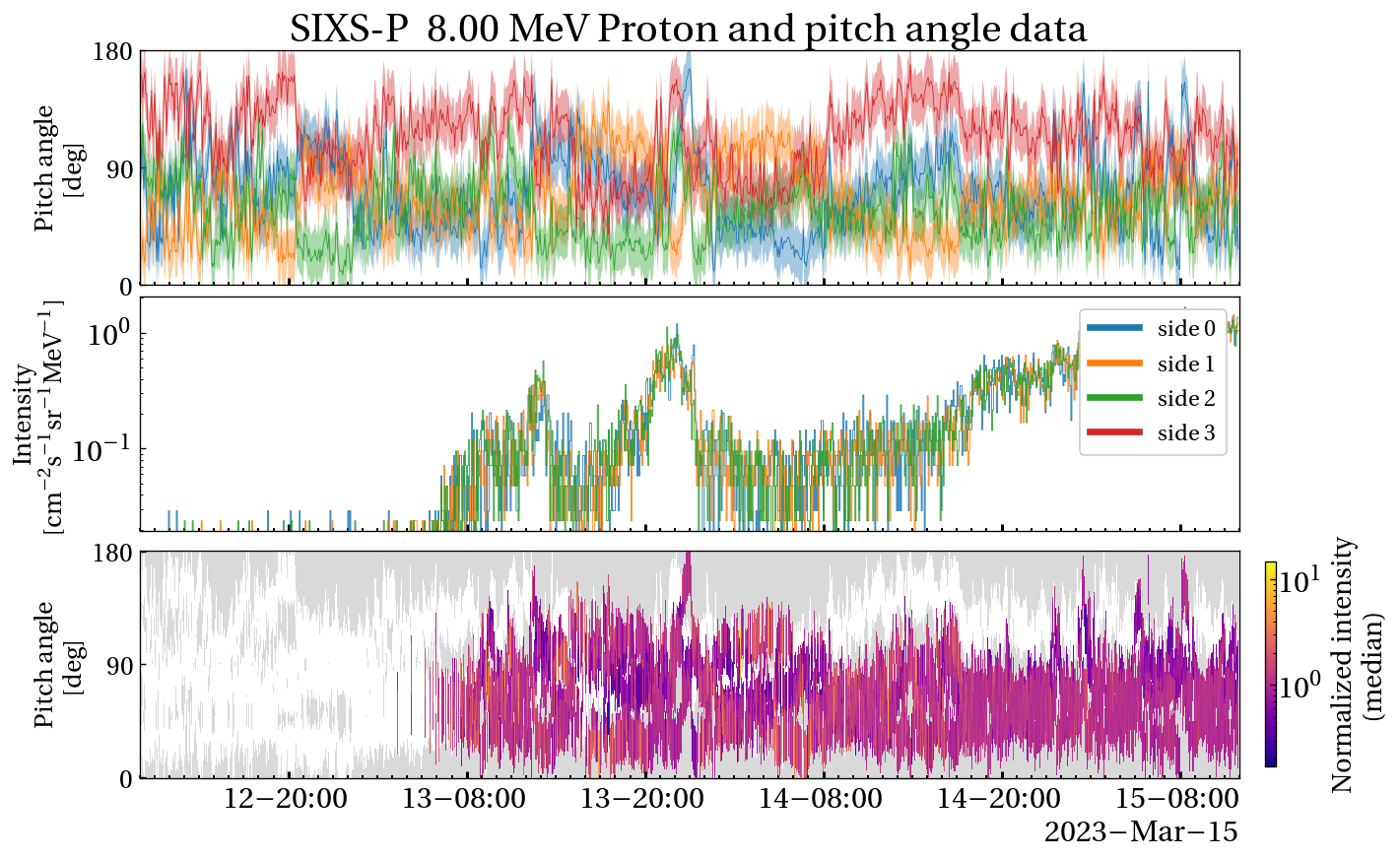}
    \caption{PADs of 0.9 MeV electrons (left) and 8 MeV protons (right) as observed by BepiColombo/SIXS. Top: pitch angles covered by the different viewing directions ('sides') of the instrument. Center: Sectored intensities of the different sides (excluding side 3 due to strong noise), bottom: PAD with normalized intensities color coded.}
    \label{fig:sixs_pad_protons}
\end{figure*}

\begin{figure*}[ht!]
    \centering
    \includegraphics[width=0.49\textwidth]{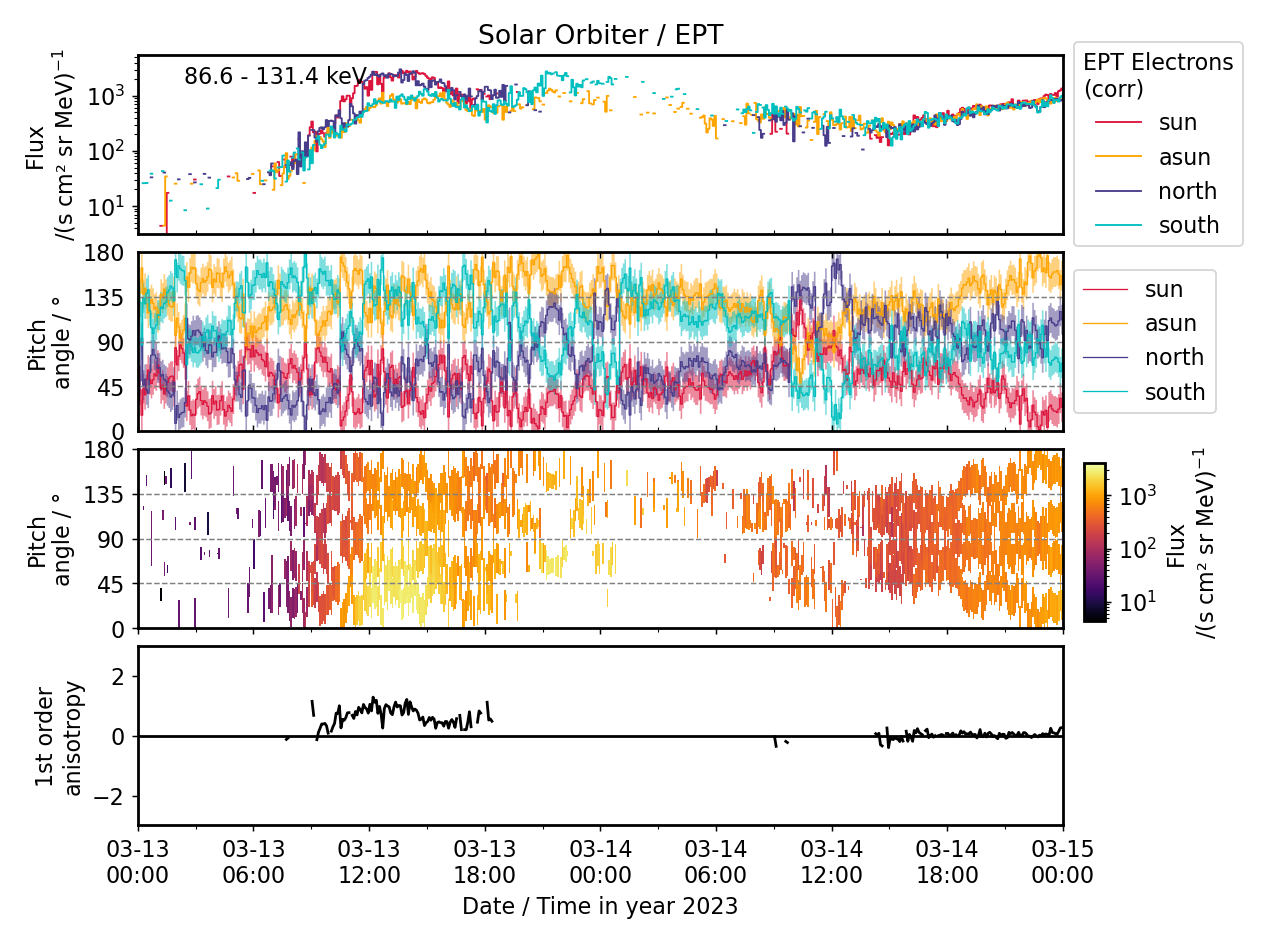}
    \includegraphics[width=0.49\textwidth]{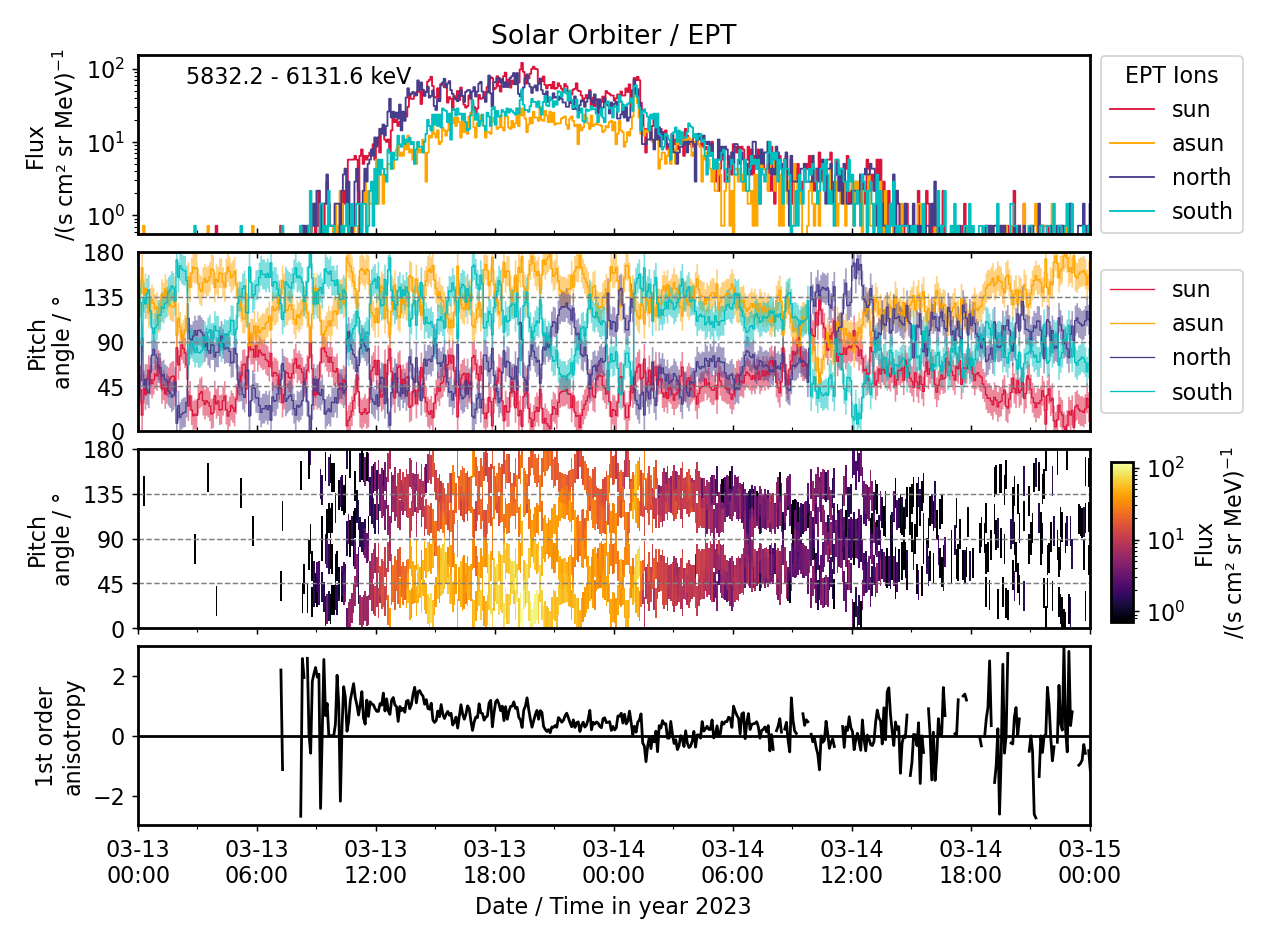}
    \caption{Energetic particle PADs as observed by Solar Orbiter's EPD/EPT. Left: $\sim100$keV electrons, right: $\sim6$MeV ions. From top to bottom: Sectored intensities as measured by EPT's four different viewing directions, pitch-angles covered by those viewing directions taking into account the opening angle of the telescopes, color-coded intensity PAD, and first-order anisotropy.}
    \label{fig:solo_pad}
\end{figure*}

\begin{figure*}[ht!]
    \centering
    \includegraphics[width=0.49\textwidth]{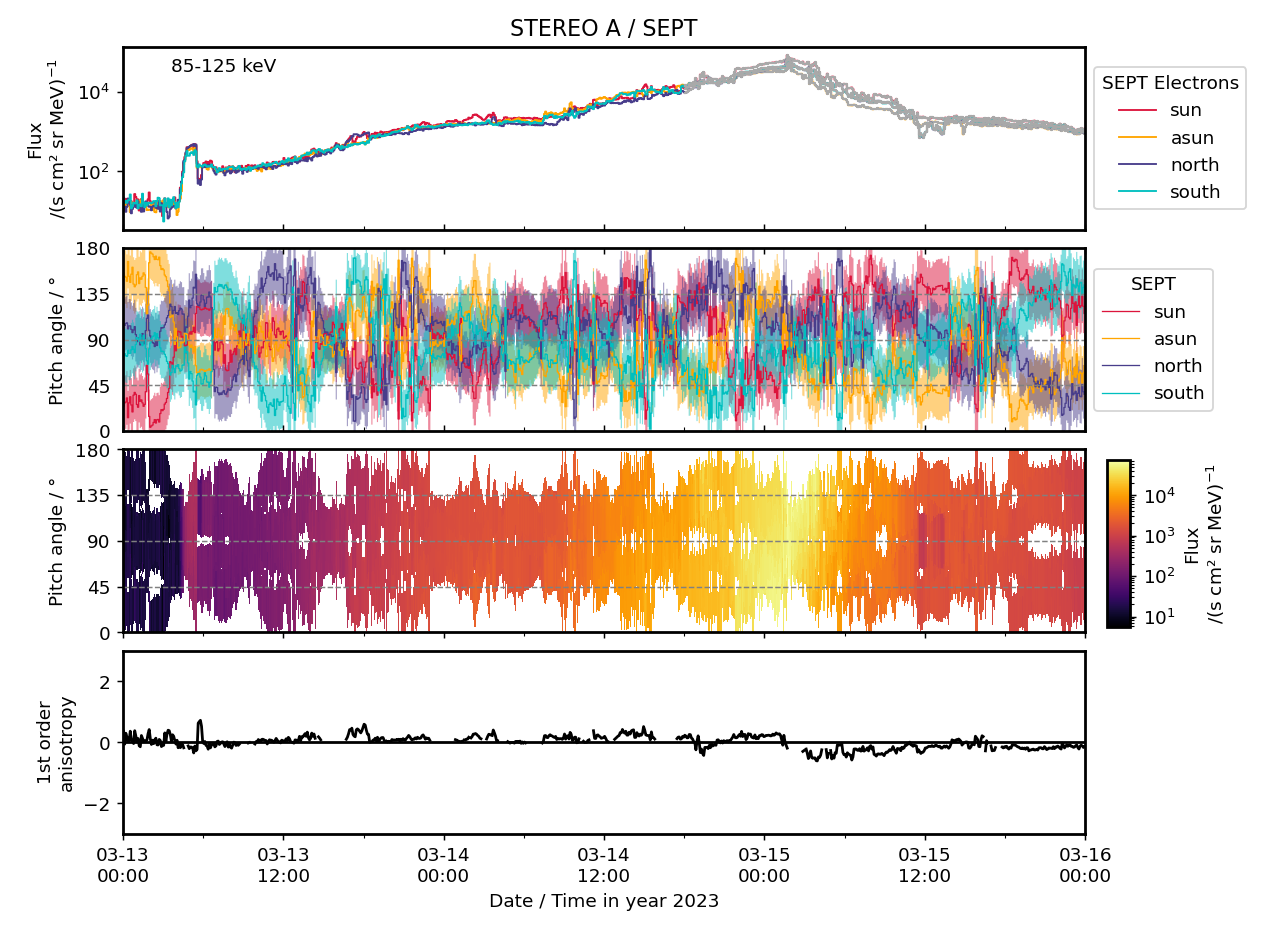}
    \includegraphics[width=0.49\textwidth]{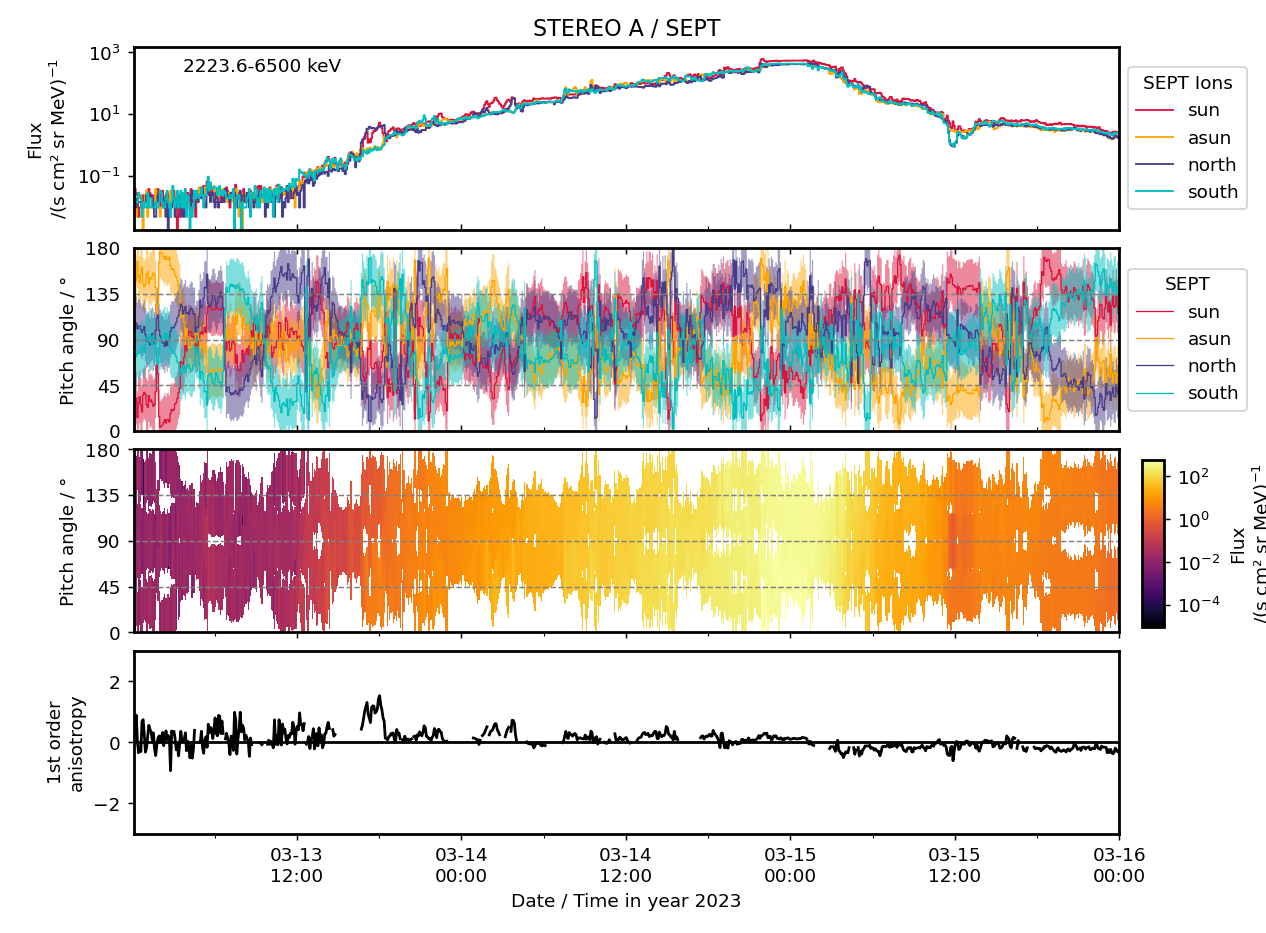}
    \caption{Energetic particle PADs as observed by STEREO A / SEPT in a similar format as in Fig.~\ref{fig:solo_pad}. Left: $\sim100$keV electrons (the gray shading of the intensity-time series in the top panel marks a period when the electron intensities are contaminated by ions and can therefore not be trusted.). Right: 2--6 MeV ions.}
    \label{fig:stereo_pad}
\end{figure*}

\begin{figure*}[ht!]
    \centering
    \includegraphics[width=0.49\textwidth]{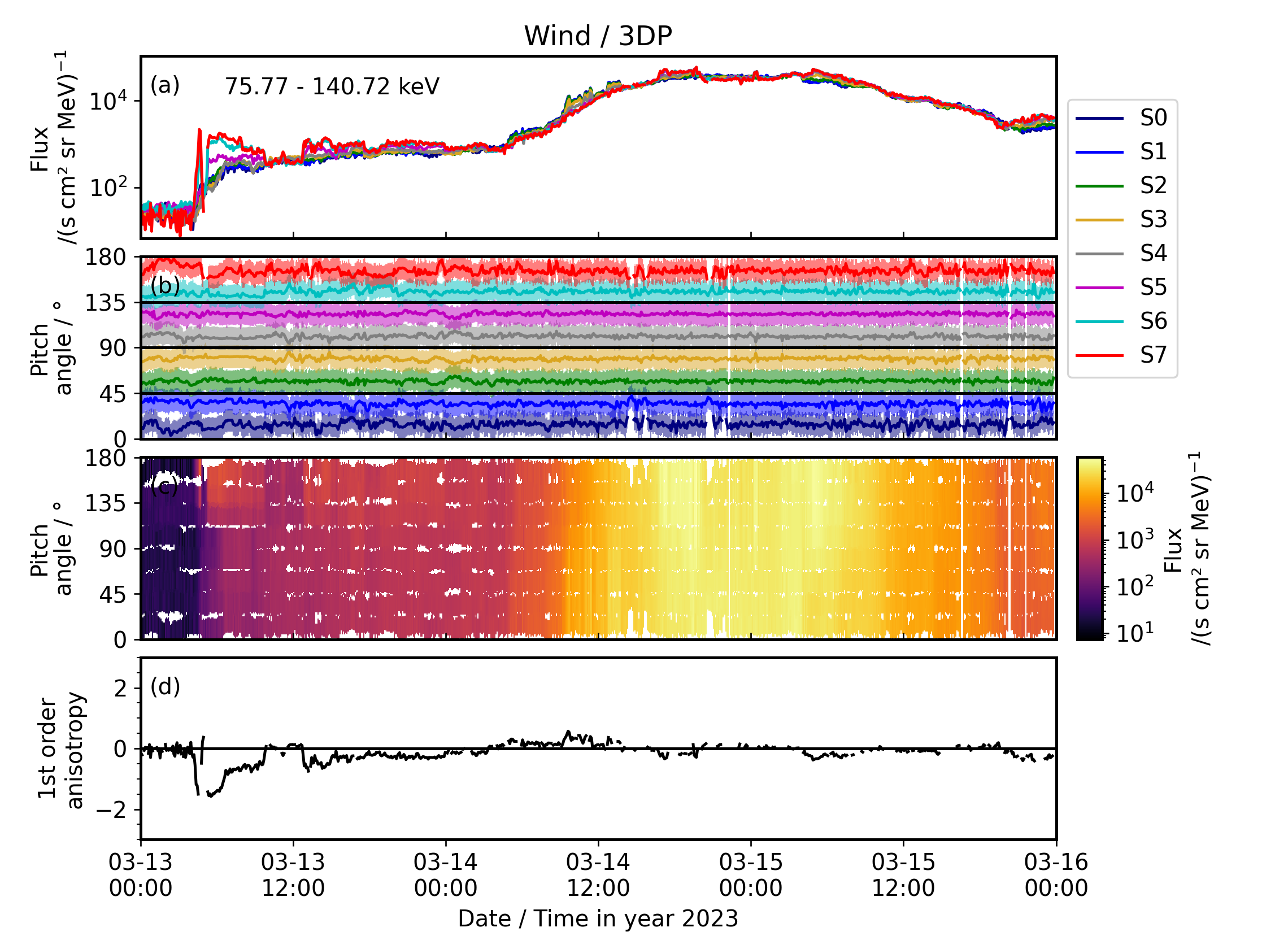}
    \includegraphics[width=0.49\textwidth]{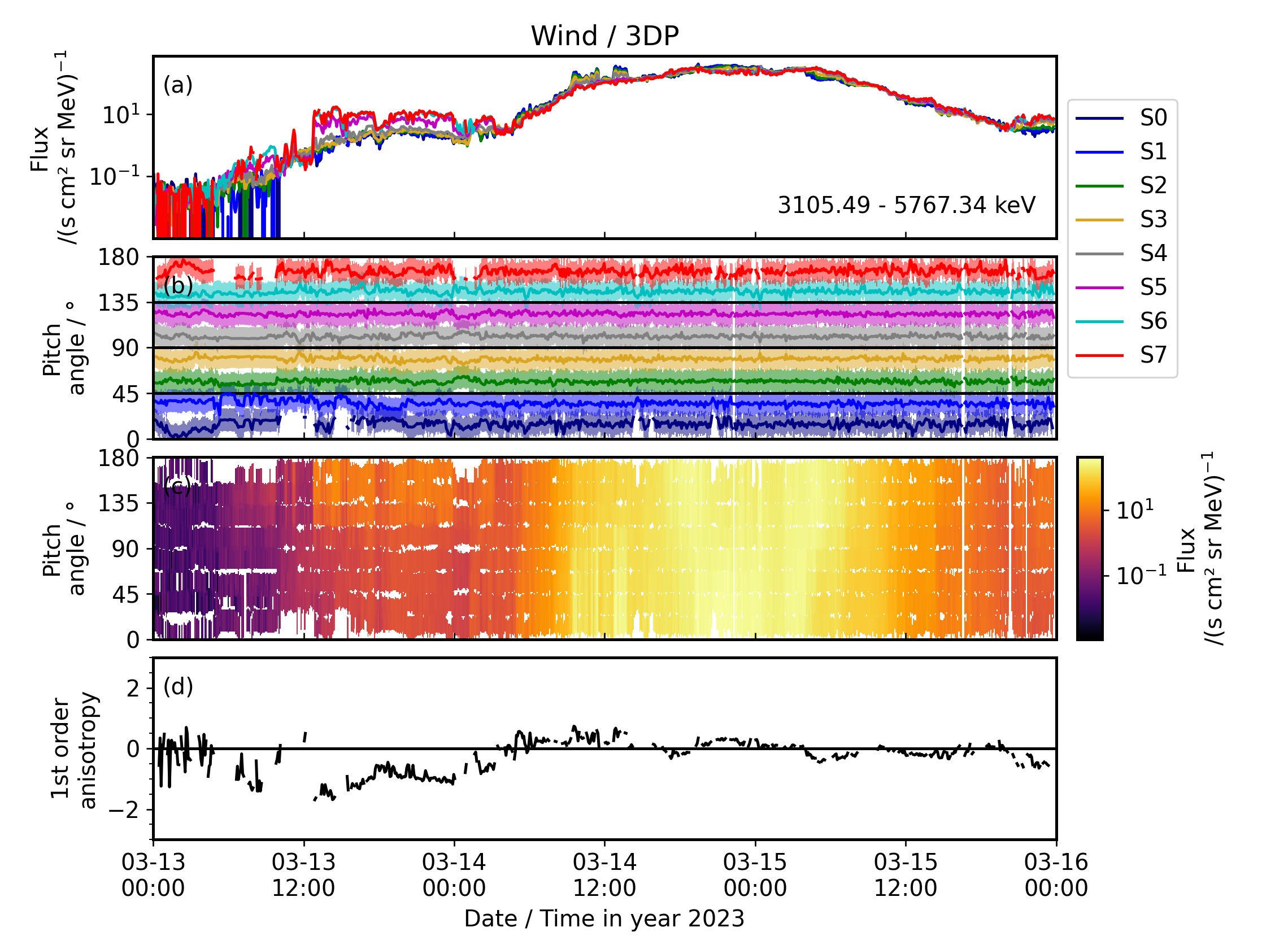}
    \caption{Energetic particle PAD plot as observed by Wind / 3DP in the same format as Fig.~\ref{fig:solo_pad}, but providing eight sectors, for $\sim100$ keV electrons (left) and 3--6 MeV protons (right).}
    \label{fig:wind_pad}
\end{figure*}

\begin{figure}[ht!]
    \centering
    \includegraphics[width=0.5\textwidth]{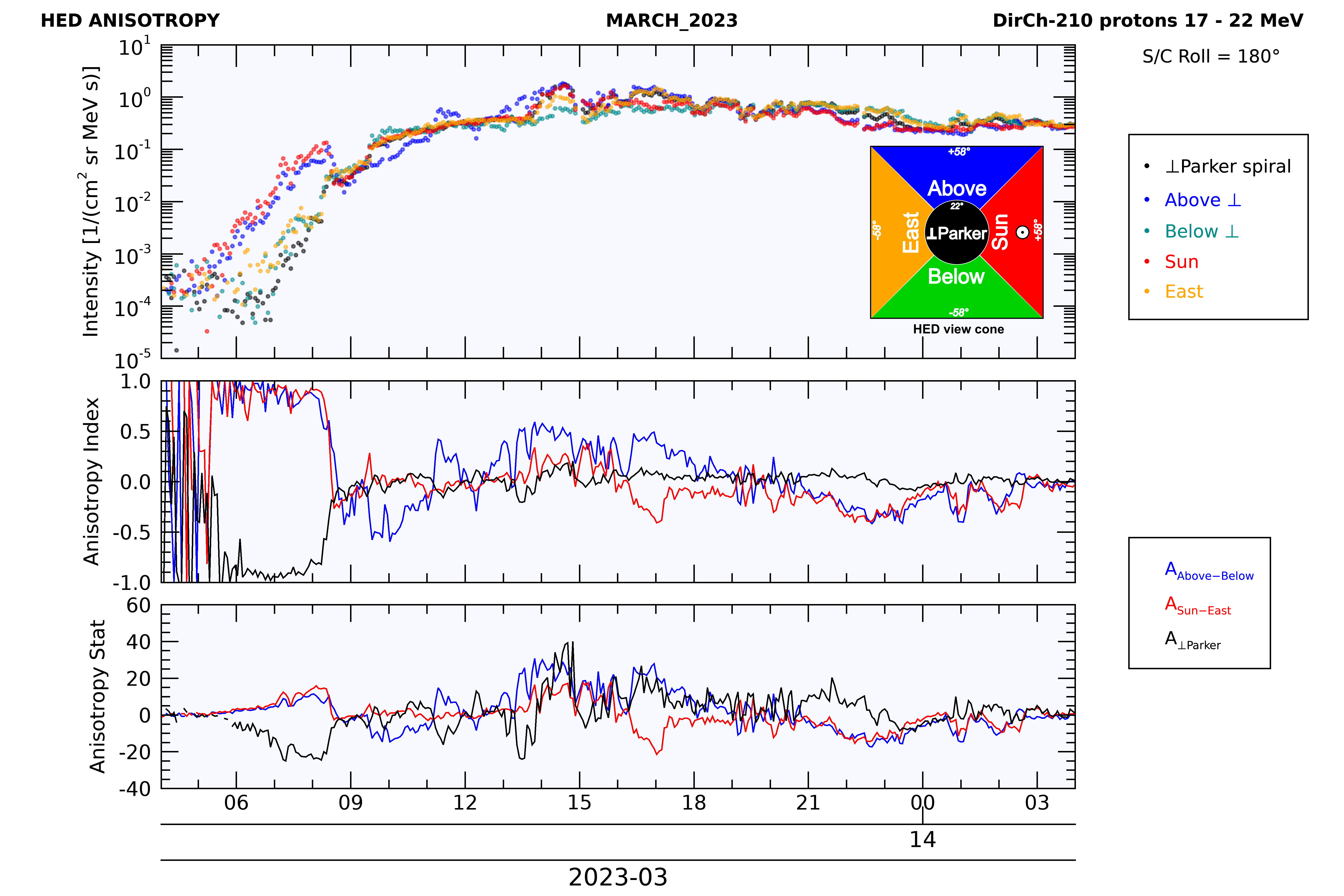}
    \caption{ERNE/HED directional intensities and corresponding anisotropy indices. The uppermost panel displays the intensity time evolution in the five selected direction sectors. The sector directions and the direction of the Sun in the square aperture of HED are depicted in the insert at the right-hand corner of the panel. During this time, the HED view cone axis was perpendicular to the nominal IMF direction (Parker spiral). The half angle of the squared aperture is 58\textdegree and the half angle of the measurement central cone is 22\textdegree. The middle panel shows the values of three anisotropy indices. These are calculated based on the differences of measured intensities (a=Above, b=Below, e=East, s=Sun, z=Zenith): $A_\mathrm{Above-Below}=\frac{I_\mathrm{a}-I_\mathrm{b}}{I_\mathrm{a}+I_\mathrm{b}}$, $A_\mathrm{Sun-East}=\frac{I_\mathrm{s}-I_\mathrm{e}}{I_\mathrm{s}+I_\mathrm{e}}$ and $A_{\perp \mathrm{Parker}}=\frac{I_\mathrm{z}-\overline{I_\mathrm{abse}}}{I_{z}+\overline{I_\mathrm{abse}}}$ where $\overline{I_\mathrm{abse}}$ is the average intensity of the indicated direction bins. The lowest panel shows the indices weighed by the count statistics.}
    \label{fig:ani_soho}
\end{figure}

In Figs.~\ref{fig:PSP_ani}--\ref{fig:ani_soho} we show PADs and, where it was possible to determine those, also first-order anisotropies \citep[][]{Bruedern2022} as observed by the multiple S/C, in order to characterize the magnetic connections to the particle injection regions at the various observer locations. Due to the different instrumentation carried by the various space missions, it is hard to find perfectly matching energy channels, being restricted to those instruments that provide different viewing directions. We therefore select the energy channels in this section for each S/C separately, trying to provide similar energy channels, but taking also particle statistics into account as well as preferring energy channels with clearer anisotropy signatures. 

\underline{Parker:} Figure~\ref{fig:PSP_ani} shows $\sim8$ MeV ion intensities measured in the three different viewing directions of EPI-Hi/LET. LET-A and B point along the nominal Parker spiral towards and away from the Sun, respectively. LET-C is oriented perpendicular to LET-A and B. As described in Sect.~\ref{subsec:seps} EPI-Hi instruments went into DT3 due to the extremely high intensities during the event and the associated IP shock crossing. We can therefore only use the pixel data to analyze anisotropies, and make again the assumption that the ion measurements are dominated by protons. We also used a low energy channel of about 8~MeV, which suffers less from those high-energy particle contributions coming from the other side of the detector. 
The top panel of Fig.~\ref{fig:PSP_ani} shows the whole event observed by Parker and the bottom panel is a zoom-in around the first half of the event, where the arrivals of the shock and the following CME ejecta are marked by a red line and blue shading, respectively. An anisotropic intensity rise is observed until the shock arrival, with LET-A measuring higher intensities than LET-B and C. The protons show a strong shock-associated peak downstream of the shock. A strong intensity drop is observed around 08:00~UT, coincident with a transition to smoother magnetic field vectors compared to the immediately preceding portion of sheath region. At the time of the CME ejecta arrival (coincident with a rise in the magnetic field magnitude and the start of a period characterized by clear bidirectional electron pitch-angle distributions), another intensity drop is observed, which shows also a stronger depletion in LET-C.

\underline{BepiColombo:} Figure~\ref{fig:sixs_pad_protons} presents the PAD of 1MeV electron (left) and 8 MeV protons (right) as observed by SIXS. The top panels show the pitch angles covered by the four sides of the instrument that are available during the cruise phase. Unfortunately, side 3 became noisy so that only three sides can be included in the intensity-time series panels in the center and in the color-coded intensity-pitch-angle distribution shown in the bottom panels. Especially during the beginning of the proton event, the covered pitch-angle range is strongly limited to almost only half a hemisphere (pitch angles from $\sim0--90$°. However, BepiColombo is mostly situated in a magnetic field with outward polarity, which means that particles arriving from the Sun direction are expected to be observed at small pitch angles, which is the hemisphere covered by the instrument. Nonetheless, no anisotropy is observed neither for electrons nor for protons, which suggests that BepiColombo was not directly connected to the SEP injection region. This is further supported by the very gradual rise of the event. Furthermore, as discussed in Sect.~\ref{subsec:seps}, the electron event observed by SIXS (see Fig.~\ref{fig:in-situ1}, right) is even more gradual and arrives delayed with respect to the protons. The depression of proton intensities around 16 UT on 13 March is associated with the arrival of the shock and CME.

\underline{Solar Orbiter:} Figure~\ref{fig:solo_pad} shows PADs and first-order anisotropies for $\sim100$keV electrons (left) and $\sim6$MeV ions (right) as observed by EPD/EPT. The pitch-angle coverage is overall good throughout the event, however, electron measurements suffer from strong ion contamination, especially during the later phase of the event. We applied an ion contamination correction as described in \citet{Jebaraj2023b} resulting in the large data gaps in Fig.~\ref{fig:solo_pad} (left), where no reliable electron measurement is possible. 
While the rise phases of the proton and electron events are very gradual small but significant anisotropies are observed especially during the peak phase of the event. The phase of significant anisotropy lasts for more than half a day for ions and at least for about six hours for electrons. This suggests that a magnetic connection to or near the SEP injection region was established at least during the later phase of the event. Furthermore, the long-lasting anisotropies mark ongoing injections of particles. most likely realized through an ongoing particle acceleration by a shock.

\underline{STEREO~A:} The pitch-angle coverage of STEREO/SEPT throughout the event (Fig.~\ref{fig:stereo_pad}) is more limited as for Solar Orbiter (Fig.~\ref{fig:solo_pad}) and Wind (Fig.~\ref{fig:wind_pad}) with pitch angles along the magnetic field (0 and 180) not being covered most of the time. However, even during short periods of better pitch-angle coverage no large anisotropies are observed, suggesting that no direct connection to the source location is present. A few slightly anisotropic periods are seen in the $\sim100$keV electron and 2--6 MeV ion PADs shown in Fig.~\ref{fig:stereo_pad}. However, these do not correspond to periods of improved pitch-angle coverage but occur rather sporadically throughout the long-lasting gradual rise phase. This suggests that the S/C encounters from time to time flux tubes, which have a better magnetic connection to the SEP injection region, potentially caused by field line random walk processes \cite[e.g.;][]{Laitinen2016}, which are also known to lead to flux dropouts in case of encountering flux tubes that do not connect to the source region \cite[e.g.,][]{Mazur2000}. We note, however, that the larger anisotropies observed within these flux tubes could also be related with less diffuse transport conditions than in the surrounding magnetic field.
At the very beginning of the electron event, a rather sharp rise of intensities is observed, which is cut off after about an hour when a magnetic structure passes the S/C (Fig.~\ref{fig:in-situ2}, right). This initial sharp rise is also observed in the high energy (60--10 MeV) protons (Fig.~\ref{fig:in-situ2}, right), which arrive at almost the same time like the electrons. Only a small anisotropy is observed. However, due to the non-complete pitch-angle coverage of SEPT at that time the real anisotropy could be underestimated. The reason for this spike-like feature could be an initially good magnetically connection with or near to a region exhibiting  a larger particle injection efficiency than during the rest of the event when the rise is very gradual. This connection is likely disrupted at the time when the magnetic structure passes the S/C.

\underline{Wind and SOHO:} At Wind we see a similar spike-like pattern during the beginning of the electron event, which shows however strong anisotropies. Figure~\ref{fig:wind_pad} shows $\sim100$keV electrons (left) and 3-6 MeV protons (right). Also the protons show significant, medium strong, anisotropies which are also long-lasting (almost two days). Although Wind is nominally situated at a large separation angle with respect to the associated solar source region (see Sect.~\ref{subsec:connectivity} for a discussion of the S/C magnetic connections to the Sun), the S/C observes stronger anisotropies than Solar Orbiter. This suggests that a good magnetic connection to the source region is established from the beginning of the event onwards. Furthermore, as for Solar Orbiter, the long-lasting anisotropies suggest a long-lasting particle injection, likely from a CME-driven shock.
Figure~\ref{fig:ani_soho} shows an estimation of 17--22 MeV proton anisotropies based on SOHO/ERNE observations. Although the instrument does not provide different viewing directions, it is possible to utilize the sectorization of the detector to evaluate if the particle distribution is anisotropic within the view cone of the instrument.
The measurements show a clear and strong anisotropy from the beginning of the event around 5:30~UT until at least 9:00 UT. The SOHO/ERNE observations therefore suggest the scenario of a good magnetic connection to the source region right from the beginning of the event, which also lasts for several hours even for $\sim20$MeV protons.


\section{SEP timing analysis of the first arriving particles} \label{app:timing}

\begin{figure*}[ht!]
    \centering
    \includegraphics[width=0.9\textwidth]{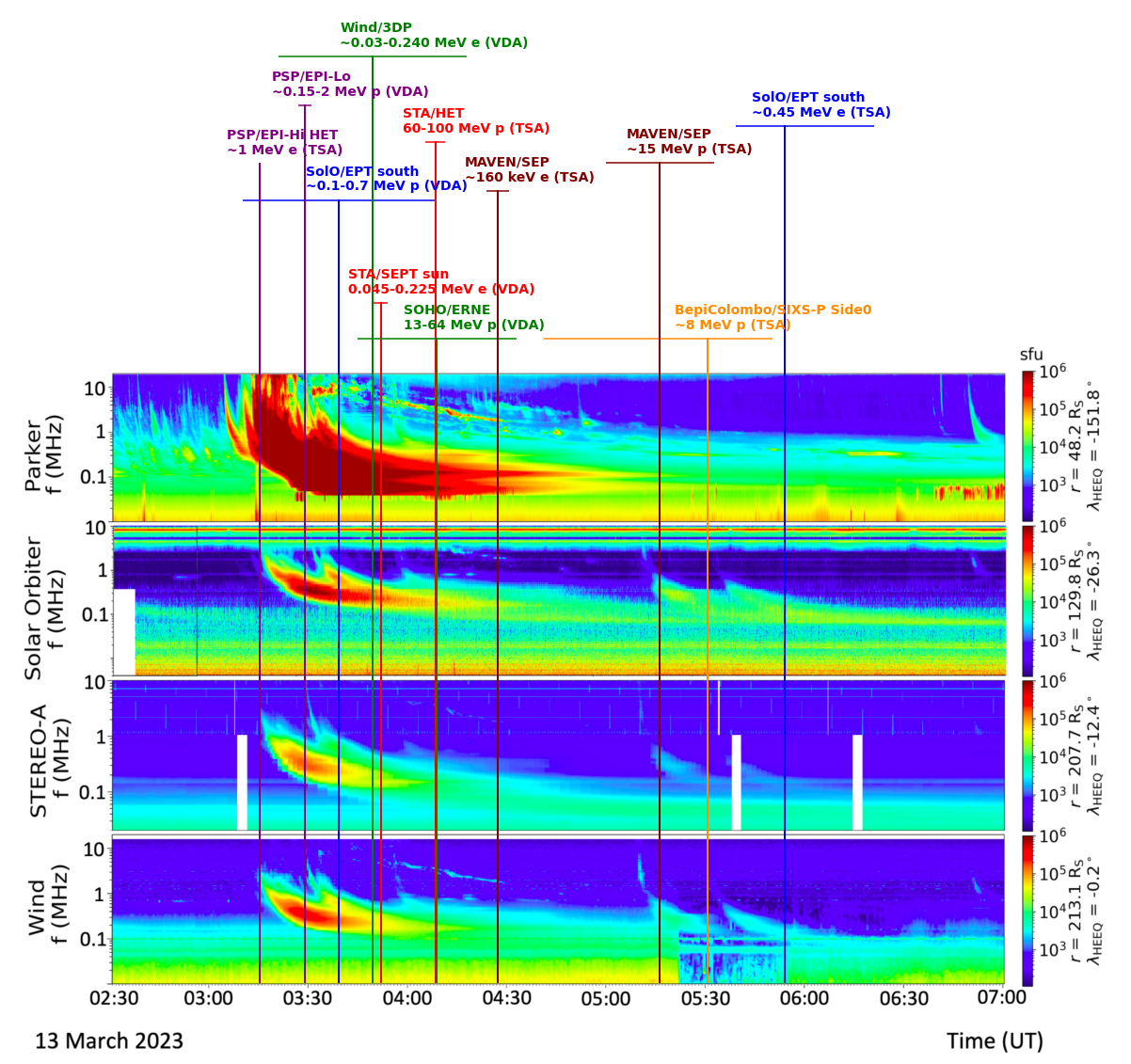}
    \caption{Inferred SEP injection times (vertical lines) including uncertainties (horizontal bars) plotted over dynamic radio spectrograms observed, from top to bottom, by Parker, Solar Orbiter, STEREO~A, and Wind. The radio data have been corrected for light travel time and are therefor shifted back to the Sun. The annotations on top of each vertical line specify the S/C, instrument, species and energy range used to infer the injection time as well as the method (VDA or TSA) for Parker (PSP, purple), Solar Orbiter (SolO, blue), STEREO~A (STA, red), BepiColombo (yellow), Wind (green), and Maven (brown).}
    \label{fig:time_line}
\end{figure*}

While this paper presents two possible drivers of the ESP event observed by inner-heliospheric S/C all around the Sun, this section analyses the SEP event, focusing on the first-arriving particles observed by the various inner-heliospheric observes.
Fig.~\ref{fig:time_line} presents a radio spectrogram as observed from top to bottom by Parker, Solar Orbiter, STEREO~A, and Wind during the time of the main eruption showing the associated type~III and type~II radio bursts. The vertical lines overlaid represent the inferred injection times of SEPs as detected by different S/C marked by different colors (see figure caption for details). In order to infer the SEP injection times, we use a Velocity Dispersion Analysis (VDA), which takes into account SEP onset times over a range of energy channels. However, often no clear velocity dispersion is observed and in those cases we perform a Time Shift Analysis (TSA) based on a single energy channel and assume a propagation path length along a nominal Parker spiral field line according to the measured solar wind speed. Onset times (except for Parker) were determined with the new hybrid Poisson-CUSUM method \citep[][]{Palmroos2024} using the PyOnset software package\footnote{https://github.com/Christian-Palmroos/PyOnset}.
As discussed already in Sect.~\ref{app:ani}, it can be challenging to find exactly matching energy channels among the various S/C. We therefore select these based on good particle statistics and a well-defined onset. In case multiple viewing directions are available, we use those, where the earliest onset times are observed. 

Figure \ref{fig:time_line} shows that most of the injection times, both for electrons and protons, were inferred to be in temporal coincidence with the main eruption as seen in radio observations. While the prompt injection times for Parker are expected due to its close proximity to the eruption site, those of the other S/C are surprising given their far longitudinal separations from the eruption sector. This allows the conclusion that the main eruption on 13 March 2023 created a prompt, circumsolar SEP event, which was detected by six well-separated observers, namely Parker, Solar Orbiter, BepiColombo, SOHO and Wind, STEREO~A, and MAVEN.

No clear velocity dispersion was observed for electrons by Parker, which is likely due to the very short travel time to Parker's small heliocentric distance and the very fast speeds of the electrons. The inferred injection time of ${\sim}1$~MeV electrons is at 03:14:29~$\pm$~00:00:10 UT, which is in agreement with the start time of the flare at 03:13~UT (cf. Sect.~\ref{sec:overview}). The $\sim$0.15--2 MeV proton VDA suggests a slightly later injection at 03:28$\pm$~00:02 UT, which could be due to the shock taking some time to form. This pattern of earlier injection times for electrons compared to protons is also observed by STEREO~A, MAVEN, and SOHO/Wind, however, the latter ones having so large uncertainties that an earlier electron injection cannot be proven. At STEREO~A a clear velocity dispersion was seen for near-relativistic electrons measured by SEPT. Applying a VDA fit to the 0.045--0.225 MeV electrons (leaving out a few outliers in the range) yields a path length $L = 2.122 \pm 0.138$~au and an injection time $t_{inj} =$~03:51~$\pm$~00:02 UT. Neither SEPT nor HET saw clear velocity dispersion for protons. TSA applied to the first arriving protons (detected by STEREO/HET at 60--100 MeV with an onset of 04:54 UT) and using the path length acquired from the electron VDA, yields an injection time of $t_{inj} =$~04:08~$\pm$~00:03 UT.

For ${\sim}161$~keV electrons reaching MAVEN at Mars, we infer an injection time of $t_{inj} =$~04:27~$\pm$~00:04 using TSA and a path length according to a nominal Parker spiral based on a solar wind speed of 350 km~s\(^{-1}\) (cf. Table~\ref{tab:coordinates}).
The uncertainty is the sum of the ${\sim}95$\% confidence interval for the ${\sim}161$~keV electrons onset time and the uncertainty related to the travel time, assuming $10\%$ uncertainty for the path length. TSA for 13--17.3 MeV protons with the same path length pointed to an injection time of $t_{inj} =$~05:16~$\pm$~00:16, with the uncertainty calculated in an identical fashion to electrons.

Near-Earth S/C observed a prompt event showing velocity dispersion both for electrons and protons, as well as anisotropic phases (cf. Sect.~\ref{app:ani}). For electrons in the range of 0.03--0.24 MeV observed with Wind/3DP, we infer an injection time of $t_{inj} =$~03:49~$\pm$~00:29 UT. For protons observed by SOHO/ERNE in the range between 13--64 MeV, we determine an inferred injection time of $t_{inj} =$~04:08~$\pm$~00:24 UT.

The timing pattern is different at Solar Orbiter and BepiColombo with earlier inferred proton injections compared to electrons. A VDA applied to 0.09--0.68 MeV protons observed by Solar Orbiter/EPT yielded a path length $L = 0.627 \pm 0.064$~au and injection time $t_{inj} =$~03:38~$\pm$~00:29 UT. As discussed in Sect.~\ref{sec:in_situ_obs} Solar Orbiter and BepiColombo observe an inverse velocity dispersion for protons, which limited the energy range usable for VDA. Electrons at Solar Orbiter did not show velocity dispersion. TSA applied to the onset time of the 0.44--0.47 MeV electrons, which were the first to reach the S/C and also had a clear onset in the time series data, and using the path length we acquired from protons, yields an injection time of $t_{inj} =$~05:53$^{+00:27}_{-00:15}$UT\footnote{We note that the hybrid Poisson-CUSUM method may yield asymmetric uncertainties}. 

As discussed in Sect.~\ref{sec:in_situ_obs} protons below 8 MeV observed by BepiColombo are likely related with an earlier eruption, and electron measurements below 1 MeV were strongly contaminated by protons. 
A TSA applied to the ${\sim}8$~MeV proton onset at $t_{0} =$~05:57 yielded an injection time $t_{inj} =$~05:30~$^{+00:20}_{-00:49}$ UT. For the path length we used a nominal Parker spiral path length based on solar wind speed of 400 km~s\(^{-1}\) as no solar wind measurements are available during BepiColombo's cruise phase. The uncertainty is a combination of the integration time used to find the onset (5 min) and an estimated $10 \%$ uncertainty related to the path length.
TSA applied to the first arriving electrons that were not contaminated by protons and had a clear onset in the time series data (${\sim}960$~keV) yields an injection time of $t_{inj} =$~15:06~$^{+00:52}_{-01:22}$UT. Both inferred injection times are rather late as compared with the eruption start time. Given, however, the rather isotropic fluxes observed by BepiColomobo/SIXS (cf.\ Sect.~\ref{app:ani}) suggest that particle diffusion was involved, which could account for the delayed particle arrivals.

\begin{figure*}[ht!]
    \centering
    \includegraphics[width=1\textwidth]{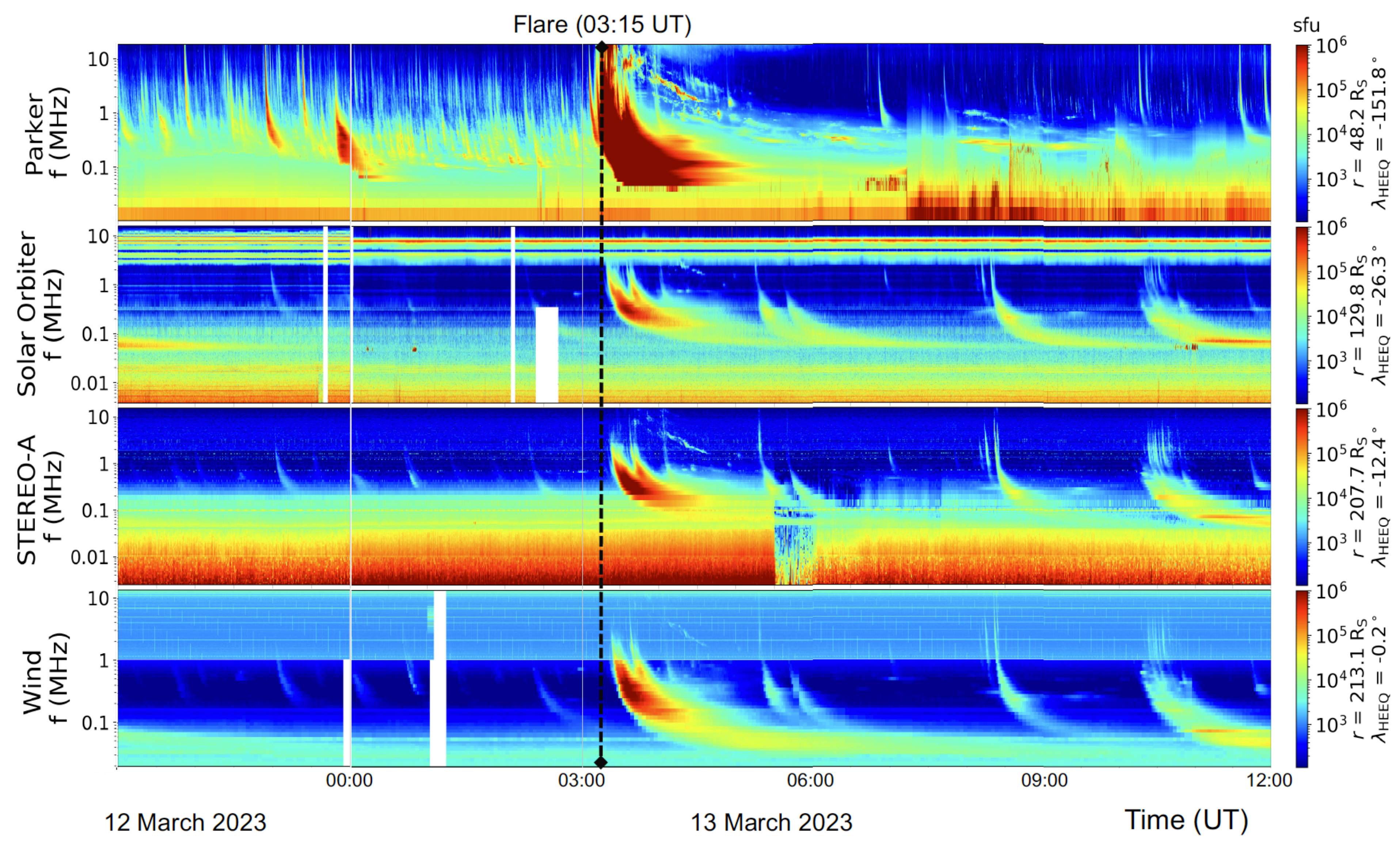}
    \caption{Dynamic radio spectrograms observed, from top to bottom, by Parker, Solar Orbiter, STEREO~A, and Wind. The vertical line marks the start time of the eruption. Similar to Fig. \ref{fig:time_line}, data have been corrected for light travel time and are therefore shifted back to the Sun.}
    \label{fig:time_line_full}
\end{figure*}

\section{CME reconstructions used as input for the EUHFORIA simulations}\label{app:cme_reconstructions}

The heliospheric conditions at the time of the particle release, in which the particles propagate, can significantly affect the SEP timing and intensity profiles. Moreover, the magnetic connectivity at the onset time can be relevant to the understanding of the SEP observations. The preconditioning of the heliosphere and the interaction of the IP structures that might be present at the onset time can actively influence this connectivity. We selected five pre-event CMEs and one \revise{post-event CME} to be included in the EUHFORIA simulation based on the following criteria: (1) CMEs erupting from 2023 March 10 to March 13, namely covering up to three days earlier than the main eruption related to the SEP event; (2) all CMEs directed to Earth whose projected speed were above 350 km s\textsuperscript{-1} based on the CDAW SOHO LASCO CME catalogue\footnote{\url{https://cdaw.gsfc.nasa.gov/CME_list/}} \citep{Yashiro2004}, and (3) all CMEs directed to any location whose projected speed were above 500 km s\textsuperscript{-1} and whose projected angular width was above 90$^{\circ}$ based on the CDAW catalogue.  The six selected CMEs are listed in Table \ref{table:GCS_fit}, where column 1 shows the date and time of the first appearance in the SOHO/LASCO/C2 field of view. 

Taking advantage of the multi-point view from STEREO~A and SOHO we performed the 3D reconstruction of the listed CMEs from $\sim$3.5 to $\sim$20 R\textsubscript{$\odot$} when possible. We fitted the CMEs with the widely-used graduated cylindrical shell model \citep[GCS;][]{Thernisien2006GCS,Thernisien2011}. The GCS model uses the geometry of what looks like a hollow croissant to fit a flux-rope structure using coronagraph images from multiple viewpoints. The tools used for the reconstruction are (1) the \textit{rtsccguicloud.pro} routine, available as part of the \textit{scraytrace} package in the SolarSoft IDL\footnote{\url{http://www.lmsal.com/solarsoft/}} library \citep{Freeland1998} and (2) \textit{PyThea}, a software package to reconstruct the 3D structure of CMEs and shock waves \citep{Kouloumvakos2022} written in Python and available online\footnote{\url{https://doi.org/10.5281/zenodo.5713659}\label{Pythea}}.

\section{Determination of in situ shock characteristics}\label{app:insitu_shock}

In situ shock arrivals were recorded at all inner-heliospheric S/C shown in Fig.~\ref{fig:solar-mach} (right). To establish a coherent picture of their origin, we assessed their in-situ characteristics. A list of the key shock parameters for each S/C is presented in Table~\ref{table:shock_params}. Likely the most important parameter is the shock normal, which was estimated using several different techniques depending on the available data \citep[][]{Paschmann00}. Since magnetic field data were available for all observers, the minimum variance analysis \citep[MVA;][]{Sonnerup98}, magnetic coplanarity theorem \citep[MCT;][]{Colburn66}, and general singular value decomposition \citep[SVD;][]{Golub13} were employed. When plasma data were available, additional methods (reduced Rankine-Hugoniot solutions) such as mixed mode \citep[MXM;][]{Paschmann00} and velocity coplanarity theorem \citep[VCT;][]{Abraham72} were also applied. It is worth noting that large errors are common due to the challenges of estimating shock parameters using single spacecraft data \citep[see][]{Paschmann00}. Such techniques assume a downstream steady state which is non-trivial to identify in data \citep[][]{Gedalin22,Gedalin22JPP}.

Because the shock normal parameter is highly variable and unstable, depending on data quality and the presence of small-scale changes in the field (AC components), we identified the mean field with \(\geq 2\sigma\) significance. For this purpose, we applied a continuous wavelet transform (Morlet) to remove the AC components and isolate the mean field. From the mean field data, the discontinuity was identified using a discrete Haar wavelet transform. For methods that consider vector fields upstream and downstream (MCT, VCT, and MXM), a window size of 20 minutes was used. The window sizes were chosen to ensure that the analysis methods considered only the steady-state upstream and downstream conditions, excluding gyro-phase bunched transmitted ions \citep[][]{Bale05,Gedalin22}.

The methods adopted here provide the most physically robust estimation of the shock parameters, considering that the Rankine-Hugoniot relations are steady-state solutions that relate only the flow across a discontinuity defined by changes in fluid parameters \citep[][]{Kennel88}. Errors in the time series data were estimated using the root mean square (rms) method and were propagated in a standard linear fashion when estimating other parameters.

Once \( \hat{\textbf{n}} \) is determined, we can estimate the shock angle, \( \theta_{Bn} \), which is the angle between the shock normal \( \hat{\textbf{n}} \) and the upstream magnetic field \( \mathbf{B}_{\mathrm{u}} \), using the following definition:

\[ \theta_{Bn} = \arccos \left( \frac{\mathbf{B}_{\mathrm{u}} \cdot \hat{\textbf{n}}}{\|\mathbf{B}_{\mathrm{u}}\| \|\hat{\textbf{n}}\|} \right) . \]

The shock speed is estimated using conservation of mass, and the Mach number (\(M_\mathrm{A}\)) is then estimated as the ratio between the speed of the upstream flow in the frame of the shock to the characteristic wave speed (\(v_\mathrm{A}\)) in a magnetized medium. In cases where the plasma measurements were not available (e.g. BepiColombo) we were not able to obtain reasonable estimations for \(M_\mathrm{A}\). Lastly, similar to \(M_\mathrm{A}\), the gas compression ratio requires us to know the upstream and downstream density. However, as a proxy, one may obtain the compression ratio for an MHD shock simply by knowing \(\theta_{Bn}\) through the Rankine Hugoniot relations. 

For instance, it is possible to determine the "expected" density compression and Mach number for the shock passing BepiColombo by using the MHD Rankine-Hugoniot relations. The exhaustive description of the Rankine-Hugoniot relations can be found in section 6.3 of \cite{Russell16}. Following their description, we start with the known quantities, namely, the geometry \(\theta_{Bn} = 41^\circ \pm 18^\circ\) and a magnetic compression ratio \(r_B = 1.4\). Since no detail regarding the change in the velocity along the \(\hat{\mathrm{\textbf{n}}}_\mathrm{RTN}\) is known, we must numerically solve this. We use the Rankine-Hugoniot solver made available by UCLA\footnote{\url{https://spacephysics.ucla.edu/MHDShocks/MHDShocksRHG.html}} and estimated the \(r_\mathrm{G}\) and \(M_\mathrm{A}\) of the shock at BepiColombo. Our results using plasma-\(\beta<1\) yielded \(r_\mathrm{G} \sim 1.5\pm0.15\) and \(M_\mathrm{A} \sim 1.4\pm0.1\).

\end{appendix}

\label{LastPage}

\end{document}